\documentclass[12pt]{article}
\usepackage[utf8]{inputenc}
\usepackage{graphicx}
\usepackage{pgf}
\usepackage{natbib}
\usepackage{color}
\usepackage{amsmath}
\usepackage{amssymb}
\usepackage{amsfonts}
\usepackage{hyperref}
\hypersetup{
    colorlinks=true,
    citecolor=blue,
    linkcolor=magenta,
    filecolor=magenta,      
        }
\usepackage[algo2e]{algorithm2e} 
\usepackage{algorithm}
\usepackage{verbatim}
\usepackage{caption}
\usepackage{subcaption}
\usepackage{enumitem}
\usepackage[left=1in,right=1in,top=1in,bottom=1in]{geometry}
\newcommand{\bc}{\boldsymbol c}

\newcommand{\btheta}{\boldsymbol\theta}

\title{Statistical inference for core-periphery structures}
\author{Eric Yanchenko\footnote{Global Connectivity Program, Akita International University, Akita, 010-1211, Japan. eyanchenko@aiu.ac.jp}, 
Srijan Sengupta\footnote{Department of Statistics, North Carolina State University, Raleigh, NC, 27606. ssengup2@ncsu.edu}, 
and Diganta Mukherjee\footnote{Sampling and Official Statistics Unit, Indian Statistical Institute, Kolkata, 700108, India. diganta@isical.ac.in}
}
\date{}

\begin{document}

\maketitle

\begin{abstract}
\noindent
Core-periphery (CP) structure is an important meso-scale network property where nodes group into a small, densely interconnected {core} and a sparse {periphery} whose members primarily connect to the core rather than to
each other.
While this structure has been observed in numerous real-world networks, there has been minimal statistical formalization of it. In this work, we develop a statistical framework for CP structures by introducing a model-agnostic and generalizable population parameter which quantifies the strength of a CP structure at the level of the data-generating mechanism. We study this parameter under four canonical random graph models and establish theoretical guarantees for label recovery, including exact label recovery. 
Next, we construct intersection tests for validating the presence and strength of a CP structure under multiple null models, and prove theoretical guarantees for type I error and power.  These tests provide a formal distinction between exogenous (or induced) and endogenous (or intrinsic) CP structure in heterogeneous networks, enabling a level of structural resolution that goes beyond merely detecting the presence of CP structure. The proposed methods show excellent performance on synthetic data, and our applications demonstrate that statistically significant CP structure is somewhat rare in real-world networks.
\end{abstract}


\setcounter{page}{1}

\clearpage

\section{Introduction}

The study of networks has exploded in popularity over the last few decades, thanks in part to their wide range of applications \citep{newman2010networks, sengupta2023statistical}. Across these various domains, one common feature that has been observed is {\it core-periphery (CP) structure} where the network can be broken down into two groups: a small, densely interconnected {\textit{core}} group and a sparse {\textit{periphery}} group whose members primarily connect to the core rather than to
each other \citep{BORGATTI2000, Csermely2013}. 
CP structures have been observed in global trade \citep{krugman1996self, magone2016core} and airport networks \citep{lordan2017analyzing, lordan2019core}, where nations with large economies and airline hubs comprise the core, respectively. 
CP structures are fundamentally hierarchical, balancing the benefits of centralization (e.g., efficient coordination) with the reach and scalability of large, diversified and distributed systems. 
Understanding and modeling such architectures is critical in domains where control, resilience and strategic redundancy are essential.
For example, if a major airline's hub airport went offline due to a storm, this would cause much greater disruption to flight schedules than if a peripheral airport was nonoperational.

There are numerous algorithms and metrics to identify CP structure in a network. The most widely used approach is the Borgatti–Everett (BE) metric \citep{BORGATTI2000}, which assumes a fixed binary template for ``ideal'' CP structure and maximizes similarity to this template.
Other methods include \cite{Rombach2017} and \cite{kojaku2017} which build off the BE metric, and \cite{Holme2005} who defines the core as the $k$-core with largest closeness centrality. 
These methods lack statistical inference, however, and produce purely \textit{descriptive} outputs: given any network, even one generated from a model with no underlying CP structure, they will still return a CP partition.
Without a rigorous statistical basis for probabilistic error bounds or significance testing of these outputs, it is impossible to distinguish true CP structure from artifacts of random noise, algorithmic uncertainty and/or exogenous factors. This makes the resulting labels potentially misleading and undermines their utility for downstream decision-making \citep{yanchenko2022coreperiphery}. 

Beyond just detection, there has been minimal formulation of CP structure from a statistical inference perspective. 
The works that do consider statistical properties tend to be fairly narrow in their scope, focusing on just one or two data-generating models \citep[e.g.,][]{Newmann2015, naik2021, gallagher2021clarified}. This means that the nuanced ways in which CP structures manifest in  many popular random graph models have not been explored. Similarly, to the knowledge of the authors, there are not any theoretical guarantees that the true CP labels will be recovered by the existing methods. Hypothesis tests for the statistical significance of a CP structure with theoretical guarantees of type-1 (false positive) and type-2 (false negative) errors are also rare. Existing approaches often rely on bootstrap procedures to compute empirical $p$-values \citep{BOYD2006, kojaku2017, elliott2020}, but these typically do not scale well to large networks. Moreover, these tests are generally limited to a single null model, usually the \citet{Erdos1959} (ER) model, which is overly simplistic. It has been shown in the context of other network properties like community structure \citep{yanchenko2021} and the small-world property \citep{lovekar2021testing} that many networks differ from the ER model without truly exhibiting the property of interest. That is, the ER null is so weak that nearly any real-world network will reject against it, regardless of true CP structure.  \citet{kojaku2017} further stress the need for null models that account for degree heterogeneity, which the ER model lacks.

In this work, we seek to address these gaps by presenting a principled and comprehensive taxonomy of CP structure from a statistical inference perspective. We begin by defining a model-agnostic population parameter that quantifies the strength of the CP structure in the data-generating process. The generality of this parameter allows statistically rigorous comparison of CP structures across networks of varying size, sparsity and heterogeneity. We focus on four canonical random graph models: the Erdős–Rényi (ER) model, stochastic block model (SBM), Chung–Lu (CL) model and degree-corrected stochastic block model (DCBM). We find that the ER model, by definition, lacks CP structure, while the SBM exhibits {\it endogenous} (or intrinsic) CP structure under certain parameter regimes. In contrast, the CL model displays only {\it exogenous} (or externally imposed) CP structure, which is induced entirely by degree heterogeneity. Finally, the DCBM exhibits both endogenous and exogenous CP structure which can interact in non-trivial ways to either reinforce or suppress the overall CP pattern. We then prove a unified result showing that the sample analogue of our proposed parameter achieves perfect recovery (strong consistency) of the true CP labels under the SBM, CL, and DCBM. These results establish CP structure on firm and rigorous statistical foundations, analogous to the treatment of community detection in \cite{bickel2009nonparametric}.

Building on this foundation, we develop intersection-based hypothesis tests to detect and validate CP structure against two null models: ER and CL. The ER null enables detection of \textit{any} statistically significant CP structure, while the CL null helps determine whether the observed CP structure \textit{exceeds} what could be explained by degree heterogeneity alone.
Through these tests, we formalize CP structure as an intersection criterion where \textit{both} conditions must be satisfied: core-core connectivity must exceed core–periphery connectivity, and core-periphery connectivity must exceed periphery-periphery connectivity.
This formulation helps distinguish true CP structure from other mesoscale patterns such as disassortative community structure. 
Together, these tests provide a simple and practical toolbox for analyzing real-world networks, offering a rich understanding of the underlying generative mechanisms and 
enabling a level of structural resolution that goes beyond merely detecting the presence of CP structure.
We prove theoretical guarantees for type I error and power for both tests. 
Finally, extensive simulations demonstrate fast and accurate identification of CP structures as well as highly precise and reliable test performance. A comprehensive analysis of thirteen real-world networks reveals a provocative finding: contrary to conventional wisdom about the ubiquity of CP structures, statistically significant CP structure appears to be relatively rare.

While this paper shares some conceptual similarities with \cite{yanchenko2021}, there are several key differences. Most importantly, the goal of this present work is broader, seeking a comprehensive taxonomy of CP structure across various random graph models, and developing an inference framework based on this taxonomy. Second, we prove the asymptotic recovery of the CP labels under three different scenarios. Finally, the hypothesis testing procedures diverge since this paper proposes an intersection test, requiring different proof techniques, and yields analytic cut-offs for multiple null models.

\section{Model parameter and label recovery}\label{sec:meth}

\subsection{Notation}
For this work,  we only consider simple, unweighted, undirected networks with no self-loops. 
Consider a network with $n$ nodes.
Let $A$ denote the $n\times n$ adjacency matrix where $A_{ij}=1$ if node $i$ and node $j$ have an edge, and 0 otherwise.
We write $A\sim P$ as shorthand for $A_{ij}\mid P_{ij}\stackrel{\text{ind.}}{\sim}\mathsf{Bernoulli}(P_{ij})$
for $1\leq i<j\leq n$. 
We define a {\it core-periphery assignment} to be a vector $\bc\in\{0,1\}^n$ such that $c_i=1$ if node $i$ is in the core and $0$ otherwise. Additionally, let \( k = \sum_{i=1}^n c_i \) denote the size of the core and \( \alpha_n = k/n \) its proportion.

\subsection{Model parameter}\label{sec:param}
To quantify the strength of the CP structure in the data-generating process, we define the parameter
\begin{equation}\label{eq:param}
    \rho(P,\bc) :=\frac{\sum_{i<j} \left(P_{ij} - \bar P\right)\Delta_{ij}}{\frac12n(n-1)
    \{\bar P(1-\bar P)\bar\Delta(1-\bar\Delta)\}^{1/2}},
\end{equation}
where $\Delta_{ij}(\bc)\equiv\Delta_{ij}=c_i + c_j - c_ic_j$, $\bar P = \frac{1}{{n \choose 2}} \sum_{i<j} P_{ij}$ is the mean edge probability and $\bar \Delta$ is the entry-wise mean of the upper triangular part of matrix $\Delta$. The numerator captures the expected number of (centered) core-core and core-periphery edges, which should be large for models with a strong CP structure. The denominator normalizes this quantity by network size, sparsity and core size, allowing for objective comparison across networks and generative models.
By construction, $0< \rho(P,\bc)\leq 1$ when the average core-core and core-periphery connection probabilities exceeds the global average.
In particular, if $P$ represents an ``idealized'' CP structure where $P_{ij}=1$ if $c_i=1$ or $c_j=1$, and 0 otherwise, then $\rho(P,\bc)=1$. 
Thus, $\rho(P,\bc)$ quantifies CP structure in a manner consistent with our intuition. Importantly, $\rho(P,\bc)$ is \textit{model-agnostic} since it is not tied to any specific class of models and applies to any edge-independent random graph model. This generality makes it a flexible foundation for a unified statistical framework.

\subsection{Statistical models and core-periphery taxonomy}
\label{sec:models}
While the CP parameter is universally applicable to any model $P$, it is instructive to understand its behavior under specific random graph models of interest. We consider the following four such models.

\begin{itemize}[leftmargin=*]
    \item Erd\"{o}s-R{\'e}nyi (ER) model: Here $P_{ij}=p$ for all $i,j$ \citep{Erdos1959}, which means that the ER model tautologically does not exhibit a CP structure \citep{BOYD2006, kojaku2017, elliott2020}. The metric defined in \eqref{eq:param} aligns with this intuition since $\rho(P,\bc)=0$ for all $\bc$ under the ER model (because in the numerator, $P_{ij} - \bar{P}$ is always zero).

    \item CP-SBM: The SBM has been thoroughly studied in the context of community structure \citep{holland1983stochastic}. 
    Since CP structure is a meso-scale property of networks,
    the SBM is also a natural model for CP structure \citep[e.g.,][]{Newmann2015, elliott2020}.
    We consider a modification of this model for CP structure, called CP-SBM, where $P_{ij} = \Omega_{c_i c_j}$ and $\Omega$ is a $2 \times 2$ symmetric blockmatrix such that
    \begin{equation}\label{eq:sbm}
    \Omega
    =
    \begin{pmatrix}
        p_{11} & p_{12}\\
        p_{21} & p_{22}
    \end{pmatrix}
    \text{ and }
    p_{11} > p_{12}=p_{21} > p_{22}.
\end{equation}
It is obvious that this model possesses inherent or \textit{endogenous} CP structure (with a slight abuse of notation where $c_i = 0$ is substituted by $c_i = 2$ for the periphery group).
To understand the behavior of $\rho(P, \bc) $ under the CP-SBM, note that $\Delta_{ij} = 1$ coincides with $P_{ij} = p_{11}$ or $p_{12}$, which makes the numerator positive since $p_{11} > p_{12} > p_{22}$.
More specifically, if (i) $\alpha_n = 0.5$ or (ii) $\alpha_n \rightarrow 0$, then (i) 
$
\rho(P, \bc) \sim
\frac{
(p_{11} - p_{22}) + 2(p_{12} - p_{22})
}{
4 \sqrt{3 \bar{P}(1 - \bar{P})}
}
$
and (ii)
$\rho(P, \bc) \sim 
\sqrt{2\alpha_n} \cdot \frac{p_{12} - p_{22}}{\sqrt{\bar{P}(1 - \bar{P})}}$.
In both cases, $\rho(P, \bc) > 0$ and increases with increasing values of $p_{11} - p_{22},$ i.e., core-core $-$ core-periphery connectivity; 
$p_{12} - p_{22}$, i.e., core-periphery $-$ periphery-periphery connectivity; and core size $\alpha_n$. These properties accord with our intuition of CP structure and further validate \eqref{eq:param} as a sensible CP model parameter.

 \item Chung-Lu (CL) model: This model allows for degree heterogeneity among nodes by introducing weight parameters $\btheta=(\theta_1,\dots,\theta_n)^\top$ and defining $P_{ij}=\theta_i\theta_j$ \citep{chung2002average, dasgupta2022scalable}. This model has no natural or endogenous core and periphery groups.     
    For a given core size $k$, however, we can rank the nodes as $\theta_{(1)}\geq\theta_{(2)}\geq\cdots\geq \theta_{(k)}\geq\cdots\theta_{(n)}$, and define $c_i = 1$ for $\theta_{(1)}, \cdots, \theta_{(k)}$ and  $c_i = 0$ for $\theta_{(k+1)}, \cdots, \theta_{(n)}$.
    Then $\rho(P,\bc)\sim \frac{\{\alpha_n^2(\bar\theta_c^2-\bar\theta^2) + \alpha_n(\bar\theta_c\bar\theta_p-\bar\theta^2)\}}{\{\bar\theta^2(1-\bar\theta^2)\bar\Delta(1-\bar\Delta)\}^{1/2}}$, where 
    \( \bar{\theta}_c \), \( \bar{\theta}_p \), and $\bar\theta$ are the mean 
\( \theta_i \) values in the core, periphery and overall network, respectively,
allowing self-loops for simplicity. If we let $\alpha_n\to0$, then the numerator becomes $\alpha_n(\bar\theta_c\bar\theta_p-\bar\theta^2)>0$ since $\bar\theta\sim \bar\theta_p$ and $\bar\theta_c>\bar\theta_p$. Thus, the CL model yields $\rho(P,\bc)>0$ which seems to indicate the presence of a CP structure, even though we said it is not endogenous. We call this {\it exogenous} (or externally induced) CP structure, in the sense that this is induced or imposed due to degree heterogeneity. A formal and statistically rigorous distinction between endogenous and exogenous CP structure is a central theme of this paper.

\item CP-DCBM: The degree corrected block model (DCBM) \citep{karrer2011stochastic} extends the SBM to allow for degree heterogeneity, i.e., $P_{ij}=\theta_i\Omega_{c_i,c_j}\theta_j$ where $\btheta$ is again a weight parameter vector of length $n$.
If $\Omega$ satisfies \eqref{eq:sbm}, then this model too possesses endogenous CP structure, so we call it CP-DCBM.
The CP-DCBM can be interpreted as a combination of the CP-SBM and the CL model.

In addition to endogenous CP structure, the CP-DCBM also possesses exogenous CP structure due to degree heterogeneity.
The strength of the overall CP structure depends on how the exogenous component interacts with the endogenous CP structure.
If core nodes have larger \( \theta_i \), the CP signal is amplified; if periphery nodes have larger \( \theta_i \), however, the exogenous component may weaken or even \emph{cancel out} the endogenous structure.
For example, if the core group vertices also have the highest $\theta_i$ values, then the highest values of $P_{ij} - \bar{P}$ correspond to $\Delta_{ij} = 1$, leading to a high value of $\rho(P, \bc)$. In this case, the exogenous component \textit{reinforces} the endogenous component and the CP structure is amplified. More generally, if $\alpha_n \rightarrow 0$, then the CP-DCBM yields
$\rho(P, \bc) \sim 
\sqrt{2\alpha_n} \cdot \frac{(\bar\theta_{1:k} p_{12} - 
\bar\theta_{k:n} p_{22})\bar\theta_{k:n}}{\sqrt{\bar{P}(1 - \bar{P})}}
$, 
where $\bar\theta_{1:k}=\frac1k\sum_{i=1}^k\theta_i$ and $\bar\theta_{k:n}=\frac1{n-k}\sum_{i=k+1}^n\theta_i$. If $\bar\theta_{1:k}p_{12}>\bar\theta_{k:n}p_{22}$, then $\rho(P,\bc)>0$. 
This condition implies, however, that simply having $p_{12}>p_{22}$ does not guarantee $\rho(P,\bc)>0$. If $p_{12}>p_{22}$ but $\btheta$ is such that $\bar\theta_{1:k}p_{12}<\bar\theta_{k:n}p_{22}$, then $\rho(P,\bc)<0$. This occurs because the exogenous component is \textit{misaligned}, suppressing the endogenous CP structure.
As a final example, if $p_{12} = \sqrt{p_{11} p_{22}}$ and $\theta_i = \sqrt{p/p_{11}}$ for the core nodes while $\theta_i = \sqrt{p/p_{22}}$ for the periphery nodes,
then $P_{ij} = p$ for all $i,j$ and this CP-DCBM is equivalent to an ER model ($\rho(P, \bc) = 0$).
In this case, the degree heterogeneity entirely cancels out the endogenous CP structure.

\end{itemize}

\noindent
To summarize, the ER model does not have a CP structure, while the CP-SBM endogenously possesses CP structure. The CL model exhibits an exogenous version of CP structure, while the CP-DCBM possesses both endogenous and exogenous CP structure which could interact in complex ways.

\begin{figure}[h]
    \centering
    \includegraphics[trim={4cm 4cm 0 4cm}, clip, width=0.5\linewidth]{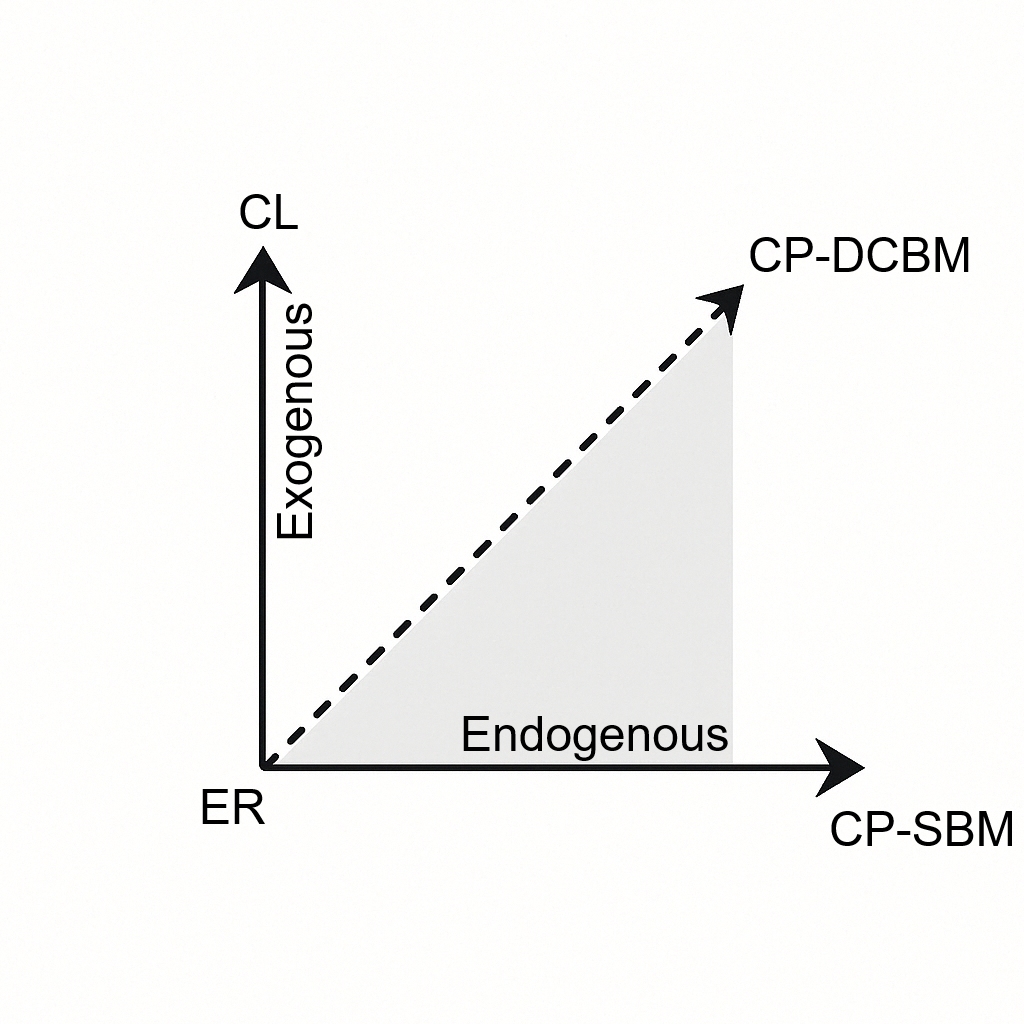}
    \caption{CP components and taxonomy: the $x$-axis represents \textit{endogenous} CP structure (i.e., intrinsic, inherent, or innate), and the $y$-axis represents \textit{exogenous} CP structure (i.e., externally induced or imposed), typically due to degree heterogeneity.
    The ER model lies at the origin, exhibiting neither endogenous nor exogenous CP structure.
    The CP-SBM has purely endogenous CP structure, whereas the CL model only exhibits exogenous CP structure.
    The CP-DCBM incorporates both endogenous and exogenous CP structure, but these two dimensions can interact in complex ways, potentially offsetting or reinforcing each other.
    The shaded region illustrates the regime under the CP-DCBM where endogenous structure dominates over exogenous effects. This scenario is rigorously formalized in Theorems 2.1, 3.1 and 3.2.}
    \label{fig:term1}
\end{figure}

Figure \ref{fig:term1} provides a visual representation of the  taxonomy of CP structure for these four models.
Of particular interest is the scenario under the CP-DCBM where the endogenous CP structure is not suppressed by degree heterogeneity.
This is the shaded region in Figure \ref{fig:term1} and will be a major focus for the theoretical results in Theorems 2.1, 3.1 and 3.2.
Across all models, the proposed metric \( \rho(P, \mathbf{c}) \) captures CP strength in a way that aligns with both theoretical intuition and empirical behavior. This interpretability and versatile nature of  $\rho(P, \mathbf{c})$ makes it an especially well-suited population parameter for statistical inference.


\subsection{Sample metric}
In practice, we do not observe $P$ or $\bc$, but only the network adjacency matrix, $A$.
Therefore, we need a sample version of the CP metric for statistical inference.
To this end, for some $\bc$, define
\begin{equation}\label{eq:stat1}
    T(A,\bc) := \frac{\sum_{i<j} (A_{ij} - \bar A) \Delta_{ij}}{\tfrac12n(n-1)\{\bar A(1-\bar A)\bar\Delta(1-\bar\Delta)\}^{1/2}}.
\end{equation}
Both the numerator and denominator of $T(A,\bc)$ serve as sample analogs of those in the population-level metric $\rho(P, \bc)$, making $T(A,\bc)$ a natural estimator of $\rho(P, \bc)$. We formalize the consistency of this estimator in Theorem 2.1. Furthermore, $T(A,\bc)$ coincides with the widely used BE metric proposed by \cite{BORGATTI2000}, thereby aligning our inference framework with established quantitative approaches in the network science literature.
In practice, the core-periphery assignment vector, $\bc$, is typically unknown, so we estimate it by maximizing $T(A,\bc)$, i.e., 
\begin{equation}
\label{eq:chat}
    \hat{\bc} = 
    \arg\max_{\bc}\{T(A,\bc)\}.
\end{equation}
We adopt the greedy algorithm from \cite{yanchenko2022divide} for the optimization in \eqref{eq:chat}. Please see the Supplemental Materials for further details (Algorithm 1).

\subsection{Consistent recovery of core-periphery labels}\label{sec:recover}
We are now ready to prove the first major theoretical contribution of this work: the sample metric in \eqref{eq:stat1} and \eqref{eq:chat} achieves consistent recovery of the CP labels. All technical details for this and the ensuing section can be found in the Supplemental Materials.

As per the convention in the literature \citep{bickel2009nonparametric, zhao12, sengupta2015spectral, senguptapabm,chakrabarty2023sonnet,bhadra2025scalable}, we introduce the parameter $\varrho_n$ to control the sparsity of the network, i.e., $A \sim P_n=\varrho_nP$ where $P$ does not depend on $n$. Additionally, we assume $k$ is known and $k/n:=\alpha_n \rightarrow 0$ as $n \rightarrow \infty$.
Our results can be extended to $\alpha_n=\mathcal O(1)$ but we focus on $\alpha_n\to0$ given that the empirical literature generally has a smaller core than periphery \citep[e.g.,][]{naik2021}. The consistency results also require the following sparsity assumption (A1) on the number of edges in the network and in the core: 
\begin{equation}
    \label{eq:A1}
    \text{\bf A1. } \alpha_n\to 0 \text{ and } n\varrho_n\alpha_n\to\infty \text{ as } n\to\infty. \hfill
\end{equation}
Let $\bc^*$ and $\hat{\bc}$ denote the true and estimated core-periphery assignment vector, respectively. By ``true'' CP labels, $\bc^*$, we mean the labels corresponding to the model descriptions in Section \ref{sec:models}, e.g., the true block model labels for the SBM. In the proof of the following theorem, however, we show that this is also equivalent to $\bc^*=\arg\max_{\bc}\{\rho(P,\bc)\}$. We quantify the estimation error as the fraction of misclassified nodes, i.e.,
\begin{equation}
    \xi_n(\hat{\bc})
    =\frac1n\sum_{i=1}^n I(\hat c_i\neq c_i^*).
\end{equation}
In the following theorem, we first characterize an upper bound for $\xi_n({\hat{\bc}})$ for each of the three models with endogenous and/or exogenous CP structure before further establishing that, under an additional assumption, the method achieves \textit{exact} recovery  or \textit{strong consistency}, i.e.,
$\mathbb P\left({\hat{\bc}} = \bc^* \right)\to1$.
To the knowledge of the authors, these are the first results establishing label recovery guarantees of any CP metric.\\

\noindent
{\sc Theorem 2.1.} {\it Let $P_n$ be a CP-SBM, CL or CP-DCBM model. Then under \eqref{eq:A1},}
\begin{equation}
\label{eq:th2.1.1}
    \mathbb P\left(\xi_n(\hat\bc)<\kappa(P)\alpha_n\sqrt{\frac{(8+\eta) \log(n\varrho_n\alpha_n)}{n\varrho_n}}\right)\to 1\text{ for any }\eta>0\text{ as }n\to\infty, \text{ where}
\end{equation}
{\it
\begin{itemize}
    \item If $P_n$ is a CP-SBM, then $\kappa(P)=\sqrt{p_{22}}/(p_{12}-p_{22})$.
    \item If $P_n$ is a CL model with $\bar\theta=\frac1n\sum_{i=1}^n\theta_n$; $\theta_{(1)}\geq\theta_{(2)}\geq\cdots\geq \theta_{(k)}\geq\cdots\theta_{(n)}$; the ``true'' CP labels are defined as $c_i^*=1$ for the $k$ vertices with largest $\theta_i$ values and $c_i^*=0$ for the remaining $(n-k)$ vertices; and under the assumption that $\theta_{(k)}\theta_{(n)}>\theta_{(k+1)}\theta_{(k+2)}$; then $\kappa(P)=\bar\theta/(\theta_{(k)}\theta_{(n)}-\theta_{(k+1)}\theta_{(k+2)})$.
    \item If $P_n$ is a CP-DCBM, let $\theta_c = \min\{\theta_i: c^*_i = 1 \}$ be the lowest degree parameter in the core set, and $\theta_{p, max} = \max\{\theta_i: c^*_i =0\}$ and $\theta_{p, min} = \min\{\theta_i: c^*_i =0\}$ be the highest and lowest degree parameters in the periphery set, respectively; and under the assumption that 
    $
         {p_{12}}{\theta_c \theta_{p, min}} > {\theta_{p, max}^2}{p_{22}};
    $
    then $\kappa(P)=\sqrt{\bar{p}}/(p_{12} \theta_c \theta_{p, min} - p_{22}\theta_{p, max}^2)$.
\end{itemize}
}
\noindent
{\it Furthermore, 
under the additional assumption that} $
{\alpha_n}/{\sqrt{\varrho_n}} = o \left({1}/{\sqrt{n \log n }} \right)
$, {\it the labels are strongly consistent for all three models, i.e.,} 
\begin{equation}
    \mathbb P\left({\hat{\bc}} = \bc^* \right)
    \rightarrow 1 \text{ as }  n\to\infty.
\end{equation}

\noindent
{\bf Remark 1.}
The error bound in \eqref{eq:th2.1.1} is $o(\alpha_n)$, which is stronger than the standard notion of \textit{weak} consistency where $\xi_n(\hat\bc) \rightarrow 0$ in probability \citep{zhao12}.
An error rate of $o(\alpha_n)$ ensures that the number of mislabeled nodes vanishes not only as a fraction of $n$, but also as a fraction of $k$, the core size.
In fact, since $\alpha_n\to0$, even the naive estimate $\hat \bc^N$ that assigns all nodes to the periphery (i.e., $\hat c_i^N=0$ for all $i$) would satisfy weak consistency, as $\xi_n(\hat {\bc}^N)=\alpha_n\to0$. Thus, the sharper error bounds in Theorem 2.1 are are crucial for meaningfully characterizing label recovery performance.\\

\noindent
{\bf Remark 2.}
Strong consistency requires that $
{\alpha_n}/{\sqrt{\varrho_n}} = o \left({1}/{\sqrt{n \log n }} \right)
$, which is equivalent to saying that $k = o \left( \sqrt{\frac{n \varrho_n}{\log n}} \right)$.
This has an intuitive interpretation that the maximum permissible size of the core for strong consistency depends on the network density.
The sparser the network, the lower the ceiling for which strong consistency is guaranteed.
When $\varrho_n = \mathcal{O}({1}/\log n)$, i.e., expected node degrees are almost linearly increasing, strong consistency is guaranteed as long as the core size is $k=o(\sqrt{n})$.
On the other hand, when $\varrho_n = \mathcal{O}({\log^2 n}/n )$, i.e., expected node degrees increase at the rate $\log^2 n$, the ceiling is only
$k=o(\sqrt{\log n})$. \\

\noindent
{\bf Remark 3.} 
We can also interpret Theorem 2.1 in light of the discussion in Section \ref{sec:models}.
The parameter $\kappa(P)$ can be interpreted as the \textit{inverse} signal-to-noise ratio (SNR):
the lower the value of $\kappa(P)$, the stronger the presence of CP structure in the network, and hence the lower the error.
The CP-SBM has endogenous CP structure, and as long as the core-periphery edge probability $p_{12}$ is higher than the periphery-periphery edge probability $p_{22}$, the proposed method successfully recovers the CP labels. As $p_{12}$ approaches $p_{22}$, however, the core-periphery edges start behaving more similarly to the periphery-periphery edges, and it becomes harder to distinguish the core from the periphery, yielding a larger error bound.

Next, recall that the CL model has exogenous CP structure due to degree heterogeneity.
Since $\theta_{(k)}\theta_{(n)}$ is the lowest core-periphery edge probability and $\theta_{(k+1)}\theta_{(k+2)}$ is the highest periphery-periphery edge probability, as long as $\theta_{(k)}\theta_{(n)} > \theta_{(k+1)}\theta_{(k+2)}$, then core-periphery connectivity is stronger than periphery-periphery connectivity, and lower values of  $\kappa(P)$ imply larger separation between the induced core and periphery, leading to more accurate label recovery.
Finally, under the CP-DCBM, degree heterogeneity can interact with the endogenous CP structure in complex ways.
The condition ${p_{12}}{\theta_c \theta_{p, min}} > {\theta_{p, max}^2}{p_{22}}$ ensures that the endogenous core is clearly separated from the periphery, i.e., degree heterogeneity does not dominate the underlying blockmodel structure.
This condition is the mathematical version of the shaded region in Figure \ref{fig:term1}.

\section{Hypothesis Testing}\label{sec:test}
In this section, we expand the statistical framework to hypothesis tests for CP structure.
From Section \ref{sec:models} and Figure \ref{fig:term1}, we know that the ER model has no CP structure, while the CP-SBM, CL, and CP-DCBM have endogenous and/or exogenous CP structure.
Our first test detects any such CP structure against the ER null.
Here, $H_0$ posits that $P$ is an ER model, while $H_1$ states that $P$ could be any of the three other models from Figure \ref{fig:term1}.

Next, recall that the CP-DCBM has endogenous CP structure which may be obfuscated by degree heterogeneity.
Thus, our second test characterizes this endogenous CP structure against the CL null, formalizing the shaded region of Figure \ref{fig:term1}. 
Specifically, $H_0$ corresponds to $P$ as a CL model, while $P$ is a CP-DCBM model with endogenous CP structure under $H_1$.
In other words, this tests whether the network has a greater CP structure than could be explained by exogenous features alone. 
Note that our framework does not include a CL vs CP-SBM test, since neither model is a special case of the other.



Our proposed hypothesis tests differ from existing tests \citep[e.g.,][]{boyd2010computing, kojaku2017, elliott2020} in at least three key ways. First, existing works focus on a single null model (ER or configuration \citep{newman2018networks}) while we propose a test for both the ER and CL null models. This gives our approach greater flexibility and yields a richer picture of the network structure. Second, we propose an intersection test based on the observation that CP structure requires two features to be present: a core-core edge density greater than that of core-periphery, and a core-periphery edge density greater than that of periphery-periphery. Without this two-part test, non-CP structures, e.g., disassortative community structure, may mistakenly reject the null hypothesis. Finally, existing tests use a bootstrap procedure whereas we derive an analytic cutoff, allowing for better scaling to large networks. Please see the Supplemental Materials for further details on the existing hypothesis tests.


\subsection{Erd\"{o}s-R{\'e}nyi null}\label{sec:test_er}
The ER model is perhaps the most natural choice for a null model and has been used in previous works \citep{kojaku2017, elliott2020}. For a network to have a CP structure, we claim that the core-core edge density must be greater than the core-periphery edge density as well as the core-periphery edge density being greater than that of the periphery-periphery. Because we must test these two conditions, we propose an intersection test. We use $T_1(A):=T(A,\hat \bc)=\max_{\bc}\{T(A,\bc)\}$, the maximized sample CP metric, and define
$$
T_2(A) = \hat{p}_{11} - \hat{p}_{12} = 
\frac{1}{{k \choose 2}} \sum_{i,j: \hat c_i = \hat c_j =1} A_{ij} -
\frac{1}{{k (n-k)}} \sum_{i,j: \hat c_i = 1, \hat c_j =0} A_{ij},
$$
i.e., the estimated difference between the probability of a core-core and core-periphery edge.
The intersection test is rejected when $T_1 > C_1$ \textit{and} $T_2 > C_2$, where $C_1$ and $C_2$ are defined in \eqref{eq:inter_er1}.
The following theorem shows that the CP-SBM, CL, and CP-DCBM all have a statistically significantly stronger CP structure than that of the ER model under specific assumptions.\\

\noindent
{\sc Theorem 3.1.} \textit{Under $H_0$, i.e., when the network is generated from the ER model, the Type I error rate converges to 0, i.e., for any $\eta>0$,}
$
    \lim_{n\to\infty} \mathbb P\left[\{T_1(A) > C_1\}\cap \{T_2(A)>C_2\}\right] < \eta,
$
\textit{where the rejection thresholds are given by}
\begin{equation}
C_1 = \sqrt{\frac{\log (n \varrho_n \alpha_n)}{n}} 
\text{ and } C_2 = \frac{2\sqrt{2} \varrho_n \log(n)}{\sqrt{k}}.
\label{eq:inter_er1}
\end{equation}
{\it Under the assumption that} 
$
\alpha_n \geq \frac{(\log (n \log n))^2}{n}
$,
\textit{ suppose that any of the following alternative hypotheses hold:}
\begin{itemize}
    \item $H_1:$ $P$ is a CP-SBM where 
$p_{11} > p_{12} = p_{21} > p_{22}$
\item $H_1:$ $P$ is a Chung-Lu model where 
$\theta_{(k)} \theta_{(n)} > \theta_{(k+1)}\theta_{(k+2)}$
and $\theta_{(k-1)}>\theta_{(k+1)}$.
\item $H_1:$ $P$ is a CP-DCBM where 
$
 \frac{p_{12}}{p_{22}} > \frac{\theta_{p, max}^2}{\theta_c \theta_{p, min}}
$
and $\frac{p_{11}}{p_{12}} > \frac{\theta_{p, max}}{\theta_c}$.

\end{itemize}
\textit{Under each of these alternatives, the power of the test goes to one, i.e., for any $\eta>0$,}
$$
    \lim_{n\to\infty}\mathbb P\left[\{T_1(A) > C_1\}\cap \{T_2(A)>C_2\}\right] > 1-\eta.
$$

\noindent
To our knowledge, this is the first hypothesis test for CP structure with an analytic cutoff, rather than relying on a bootstrap procedure, as well as the first to provide formal statistical guarantees.\\

\noindent
{\bf Remark 4.}
The proof relies on {large-deviation bounds}, {union bounds over labelings}, and {asymptotic control of misclassification rates} to establish the vanishing Type I error and asymptotic power of the test.
Under  $H_0$, we first prove that the probability of exceeding the threshold for test statistic $T_1(A)$ vanishes asymptotically by leveraging {Lemma 1} (from proof of Theorem 2.1), which establishes an upper bound on the deviation of $T(A, \hat{\bc})$ from the expected correlation measure $\rho(P_n, \hat{\bc})$. Since $\rho(P_n, \bc) = 0$ for all possible assignments under $H_0$, choosing  $C_1=\varepsilon_n$ ensures that $\mathbb{P}[T_1 > C_1] \to 0$. 
For $T_2(A)$, the proof utilizes {Bernstein’s inequality} to control the deviation of empirical edge probabilities from their expectations.

Under $H_1$, we first show that for  $T_1(A)$, we have $T(A, \hat{\bc}) \geq T(A, \bc^*)$, where $\bc^*$ is the true community structure. A lower bound on $\rho(P_n, \bc^*)$ is obtained by decomposing it into dominant terms, and using {Assumption A1}, it is shown that $\rho(P_n, \bc^*) > 2\varepsilon_n$. This guarantees that $\mathbb{P}[T_1 > C_1] \to 1$. Note that even though $\rho(P_n, \bc^*)$ is a correlation coefficient, because of the assumptions on the data-generating model, it goes to 0 instead of being $\mathcal O(1)$.
For $T_2(A)$, the proof carefully handles two sources of error: {statistical fluctuations} and {mislabeling errors}. The statistical fluctuations are controlled using Bernstein’s inequality as before, while the mislabeling errors are addressed using bounds on $\xi_n(\hat{\bc})$, the fraction of misclassified nodes. The analysis relies on {Theorem 2.1}, which ensures that $\xi_n(\hat{\bc}) = o(\alpha_n)$ with high probability. This ensures that the true difference in probabilities, $\varrho_n (p_{11} - p_{12})$, dominates the error terms, leading to $\mathbb{P}[T_2 > C_2] \to 1$.\\

\noindent
{\bf Remark 5.}
On a practical note, when implementing this test, since
 $\varrho_n$ is unobserved, we recommend using
$\hat{p} = \frac{1}{{n^2}} \sum_{i,j} A_{ij}$
as an empirical proxy. Additionally, $\log(n\varrho_n\alpha_n)=\log(k\varrho_n)$ and $C_1$ are only defined if $k\varrho_n>1$. If $k\varrho_n<1$, the test cannot be used as either the network is too sparse or the core is too small for the hypothesis testing procedure to be reliable.

\subsection{Chung-Lu null}

Using a CL null model, we again propose an intersection test using the same test statistics, $T_1(A)$ and $T_2(A)$, but modify the rejection threshold for $T_1(A)$.
Under the null, we estimate the degree parameters as
$
\hat {\theta}_i = {d_i}/{\sqrt{2m}},
$
where $d_i=\frac1n\sum_j A_{ij}$ is the degree of node $i$ and $m=\sum_{i<j}A_{ij}$ is the total number of edges.
We estimate the null probabilities as
$
\hat P_{ij} = \hat {\theta}_i \hat {\theta}_j = {d_i d_j}/({{2m}}).
$
The CL model is a special case of the CP-DCBM when $p_{11}=p_{12}=p_{22}$, so we test
\begin{equation}\label{eq:cl}
    H_0: \boldsymbol{\theta},\ p_{11}=p_{12}=p_{22}\text{ (CL model) vs. }H_1: \boldsymbol{\theta},\ p_{11}>p_{12}>p_{22} \text{ (CP-DCBM)}.
\end{equation}
The following theorem formalizes the statistical consistency of the test.\\


\noindent
{\sc Theorem 3.2.} \textit{Under $H_0$, i.e., when the network is generated from the CL model, the Type I error rate converges to 0, i.e., for any $\eta>0$,}
$
    \lim_{n\to\infty} \mathbb P\left[\{T_1(A) > C_1\}\cap \{T_2(A)>C_2\}\right] < \eta,
$
\textit{where the rejection thresholds are given by}
\begin{equation}
    C_1 = \rho(\hat{P}, \hat c) + \tilde \epsilon_n + \varepsilon'_n,
    \text{ where }
\tilde{\epsilon}_n = 
\frac{\sqrt{\alpha_n} \log (n \alpha_n)}{n^{1.5}\sqrt{\varrho_n}}
, \text{ and }
\varepsilon'_n =\frac{\sqrt{\log (n \alpha_n \varrho_n)}}{n}.    
    \label{eq:inter_cl}
\end{equation}
{\it Under $H_1$, i.e., when the network is generated from the CP-DCBM, 
consider the following three assumptions.
\begin{enumerate}
    \item $\varepsilon'_n + \tilde \epsilon_n = o ({\alpha_n^{1.5} \varrho_n^{0.5}})$,
where 
$
\tilde{\epsilon}_n = 
\frac{\sqrt{\alpha_n} \log (n \alpha_n)}{n^{1.5}\sqrt{\varrho_n}},
\varepsilon'_n =\frac{\sqrt{\log (n \alpha_n \varrho_n)}}{n}
$.
\item There exists $\epsilon >0$ such that 
$
\frac{p^2_{12}-p_{11}p_{22}}{n^4 \alpha^2_n \varrho_n}
\left(
S_{10}^2 + S_{01}^2 - S_{11}S_{00} 
\right)
> \epsilon
$,
where
$
S_{kl} = \sum_{{(i,j):c^*_i = k,c^*_j = l}} \theta_i \theta_j,
$
for $k, l = 0,1$.
\item Similar to Theorem 2.1, let $\theta_c = \min\{\theta_i: c^*_i = 1 \}$ be the lowest degree parameter in the core set, and $\theta_{p, max} = \max\{\theta_i: c^*_i =0\}$ and $\theta_{p, min} = \min\{\theta_i: c^*_i =0\}$ be the highest and lowest degree parameters in the periphery set, respectively.
Then $
 \frac{p_{12}}{p_{22}} > \frac{\theta_{p, max}^2}{\theta_c \theta_{p, min}}
$.
\end{enumerate}
Under these three assumptions, the power of the test goes to one, i.e., for any $\eta>0$,}
$$
    \lim_{n\to\infty}\mathbb P\left[\{T_1(A) > C_1\}\cap \{T_2(A)>C_2\}\right] > 1-\eta.
$$

\noindent
{\bf Remark 6.}
The proof is similar to that of Theorem 3.1, but requires extra care thanks to the added CL model parameters. Comparing the $C_1$ cutoffs from the ER and CL null is enlightening as to the effect of the CL null. First, notice that $\epsilon_n'=\sqrt n C_1$ for the $C_1$ cutoff from the ER null model. Additionally, $\rho(\hat P,\hat c)>0$ under the CL null model, but 0 under the ER model. So it is easy to see that the rejection threshold is much larger for the CL null compared to that of the ER null. This finding is as we might expect since the CL null can be considered a more complicated null model, making rejection more difficult. Indeed, rejecting the CL null implies that the ER null is also rejected, but the opposite does not always hold.\\

\noindent 
{\bf Remark 7.}
For practitioners, we first recommend testing against the ER null. If this test is rejected, the user should next test against the CL null. If this test is not rejected, it means that the network exhibits a CL model structure and/or the observed CP structure is merely a result of degree heterogeneity.
On the other hand, if the CL null is rejected, then the network has a stronger CP structure than what could be induced by degree heterogeneity alone, and is therefore likely to be endogenous.
Together, these tests provide a rich understanding of the underlying CP structure and 
enable a level of interpretation and structural resolution that goes beyond merely detecting the presence of CP structure.

\section{Simulations}
\subsection{Detection}
In the first set of simulations, we compare the Algorithm 1 against competing methods for determining the core of a network. We compare the BE metric with several well-known metrics: \texttt{Rombach} \citep{Rombach2017}, \texttt{Curcuringu} \citep{cucuringu2016detection} and \texttt{BayesSBM}. \texttt{BayesSBM} fits the stochastic block model using a Gibbs sampler and in this way is a similar to \cite{Zhang2014} and \cite{gallagher2021clarified}. All other methods use the \texttt{cpnet} Python package and \texttt{Reticulate} package to run in \texttt{R}. Please see later in the Supplemental Materials for more details on these competing methods. In particular, the method from \citep{Rombach2017} has a comparative advantage since it is given knowledge of the true value of $k$, so its performance should be taken with a grain of salt. Again, we generate networks, apply the detection algorithms and report the classification accuracy and run-time, while varying $\tilde\rho(P)=\max_{\bc}\{\rho(P,\bc)\}$ and $n$. Unless otherwise stated, we fix $k=0.1n$, i.e., 10\% of the nodes are in the core.

\subsubsection{Stochastic block model}
\begin{figure}
     \centering
     \begin{subfigure}[b]{0.90\textwidth}
         \centering
         \includegraphics[width=\textwidth]{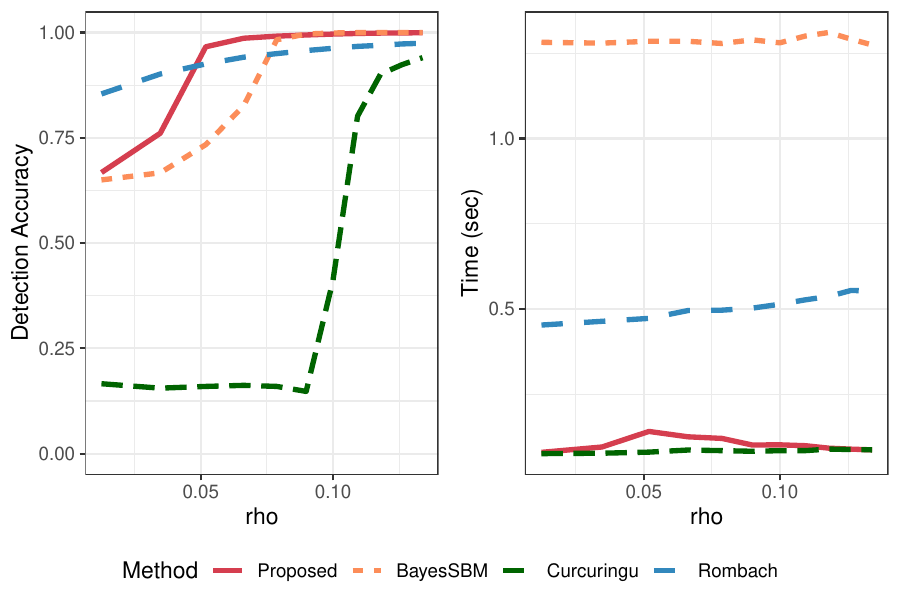}
         \caption{Detection accuracy and computing time for networks generated from a CP-SBM with increasing $\tilde\rho(P)$.}
         \label{fig:SBM_class_rho}
     \end{subfigure}
     \\
     \begin{subfigure}[b]{0.90\textwidth}
         \centering
         \includegraphics[width=\textwidth]{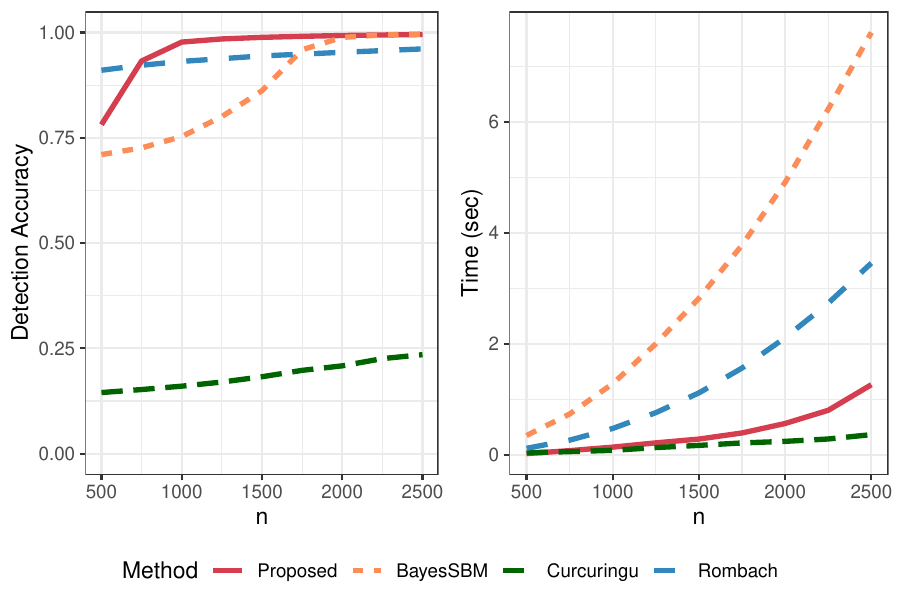}
         \caption{Detection accuracy and computing time for networks generated from a CP-SBM with increasing $n$.}
         \label{fig:SBM_class_n}
     \end{subfigure}
        \caption{Core-periphery identification results with CP-SBM-generated networks.}
        \label{fig:SBM_class}
\end{figure}

We first generate networks from a CP-SBM. Let $n=1000$ and fix $p_{22}=n^{-1}=0.001$. Then we vary $p_{12}=0.002, 0.0044,\dots,0.026$ and $p_{11}=2p_{12}$ which induces $\tilde\rho(P)\in(0.01,0.13)$. The results are in Figure \ref{fig:SBM_class}. Initially, \texttt{Rombach} has the largest classification accuracy, but for all $\tilde\rho(P)\geq 0.05$, the proposed algorithm is the most accurate method. \texttt{BayesSBM} yields perfect accuracy around $\tilde\rho(P)=0.08$ but does not perform as well as Algorithm 1. The \texttt{Rombach} metric begins strong, but plateaus without reaching 100\% accuracy. \texttt{Curcuringu} starts with a very low accuracy until $\tilde\rho(P)>0.08$ at which point it quickly increases. In terms of computation time, Algorithm 1 and \texttt{Curcuringu} are significantly faster than the other two approaches. We then fix $p_{11}=0.015$, $p_{12}=0.0075$ and $p_{22}=0.001$ ($\tilde\rho(P)=0.06$) and vary $n=500,750,\dots,2500$. The results are in Figure \ref{fig:SBM_class_n}. Again, the BE metric clearly performs the best in terms of both classification accuracy, and has a comparable computation time to that of \texttt{Curcuringu}. \texttt{BayesSBM} is able to reach 100\% classification accuracy while \texttt{Rombach} again plateaus before that point.

\subsubsection{Chung-Lu model}
\begin{figure}
     \centering
     \begin{subfigure}[b]{0.90\textwidth}
         \centering
         \includegraphics[width=\textwidth]{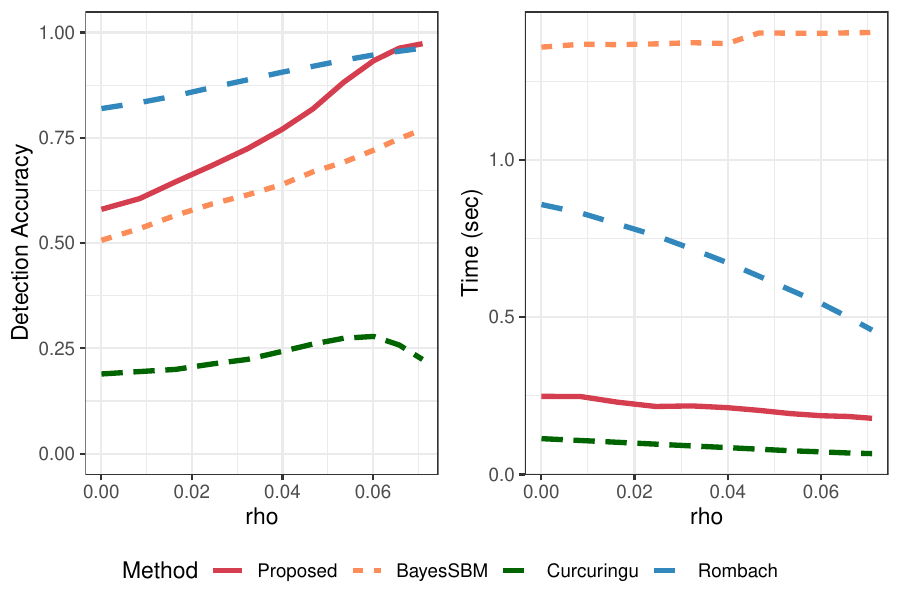}
         \caption{Detection accuracy and computing time for networks generated from a CL model with increasing $\tilde\rho(P)$.}
         \label{fig:CL_class_rho}
     \end{subfigure}
     \\
     \begin{subfigure}[b]{0.90\textwidth}
         \centering
         \includegraphics[width=\textwidth]{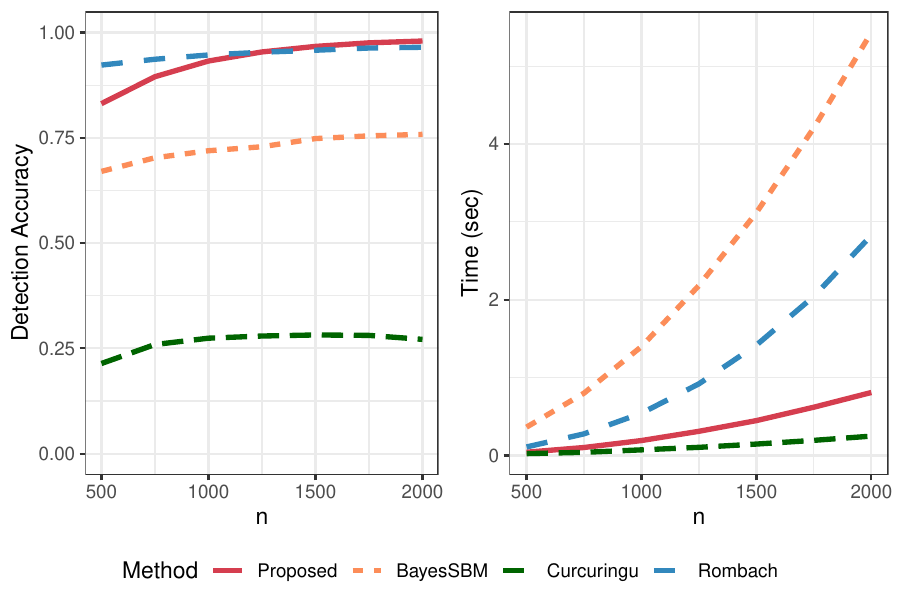}
         \caption{Detection accuracy and computing time for networks generated from a CL model with increasing $n$.}
         \label{fig:CL_class_n}
     \end{subfigure}
        \caption{Core-periphery identification results with CL-generated networks.}
        \label{fig:CL_class}
\end{figure}

Next, we compare the algorithms on networks generated from a CL model. Let $n=1000$ and without loss of generality, let $\theta_1,\dots,\theta_k$ be the weight parameters for the core nodes, and $\theta_{k+1},\dots,\theta_n$ for the periphery nodes. We sample $\theta_1,\dots,\theta_k\stackrel{\text{iid.}}{\sim}\mathsf{Uniform}(0.1+\delta/2, 0.2+\delta/2)$ and $\theta_{k+1},\dots,\theta_n\stackrel{\text{iid.}}{\sim}\mathsf{Uniform}(0.1-\delta/2, 0.2-\delta/2)$ for $\delta=0.00, 0.03, \dots, 0.10$. As $\delta$ increases, the separation between the weight parameters in the core and periphery also increases, which should make identifying the core easier. The results are in Figure \ref{fig:CL_class_rho}.\footnote{These simulations were run on a different computer than the other settings, so the computation times may not be comparable across sub-sections.} Initially, \texttt{Rombach} has the best detection accuracy, followed by the proposed method, but they are comparable for $\tilde\rho(P)\geq0.06$. \texttt{BayesSBM} performs slightly worse than the proposed method, while \texttt{Curcuringu} has consistently the lowest accuracy. In Figure \ref{fig:CL_class_n}, we fix $\delta=0.16$ ($\tilde\rho(P)=0.06$) and vary $n=500,750,\dots,2000$. Again, the proposed method and \texttt{Rombach} have similar detection accuracy for all $n$. In both settings, the proposed method and \texttt{Curcuringu} are noticeably faster than the other two methods.

\subsubsection{Degree-corrected block model}
\begin{figure}
     \centering
     \begin{subfigure}[b]{0.90\textwidth}
         \centering
         \includegraphics[width=\textwidth]{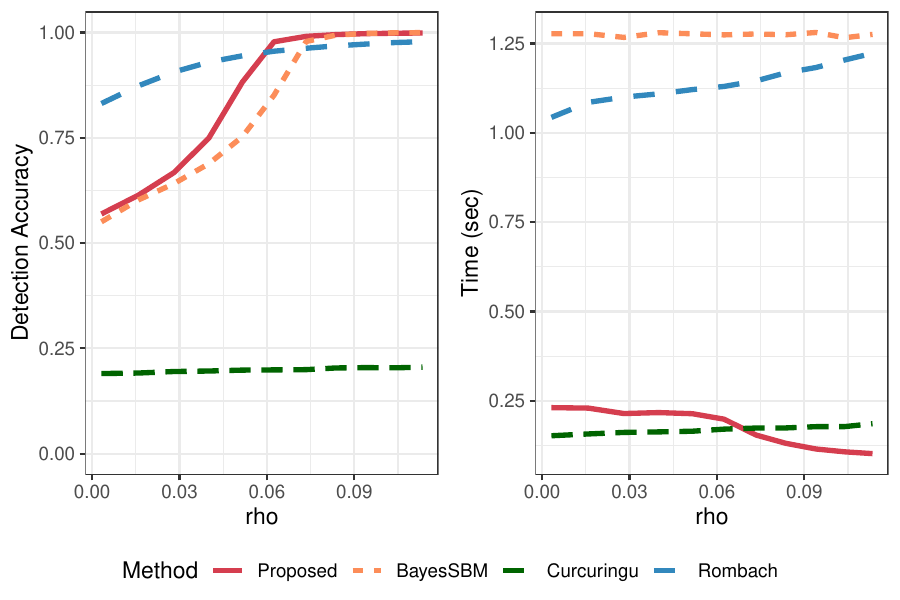}
         \caption{Detection accuracy and computing time for networks generated from a CP-DCBM with increasing $\tilde\rho(P)$.}
         \label{fig:DCBM_class_rho}
     \end{subfigure}
     \\
     \begin{subfigure}[b]{0.90\textwidth}
         \centering
         \includegraphics[width=\textwidth]{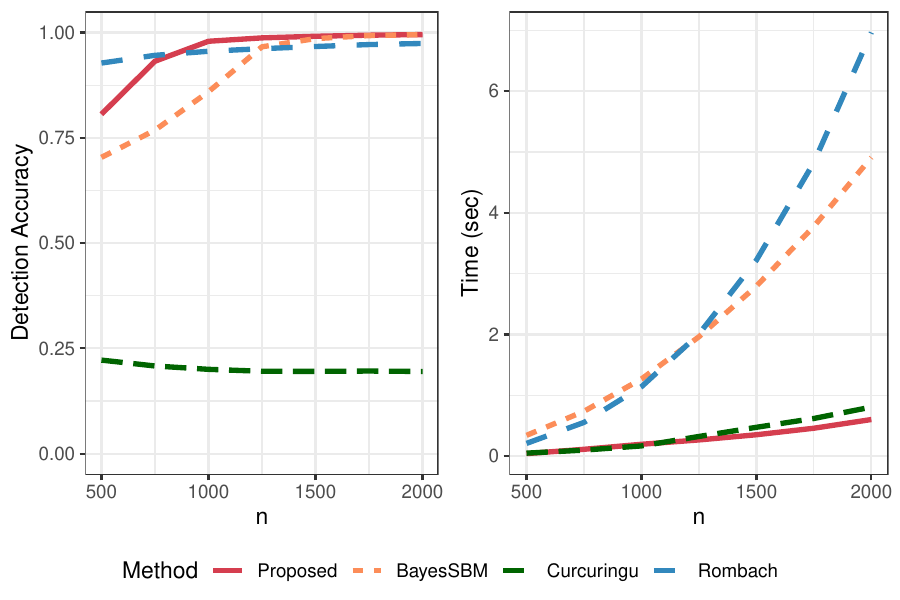}
         \caption{Detection accuracy and computing time for networks generated from a CP-DCBM with increasing $n$.}
         \label{fig:DCBM_class_n}
     \end{subfigure}
        \caption{Core-periphery identification results with CP-DCBM-generated networks.}
        \label{fig:DCBM_class}
\end{figure}

We also compare the algorithms on networks generated from a CP-DCBM. Let $n=1000$ and generate $\theta_i\stackrel{\text{iid.}}{\sim}\mathsf{Uniform}(0.6, 0.8)$. We set $p_{22}=0.001$ and vary $p_{12}=0.002,\dots, 0.08$ with $p_{11}=2p_{12}$ such that $\tilde\rho(P)\in(0.00, 0.11)$. The results are in Figure \ref{fig:DCBM_class_rho}. While the proposed algorithm and \texttt{BayesSBM} have similar classification accuracy, \texttt{Rombach} performs well at first before being overtaken around $\tilde\rho(P)=0.06$. The accuracy for \texttt{Curcuringu} is low for all parameter settings. In Figure \ref{fig:DCBM_class_n}, we report the results for $p_{11}=0.01$, $p_{12}=0.005$ and $p_{22}=0.001$ ($\tilde\rho(P)=0.06$) with $n=500,750,\dots,2000$. The trends are similar to the other settings; Algorithm 1 has the highest classification accuracy for all $n\geq1000$. \texttt{BayesSBM}'s accuracy increases with $n$ while those of \texttt{Curcuringu} and \texttt{Rombach} are roughly constant and do not reach 100\%. Again, \texttt{Curcuringu} and the proposed method consistently have the fastest computing time.

\subsection{Hypothesis testing}
We turn our attention to analyzing the performance of the proposed hypothesis tests in Theorems 3.1 and 3.2. As there are no other existing tests for CP structure with statistical guarantees, we cannot make a fair comparison with our method, so we only consider the proposed tests. We conduct simulations testing against the ER and CL null models and various alternative models.

\subsubsection{ER vs. CP-SBM}
\begin{figure}
     \centering
     \begin{subfigure}[b]{0.49\textwidth}
         \centering
         \includegraphics[width=\textwidth]{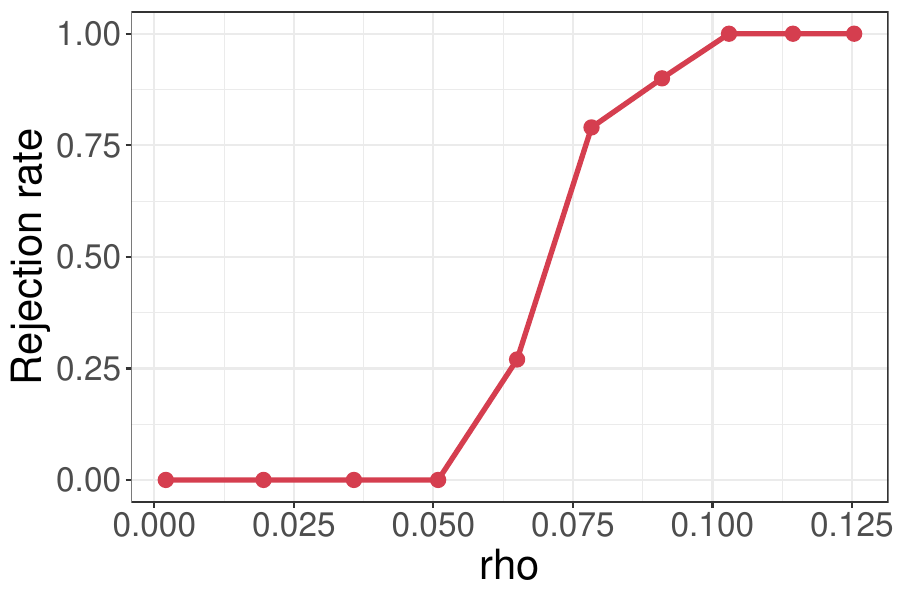}
         \caption{CP-SBM networks with varying $\tilde\rho(P)$.}
         \label{fig:SBM_rho}
     \end{subfigure}
     \begin{subfigure}[b]{0.49\textwidth}
         \centering
         \includegraphics[width=\textwidth]{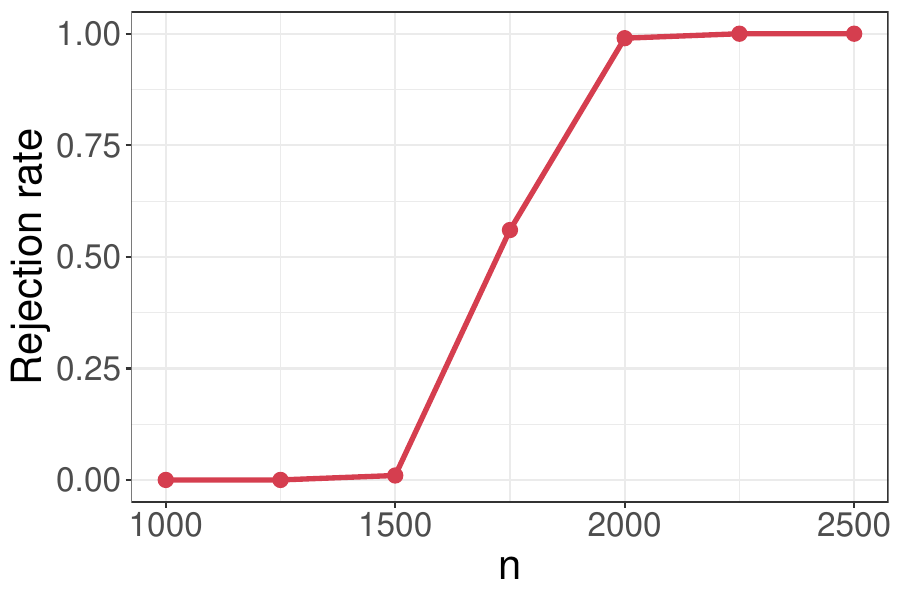}
         \caption{CP-SBM networks with varying $n$.}
         \label{fig:SBM_n}
     \end{subfigure}
     \\
     \begin{subfigure}[b]{0.49\textwidth}
         \centering
         \includegraphics[width=\textwidth]{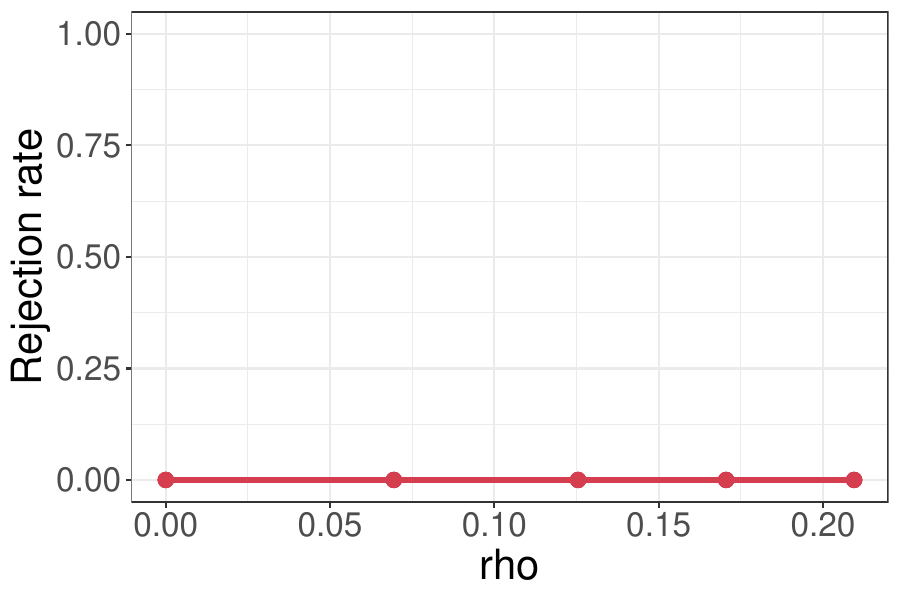}
         \caption{SBM networks with disassortative community structure (no CP).}
         \label{fig:SBM_dis}
     \end{subfigure}
        \caption{Rejection rates for networks generated from the CP-SBM being tested against the ER null.}
        \label{fig:ER_SBM}
\end{figure}

We leverage Theorem 3.1 to test the CP-SBM alternative model against an ER null. In particular, let $n=1000$ and $p_{22}=0.01$. We vary $p_{12}=0.01, \dots, 0.05$ of length 10 and set $p_{11}=2p_{12}$, implying $\tilde\rho(P)\in(0.00, 0.13)$, and generate networks from a CP-SBM. For $p_{12}>0.01$, all networks have a CP structure, so we expect a large rejection rate. The results are in Figure \ref{fig:SBM_rho}. As the strength of the CP structure in the network increases (as measured by $\tilde \rho(P)$), the rejection rate of the test also increases. Next, we fix $p_{11}=0.05, p_{12}=0.02$ and $p_{22}=0.01$, which implies $\tilde\rho(P)=0.04$, and vary $n=1000,1250,\dots,2500$. These networks also have a CP structure so we expect a large rejection rate. The results are in Figure \ref{fig:SBM_n}. As $n$ increases, so too does the power of the test. Since our theoretical results hold asymptotically, it is sensible that the test performs better for large $n$. Finally, we generate networks {\it without} a CP structure to ensure that the test maintains a small rejection rate. In particular, we generate networks with disassortative community structure where $p_{11}=p_{22}=0.01$ and $p_{12}=0.01, \dots,0.10$ of length 5, and fix $n=1000$. These parameters yield $\tilde\rho\in(0.00, 0.21)$. The results are in Figure \ref{fig:SBM_dis}. The proposed test consistently fails to reject these networks since they do not have a CP structure.

\subsubsection{ER vs. CL}
\begin{figure}
     \centering
     \begin{subfigure}[b]{0.49\textwidth}
         \centering
         \includegraphics[width=\textwidth]{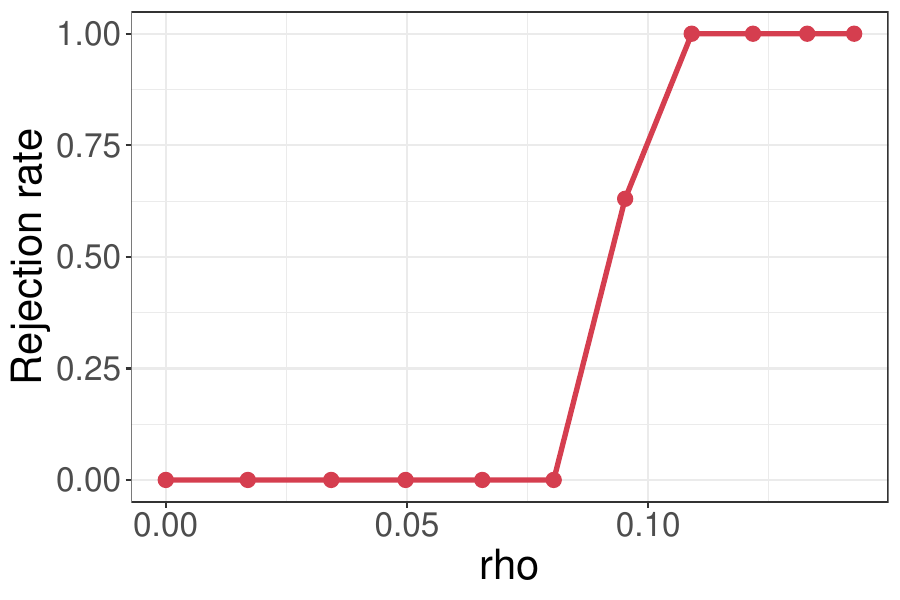}
         \caption{CL networks with varying $\tilde\rho(P)$.}
         \label{fig:ER_CL_rho}
     \end{subfigure}
     \begin{subfigure}[b]{0.49\textwidth}
         \centering
         \includegraphics[width=\textwidth]{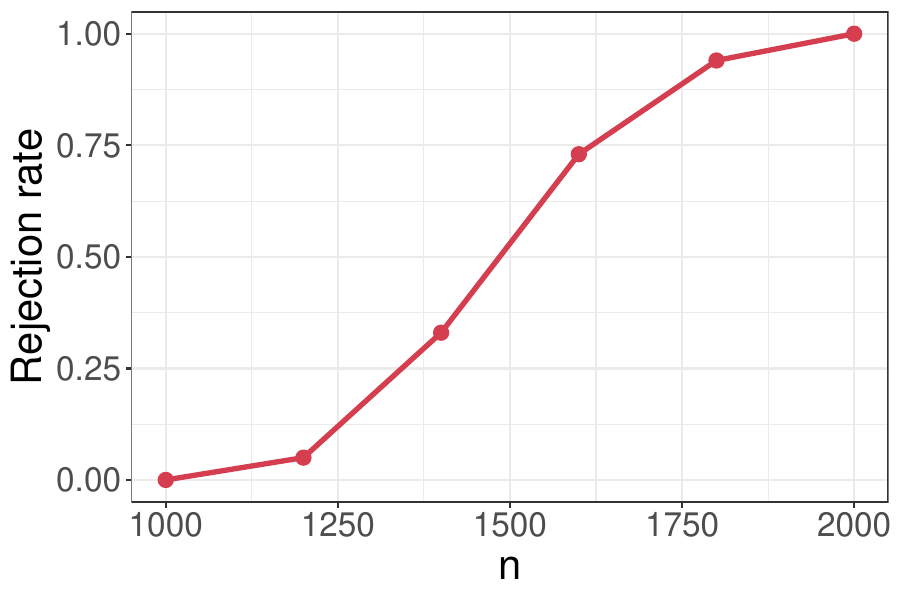}
         \caption{CL networks with varying $n$.}
         \label{fig:ER_CL_n}
     \end{subfigure}
     \\
     \begin{subfigure}[b]{0.49\textwidth}
         \centering
         \includegraphics[width=\textwidth]{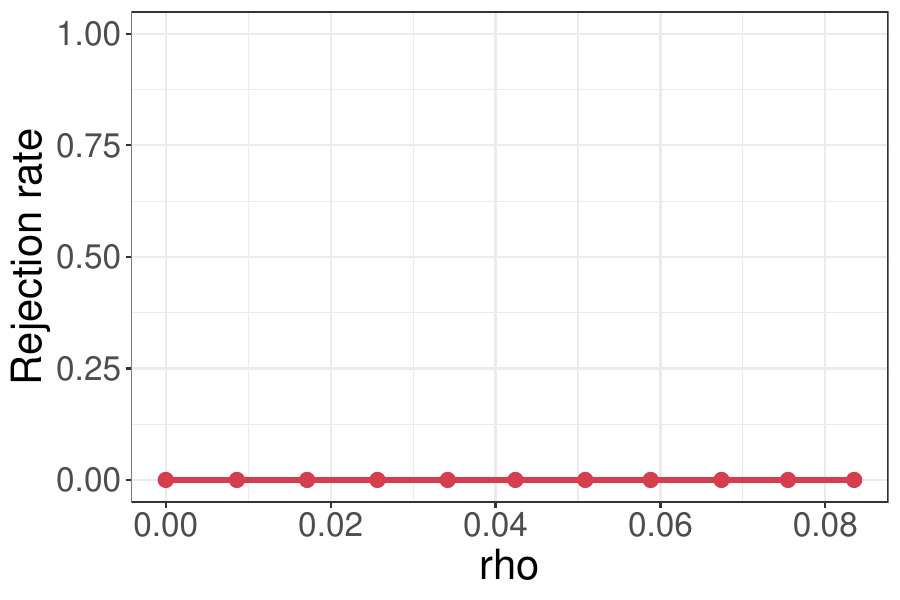}
         \caption{CL networks with Theorem 3.1 assumption not met.}
         \label{fig:ER_CL_null}
     \end{subfigure}
        \caption{Rejection rates for networks generated from the CL model being tested against the ER null.}
        \label{fig:ER_CL}
\end{figure}

We can also use Theorem 3.1 to test the ER null model against a CL model with CP structure. To generate the CL networks, we again let $n=1000$ and sample $\theta_1,\dots,\theta_k\stackrel{\text{iid.}}{\sim}\mathsf{Uniform}(0.2+\delta/2, 0.4+\delta/2)$ and $\theta_{k+1},\dots,\theta_n\stackrel{\text{iid.}}{\sim}\mathsf{Uniform}(0.2-\delta/2, 0.4-\delta/2)$ for $\delta=0.00, 0.04, \dots, 0.40$. The rejection rates are in Figure \ref{fig:ER_CL_rho}. We see that rejection rates increase around $\tilde\rho(P)=0.08$ and reach a power of by $\tilde\rho(P)=0.11$. Next, we fix $\delta=0.22$ and vary $n=1000,1200,\dots,2000$. These results are in Figure \ref{fig:ER_CL_n} and show an increasing rejection rate as $n$ increases. Finally, we generate CL networks where the assumption for Theorem 3.1 is not met. We fix $n=1000$, let $\theta_i\stackrel{\text{iid.}}{\sim}\mathsf{Uniform}(0.3-\delta/2,0.3+\delta/2)$ for $\delta=0.00, 0.04,\dots, 0.40$ and assign the $k$ nodes with largest value of $\theta_i$ to the core. The results are in Figure \ref{fig:ER_CL_null}. For all values of $\tilde\rho(P)$, the rejection rate is 0. The reason for this can be seen in Table \ref{tab:er_cl_null}. Here, we report the average value of $\theta_{(k)}\theta_{(n)}$ and $\theta_{(k+1)}\theta_{(k+2)}$ for each value of $\delta$. Recall that one of the assumptions for the test in Theorem 3.1 is that $\theta_{(k)}\theta_{(n)} >\theta_{(k+1)}\theta_{(k+2)}$. Looking at this table, however, it is clear that this assumption is not being met, leading to the low rejection rate. This result empirically validates the theoretical condition of Theorem 3.1

\begin{table}[]
    \centering
    \begin{tabular}{l|ccccccccccc}
        $\delta$ & 0.00 & 0.04 & 0.08 & 0.12 & 0.16 & 0.20 &
        0.24 & 0.28 & 0.32 & 0.36 & 0.40\\\hline
        $\theta_{(k)}\theta_{(n)}$ & 0.09 & 0.09 & 0.09 & 0.08 & 0.08 & 0.08 & 0.07 & 0.07 & 0.06 & 0.05 & 0.05\\
        $\theta_{(k+1)}\theta_{(k+2)}$ & 0.09 & 0.10 & 0.11 & 0.12 & 0.13 & 0.14 & 0.16 & 0.17 & 0.18 & 0.20 & 0.21
    \end{tabular}
    \caption{Average value of certain weight parameters when testing ER vs.~CL model with $\theta_i\stackrel{\text{iid.}}{\sim}\mathsf{Uniform}(0.3-\delta/2,0.3+\delta/2)$ for $\delta=0.00, 0.04,\dots, 0.40$.}
    \label{tab:er_cl_null}
\end{table}

\subsubsection{ER vs. CP-DCBM}
We conclude testing of the ER null model by using the CP-DCBM as the alternative model. We first fix $n=1000$ and generate network from a CP-DCBM with a CP structure where $\theta_i\sim\mathsf{Uniform}(0.6, 0.8)$, $p_{22}=0.05$ and vary $p_{12}=0.05,\dots, 0.25$ of length 10 with $p_{11}=2p_{12}$. This implies that $\tilde\rho(P)\in(0.00, 0.20)$. When $p_{12}>p_{22}$, by Theorem 3.1, we expect to see a large rejection rate. The results are in Figure \ref{fig:DCBM_rho} and we can see that the rejection rate increases with increasing $\tilde\rho(P)$. Next, we fix $p_{11}=2p_{12}=0.24$ and $p_{22}=0.05$ which yields $\tilde\rho(P)=0.08$, and vary $n=1000,1200,\dots,2000$. The results are in Figure \ref{fig:DCBM_n} which demonstrate an increasing rejection rate with $n$. Lastly, we generate disassortative DCBM community structure (no CP) to ensure the test yields a small rejection rate. We let $\theta_i\stackrel{\text{iid.}}{\sim}\mathsf{Uniform}(0.6,0.8)$, $p_{11}=0.05$, $p_{12}=0.10$ and $p_{22}=0.05$ such that and $\tilde\rho(P)=0.05$. We increase $n=1000,2000,\dots,5000$ with results in Figure \ref{fig:DCBM_dis}. We can see that even for large $n$, the rejection rate is still 0.

\begin{figure}
     \centering
    \begin{subfigure}[b]{0.49\textwidth}
         \centering
         \includegraphics[width=\textwidth]{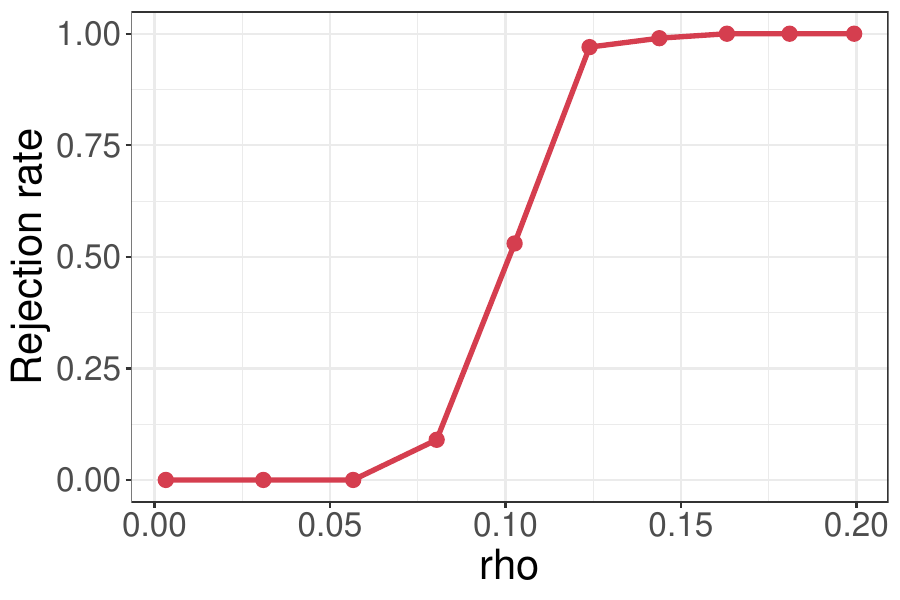}
         \caption{CP-DCBM networks with varying $\tilde\rho(P)$.}
         \label{fig:ER_DCBM_rho}
     \end{subfigure}
     \begin{subfigure}[b]{0.49\textwidth}
         \centering
         \includegraphics[width=\textwidth]{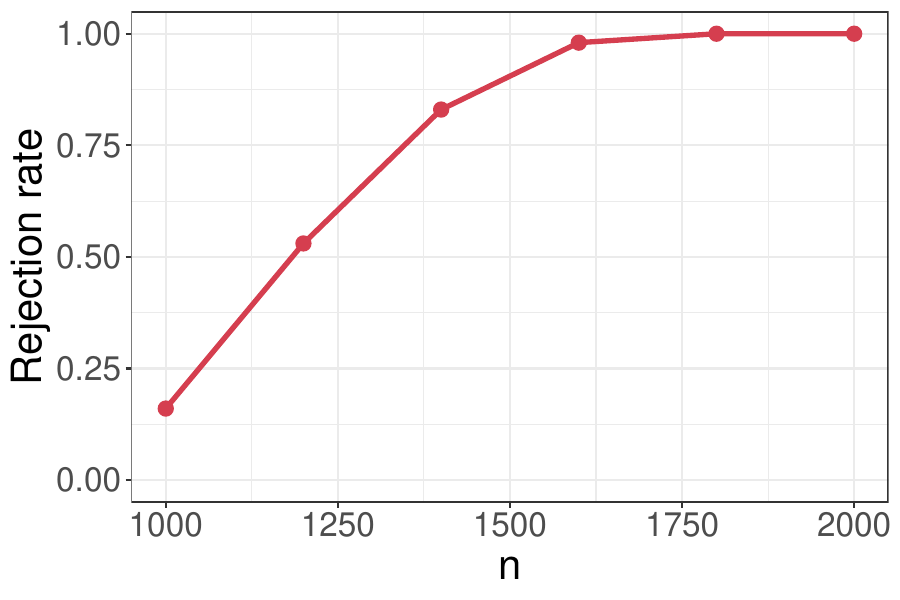}
         \caption{CP-DCBM networks with varying $n$.}
         \label{fig:ER_DCBM_n}
     \end{subfigure}
     \\
    \begin{subfigure}[b]{0.49\textwidth}
         \centering
         \includegraphics[width=\textwidth]{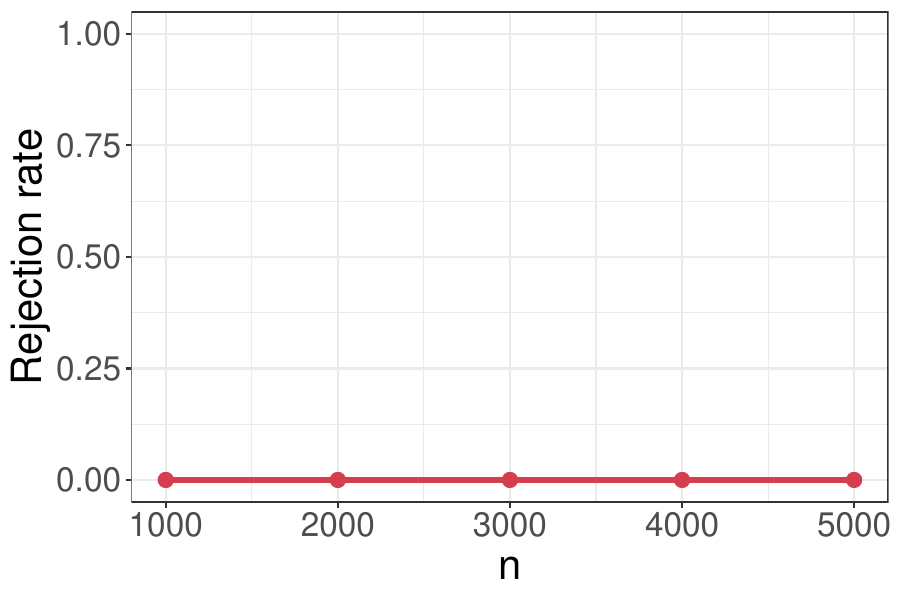}
         \caption{DCBM networks with disassortative community structure (no CP).}
         \label{fig:ER_DCBM_dis}
     \end{subfigure}
        \caption{Rejection rates for networks generated from the CP-DCBM being tested against the ER null.}
        \label{fig:ER_DCBM}
\end{figure}

\subsubsection{CL vs. CP-DCBM}
\begin{figure}
     \centering
    \begin{subfigure}[b]{0.49\textwidth}
         \centering
         \includegraphics[width=\textwidth]{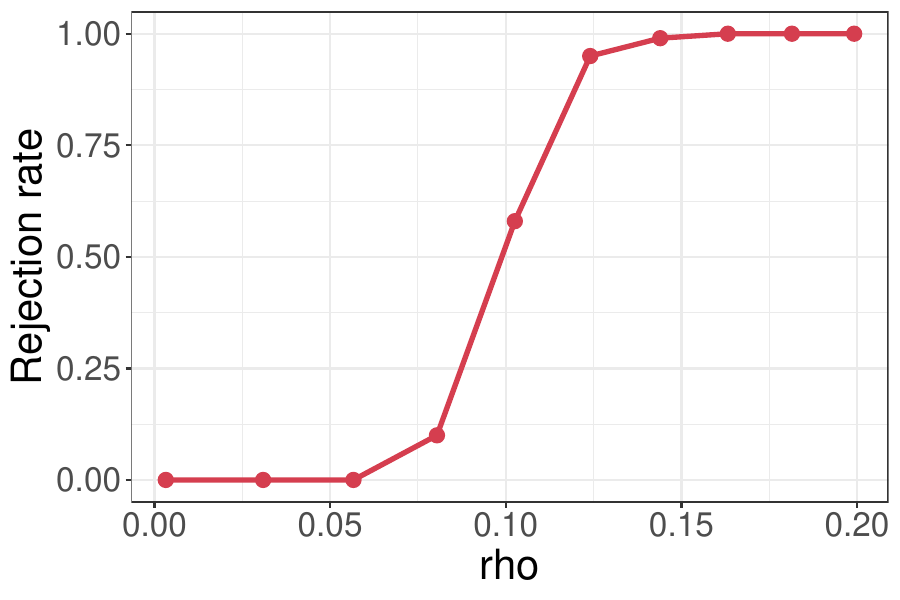}
         \caption{CP-DCBM networks with varying $\tilde\rho(P)$.}
         \label{fig:DCBM_rho}
     \end{subfigure}
     \begin{subfigure}[b]{0.49\textwidth}
         \centering
         \includegraphics[width=\textwidth]{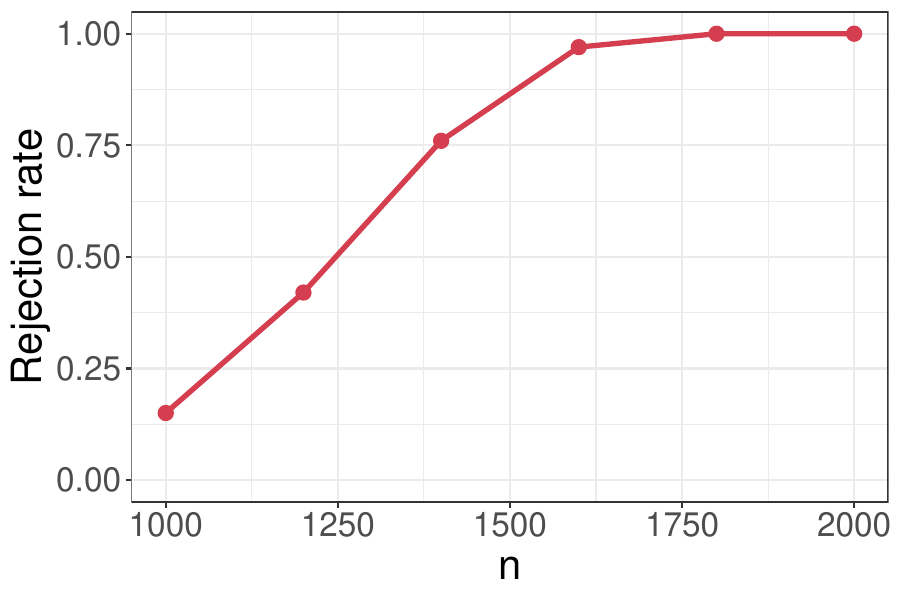}
         \caption{CP-DCBM networks with varying $n$.}
         \label{fig:DCBM_n}
     \end{subfigure}
     \\
    \begin{subfigure}[b]{0.49\textwidth}
         \centering
         \includegraphics[width=\textwidth]{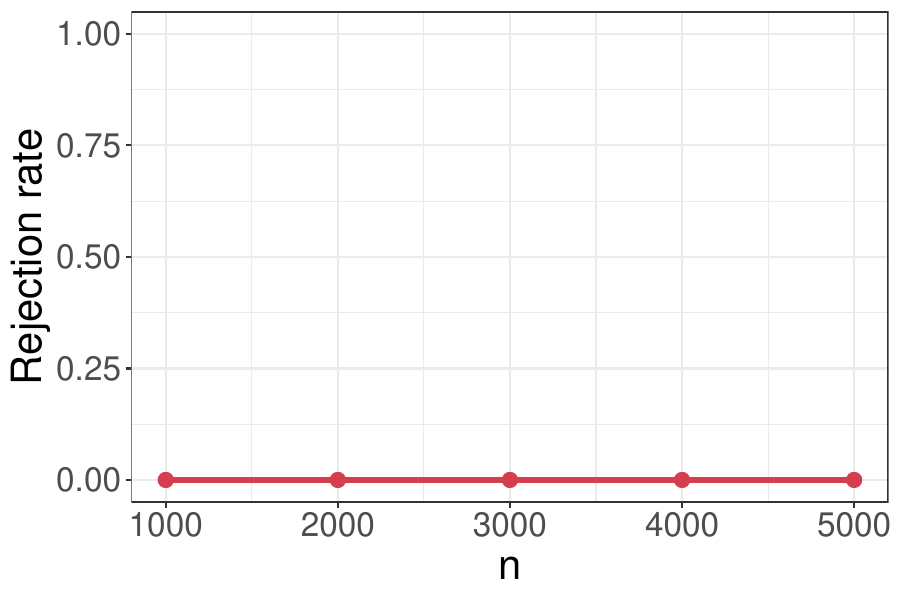}
         \caption{DCBM networks with disassortative community structure (no CP).}
         \label{fig:DCBM_dis}
     \end{subfigure}
        \caption{Rejection rates for testing the CL null against the CP-DCBM.}
        \label{fig:CL_DCBM}
\end{figure}

Finally, we test the CL null model against the CP-DCBM alternative model using the thresholds derived in Theorem 3.2. We use the same parameter settings as in the ER vs.~CP-DCBM scenario with results in Figure \ref{fig:DCBM_rho}. The results are quite similar to those using the ER null. We do notice, however, that the rejection rate increases slightly slower with $n$ using the CL null.

\section{Real-world data}

Finally, we apply the proposed hypothesis tests to thirteen real-world networks. Please see the Supplemental Materials for details on these datasets. For each network, we apply Algorithm 1 to find the CP labels $\hat{\bc}$ and test statistics $T_1(A)$ and $T_2(A)$, and apply the hypothesis tests from Theorems 3.1 and 3.2 to test for statistically significant CP structure. The testing results are reported in Table \ref{tab:realData}, with the numeric values in the SI.

All thirteen networks yield $T_1(A)>C_1$ for the ER null, but only seven networks reject the intersection test, including all of the biological networks. This means that these networks have a greater CP structure than we would otherwise expect from a network generated from the ER model, and also highlight the importance of the intersection test. Had we only considered the first part of the test based on $T_1(A)$, we would have concluded that all networks had a CP structure. By explicitly checking $T_2(A)>C_2$, however, we find that many networks do not reject the null hypothesis. This indicates that these networks may possess disassortative community structure, instead of a CP structure. Thus, our test not only gives a rejection result, but also indicates why the test failed to reject. 

Theorem 3.1 tells us that when the ER null is rejected, it could be because of a CP-SBM, CL or CP-DCBM model, so it is also important to test against the CL null. Of the seven networks which rejected the ER null hypothesis, only four now reject the CL null model (BritishMP, PolBlogs, CopenBT and Biological 1). For the other three networks (Facebook, Biological 2 and 3), this means that they do not possess an endogenous CP structure, and their previous rejection was likely due to degree heterogeneity. We stress that failing to reject the CL null does not necessarily imply that the network was generated from a CL model. Rather, it means that the data-generating mechanism has weaker CP structure than a CL model, but this could arise from a variety of models. Thus, the test cannot determine what model generated the network, but only compare the strength of CP structure in the data-generating models.

\begin{table}[]
    \centering
    \begin{tabular}{l|ccccc|ccc|ccc}
      \multicolumn{1}{c}{} &
      \multicolumn{5}{c}{} &
      \multicolumn{3}{c}{Erdos-Renyi} &
      \multicolumn{3}{c}{Chung-Lu}\\
          Network & $n$ & $\hat p$ & $\alpha_n$ & $T_1$ & $T_2$ & $T_1 > C_1$ & $T_2>C_2$ &  $\cap$ & $T_1>C_1$ & $T_2>C_2$ & $\cap$ \\\hline
        \sc{Friendship}&&&&&&&&&&&\\
         UK faculty & $81$ & $0.18$ & $0.20$ & $0.26$ & $0.34$ & $\checkmark$ & & & &\\
         Facebook & $4039$ & $0.01$ & $0.07$ & $0.10$& $0.17$ & $\checkmark$ & $\checkmark$ & $\checkmark$ & & $\checkmark$ & \\
         &&&&&&&&&&\\
        \sc{Comm.}&&&&&&&&&&&\\
        Email & $167$ & $0.24$ & $0.13$ & $0.42$& $0.41$ & $\checkmark$ & & & $\checkmark$ & & \\
         BritishMP & $381$ & $0.08$ & $0.16$ & $0.25$& $0.40$ & $\checkmark$ & $\checkmark$ & $\checkmark$ & $\checkmark$ & $\checkmark$ & $\checkmark$\\
         Congress & $475$ & $0.09$ & $0.14$ & $0.18$ & $0.16$ & $\checkmark$ & & & $\checkmark$ & &\\&&&&&&&&&&\\
        \sc{Citation}&&&&&&&&&&&\\
         PolBlogs & $1224$ & $0.02$ & $0.07$ & $0.21$& $0.23$ & $\checkmark$ & $\checkmark$ & $\checkmark$ & $\checkmark$ & $\checkmark$ & $\checkmark$\\&&&&&&&&&&\\
        \sc{Collab.}&&&&&&&&&&&\\
         DBLP & $2203$ & $0.47$ & $0.24$ &$ 0.26$& $0.23$ & $\checkmark$ & & & $\checkmark$ \\&&&&&&&&&&\\
        \sc{Contact}&&&&&&&&&&&\\
         Hospital & $75$ & $0.41$ & $0.29$ & $0.44$& 0.38 & $\checkmark$ & & &$\checkmark$ &  \\
         School & $252$ & $0.29$ & $0.21$ & $0.22$ & $0.34$ & $\checkmark$&&&\\
         CopenBT & $672$ & $0.09$ & 0.19 & 0.17& 0.19 & $\checkmark$ & $\checkmark$ & $\checkmark$ & $\checkmark$ & $\checkmark$ & $\checkmark$ \\&&&&&&&&&&\\
        \sc{Biological}&&&&&&&&&&&\\
        Biological 1 & $2220$ & $0.02$ & $0.18$ & $0.13$ & $0.10$ & $\checkmark$ & $\checkmark$ & $\checkmark$ & $\checkmark$ & $\checkmark$ & $\checkmark$\\
        Biological 2 & $3289$ & $0.02$ & $0.08$ & $0.13$ & $0.27$ & $\checkmark$ & $\checkmark$ & $\checkmark$ &  & $\checkmark$ & \\
        Biological 3 & $4040$ & $0.01$ & $0.09$ & $0.10$ & $0.13$ & $\checkmark$ & $\checkmark$ & $\checkmark$ &  & $\checkmark$ &
    \end{tabular}
    \caption{Real-data analysis results where $n$ is the number of nodes, $\hat p$ is the average edge density and $\alpha_n$ is the proportion of nodes assigned to the core. The remaining values are defined in Theorem 3.1 and 3.2.}
    \label{tab:realData}
\end{table}

\section{Discussion}\label{sec:conc}
In this work, we provided a principled treatment of core-periphery structure from a statistical inference perspective. 
Our model-agnostic population parameter quantifies the underlying CP structure and allows for a taxonomy of endogenous vs. exogenous CP structure.
We proved novel theoretical results showing exact recovery of the true CP labels while also proposing hypothesis tests against the ER and CL null models. Throughout the work, we highlighted the nuances of distinguishing endogenous and exogenous CP structure, a difficulty that most clearly arises for the CL model. In general, we leave it to the user to determine whether the exogenous CP structures present in CL networks should be considered \textit{true} CP structure. Indeed, in the real data analysis, we found that about half the networks rejected the ER null model but only half of these also rejected the CL null. This result can be interpreted in one of two ways. If we believe that the CL model is a reasonable null model, then this means that CP structure is relatively rare in real-world networks. Conversely, these results can reinforce a finding in \cite{Kojaku2018} which is that the ER null is too easy to reject, whereas the CL null is too difficult. Under this interpretation, many real-world networks do have a CP structure, but the CL null is too high of a threshold to clear, thus leading to many false negatives. These results also show that developing other meaningful null models to represent ``no CP structure'' is an important open research problem. Future work could also include extending our methods to more complicated CP structures, i.e., more than two blocks \citep{Kojaku2018} or layered structures \citep{gallagher2021clarified}.

\paragraph{Conflicting Interests} The authors have no conflicting interests to declare.

\paragraph{Code and data sharing} The code for all methods is available on the author's GitHub: \url{https://github.com/eyanchenko/CPinference}. All datasets are freely available online.

\paragraph{Funding} {Srijan Sengupta's work was supported in part by the National Science Foundation grant DMS-2413327 and the NIH grant 7R01LM013309.}

\bibliographystyle{apalike}
\bibliography{network-reference}

\clearpage

\begin{center}
    {\Large Supplemental Materials}
\end{center}

\begin{center}
    Table of Contents
\end{center}

\begin{enumerate}
    \item[7.] Greedy algorithm for CP identification
    \item[8.] Discussion of existing hypothesis tests
    \item[9.] Competing methods description
    \item[10.] Real-data analysis results
    \item[11.] Proofs for label recovery
    \item[12.] Proofs for hypothesis test
\end{enumerate}

\clearpage

\section{Algorithm}
We adopt the greedy algorithm of \cite{yanchenko2022divide} to maximize the test statistic. The algorithm works as follows. First, we randomly generate initial CP labels. Then one at a time and for each node $i$, we swap the label of the node and compute the test statistic with these new labels. The new label is kept if that test statistic is larger than before, otherwise we keep the original labels. This process repeats until the labels are unchanged for an entire cycle of all $n$ nodes. Note that the node order is randomly shuffled each iteration and multiple starting labels can be run simultaneously to avoid finding a local max. The steps are outlined in Algorithm 1.

\begin{algorithm}[H]
\SetAlgoLined
\KwResult{Core-periphery labels $\bc$}
 {\bf Input: } $n\times n$ adjacency matrix $A$\;
 
 Initialize labels $\bc$\;
 
 $run = 1$\;
 
 \While{$run > 0$}{
 
 $run =0$\;
 
 Randomly order nodes\;
 
  \For{$i$ in $1,\dots,n$}{
  Swap label of node $i$: $\bc^*=\bc$, $c^*_i=1-c_i$\;
  
  \If{$T(A, \bc^*) > T(A, \bc)$}{
  
  $\bc\longleftarrow\bc^*$\;
  
  $run = 1$
 }
   
  }
 }
 \caption{Greedy}
 \label{alg:greedy}
\end{algorithm}

\section{Existing hypothesis tests}
We briefly describe three existing hypothesis tests for CP structure found in the literature. \cite{boyd2010computing} adopt the BE metric to quantify the strength of the CP structure in the original network. They then propose a bootstrap test, generating bootstrap networks by re-wiring edges while preserving the total number of edges as the original network. The BE metric is computed on these bootstrap samples and an empirical $p$-value is then computed. Note that re-wiring edges and preserving the total number of edges is analogous to the ER null as the ER model preserves the {\it expected} number of edges on the network. \cite{elliott2020} extend the BE definition of CP structure to directed networks. This metric is then used to carry out a bootstrap testing procedure where the bootstrap networks are generated from an ER model. Lastly, in \cite{kojaku2017}, the authors show that if a configuration null model is used, then it is impossible to generate networks with a CP structure. Thus, they propose introducing at least one more block into the network. By defining a modularity-like metric to quantify the CP structure in the network, the authors generate configuration model bootstrap samples to carry out the hypothesis test.

\section{Methods}
We explain the details of the \texttt{Rombach} and \texttt{BayesSBM} methods used in the simulation study.

\subsubsection*{Rombach}
This method comes from \cite{Rombach2017} and the output is a scalar coreness $C_i$ for each node $i=1,\dots,n$. To convert these to binary labels, we sort the nodes from largest to smallest based on $C_i$. Then we choose the nodes with $k$ largest coreness value where $k$ is the true core size. This gives this method a comparative advantage as it has partial knowledge of the data-generating mechanism (known $k$).

\subsubsection*{BayesSBM}
This method computes the core labels using a Bayesian paradigm. The network $A$ is assumed to be from a two-block SBM with likelihood,
$$
    L(A|\boldsymbol{p},\bc)
    =\prod_{i<j}p_{c_ic_j}^{A_{ij}}(1-p_{c_ic_j})^{1-A_{ij}}.
$$
We can approximate this as a Poisson likelihood such that
$$
    L(A|\boldsymbol{p},\bc)
    =\prod_{i<j}\frac{e^{-p_{c_ic_j}}A_{ij}^{p_{c_ic_J}}}{A_{ij}}.
$$
Then we place a uniform prior on $\bc$, i.e., 
$$
    \pi(\bc)=\frac{1}{2^n}
$$
and let 
$$
    \pi(\boldsymbol{p})
    \propto \mathbb I(p_{11}>p_{12}>p_{22})
$$
which enforces a CP structure into the fit, similar to \cite{gallagher2021clarified}. Then we use Gibbs sampling to approximate the posterior of $\boldsymbol{p}$ and $\bc$ and assign nodes to the core if their posterior probability of being in the core is greater than 50\%, i.e., $P(c_i=1|A,\boldsymbol{p})>0.5$.

\section{Real-data analysis}

\subsection*{Data sets}
In order to examine the proposed test's properties under diverse settings, we chose networks of various sizes, densities and from various domains including: friendship \citep{nepusz2008fuzzy, leskovec2012learning}, communication \citep{michalski2011matching, greene2013producing, fink2023centrality}, citation \citep{adamic::2005aa}, collaboration \citep{gao2009graph, ji2010graph}, contact \citep{Vanhems:2013, stehle2011high, sapiezynski2019interaction} and biological networks \citep{cho2014wormnet}. All data sets were download using \citep{snapnets, nr, csardi2013package}. The networks and summary statistics are listed in Table \ref{tab:realData}, where we remove all edge weights and direction as well as self-loops. Additionally, the DBLP network was pre-processed to only consider two groups of researchers \citep{senguptapabm}.

\subsection*{Results}
In Table \ref{tab:realData2}, we report the observed values of the test statistics for the 13 real-world networks from Section 4, as well as the cutoff and rejection decision. This table provides more details than that in the main text, as it shows not only the binary rejection decision, but also the raw values. 

\begin{table}[]
    \centering
    \begin{tabular}{l|cc|cc|cc}
      \multicolumn{1}{c}{} &
      \multicolumn{2}{c}{} &
      \multicolumn{2}{c}{Erdos-Renyi} &
      \multicolumn{2}{c}{Chung-Lu}\\
          Network & $T_1$ & $T_2$ & $C_1$ & $C_2$ & $C_1$ & $C_2$ \\\hline
        \sc{Friendship}&&&&\\
         UK faculty &  $0.262$ & $0.340$ & $0.114$ & $0.553$ & $0.262$ & $0.553$ \\
         Facebook & $0.097$& $0.170$ & $0.017$ & $0.015$ & $0.103$ & $0.015$ \\
         &&&&\\
        \sc{Communication}&&&&\\
        Email & $0.423$& $0.411$ & $0.099$ & $0.724$ & $0.382$ & $0.724$ \\
         BritishMP & $0.248$& $0.400$ & $0.065$ & $0.181$ & $0.247$ & $0.181$\\
         Congress & $0.185$ & $0.164$ & $0.062$ & $0.193$ & $0.176$ & $0.193$ \\&&&&\\
        \sc{Citation}&&&&&\\
         PolBlogs & $0.207$& $0.232$ & $0.024$ & $0.047$ & $0.189$ & $0.047$ \\&&&&\\
        \sc{Collaboration}&&&&&\\
         DBLP &$ 0.261$& $0.231$ & $0.050$ & $0.448$ & $0.246$ & $0.448$ \\&&&&&\\
        \sc{Contact}&&&&&&\\
         Hospital & $0.441$& $0.380$ & $0.171$ & $1.069$ & $0.398$ & $1.069$ \\
         School & $0.218$ & $0.341$ & $0.106$ & $0.614$ & $0.229$ & $0.614$ \\
         CopenBT  & $0.170$ & $0.185$ & $0.061$ & $0.154$ & $0.169$ & $0.154$  \\&&&&&\\
        \sc{Biological}&&&&&&\\
        Biological 1 & $0.126$ & $0.103$ & $0.031$ & $0.024$ & $0.124$ & $0.024$ \\
        Biological 2 & $0.127$ & $0.276$ & $0.021$ & $0.022$ & $0.137$ & $0.022$ \\
        Biological 3 & $0.101$ & $0.134$ & $0.018$ & $0.011$ & $0.106$ & $0.011$
    \end{tabular}
    \caption{Real-data analysis results where all values are defined in Theorem 3.1 and 3.2.}
    \label{tab:realData2}
\end{table}
\clearpage

\section{Technical proofs for label recovery}
We state and prove four lemmas in this section. The first lemma is a general result that will be used to prove the subsequent results.
Lemmas 2-4 are supporting results for the CL model, the CP-SBM, and the CP-DCBM, respectively.
Finally, we state and prove Theorem 2.1.

For the results in this section, we will use the identity that
$$
\rho(P,\bc) :=\frac{\sum_{i<j} \left(P_{ij} - \bar P\right)\Delta_{ij}}{\frac12n(n-1)
    \{\bar P(1-\bar P)\bar\Delta(1-\bar\Delta)\}^{1/2}}
    =
    \frac{\sum_{i<j} \left(P_{ij} - \bar P\right)(\Delta_{ij}-\bar\Delta)}{\frac12n(n-1)
    \{\bar P(1-\bar P)\bar\Delta(1-\bar\Delta)\}^{1/2}},
$$
since
$$
\sum_{i<j} \left(P_{ij} - \bar P\right) 
=
\bar\Delta \sum_{i<j} \left(P_{ij} - \bar P\right) = 0,
$$
and, similarly,
$$
 T(A,\bc) := \frac{\sum_{i<j} (A_{ij} - \bar A) \Delta_{ij}}{\tfrac12n(n-1)\{\bar A(1-\bar A)\bar\Delta(1-\bar\Delta)\}^{1/2}} =
 T(A,\bc) := \frac{\sum_{i<j} (A_{ij} - \bar A) (\Delta_{ij}-\bar\Delta)}{\tfrac12n(n-1)\{\bar A(1-\bar A)\bar\Delta(1-\bar\Delta)\}^{1/2}}.
$$

\subsection*{Lemma 1.}
{\sc Lemma 1.} {\it Let $\varepsilon_n=\alpha_n^{1/2}\varrho_n^{1/2}
\sqrt{\frac{\log(n\varrho_n\alpha_n)}{n\varrho_n\alpha_n}}$.
From the definitions of $\alpha_n$ and $\varrho_n$ and assumption A1, clearly $\varepsilon_n\to 0$ as $n\to\infty$. Then  under A1 we have}
$$
    \mathbb P(\max_{\bc}\{|T(A_n,\bc)-\rho(P_n,\bc)|\}> \varepsilon_n)
    \to 0
    \text{ as } n\to\infty.
$$

 {\it Proof.}
Let $\hat{\bc}=\arg\max_{\bc}\{T(A,\bc)\}$. Recall also that $A \sim \varrho_n P$ and $k/n=\mathcal O(\alpha_n)$.

Now, consider some fixed $\bc$ and let $X_n(\bc)=\sum_{i<j}A_{ij}(\Delta_{ij}-\bar\Delta_n)$ and $Y_n=\sum_{i<j}A_{ij}$. Then
$$
    T(A_n,\bc) = \frac{X_n}{\sqrt{Y_n({n\choose 2}-Y_n)}}\frac{1}{\sqrt{\bar\Delta_n(1-\bar\Delta_n)}}
    := f(X_n, Y_n). 
$$
We will prove the desired result using two pieces. First, we show that $X_n(\bc)$ and $Y_n$ are ``close'' to their respective expectations, $\mu_x(\bc)$ and $\mu_y$, for all $\bc$. Then we will show that $f(x,y)$ is ``close'' to $f(x^*,y^*)$ when $x\approx x^*$ and $y\approx y^*$.
\newline

\noindent
First, Bernstein's inequality states that if $\mathsf{E}(X_i)=0$ and $|X_i|\leq M$ for $i=1, \ldots, n$, then
$$
    \mathbb P\left(\sum_{i=1}^n X_i >t\right)
    \leq \exp\left(-\frac{\frac12t^2}{\sum_{i=1}^n \mathsf{E}(X_i^2)+\frac13 Mt}\right).
$$
Thus, 
$$
    \mathbb P\left(|X_n(\bc)-\mu_x(\bc)| >t_1\right)
    \leq 2\exp\left(-\frac{\frac12t_1^2}{\sum_{i<j} P_{ij}(1-P_{ij})(\Delta_{ij}-\bar\Delta_n)^2+\frac13t_1}\right)
    \sim \exp\left(-\frac{t_1^2}{\varrho_n\alpha_n n^2 + t_1}\right)
$$
since $P_{ij}(1-P_{ij})=\mathcal O(\varrho_n)$ and $\sum_{i<j}(\Delta_{ij}-\bar\Delta_n)^2=\mathcal O(\alpha_nn^2)$. Additionally, 
$$
    \mathbb P\left(|Y_n-\mu_y| >t_2\right)
    \leq 2\exp\left(-\frac{\frac12t_2^2}{\sum_{i<j} P_{ij}(1-P_{ij})+\frac13t_2}\right)
    \sim \exp\left(-\frac{t_2^2}{\varrho_n n^2 + t_2}\right).
$$
Let $\gamma_n = \sqrt{\frac{\log(n\varrho_n\alpha_n)}{n\varrho_n\alpha_n}}$ and $t_1=\varrho_n\alpha_nn^2\gamma_n$. 
Note that from assumption A1, we have $\gamma_n \to 0$ and $n \varrho_n\alpha_n\gamma_n^2 \to \infty$ as $ n \to \infty$.
Then taking the union over $2^n$ CP assignments,
$$
    \mathbb P(\max_{\bc}|X_n-\mu_x(\bc)| > \varrho_n\alpha_nn^2\gamma_n)
    \leq 2^{n+1}\exp\left(-\frac{\varrho_n^2\alpha_n^2 n^4\gamma_n^2}{2\varrho_n\alpha_nn^2}\right) 
    \leq \exp(n\log 2 - \varrho_n\alpha_n n^2\gamma_n^2)
    \to 0
$$
since $n\varrho_n\alpha_n\gamma_n^2\to\infty$.
Additionally, $Y_n$ does not depend on $\bc$ so letting $t_2=\varrho_n n^2\gamma_n$,
$$
    \mathbb P(|Y_n-\mu_y| > \varrho_n n^2\gamma_n)
    \leq \exp\left(-\frac{\varrho_n^2n^4\gamma_n^2}{\varrho_nn^2 + \varrho_n n^2\gamma_n}\right) 
    \leq \exp(- \varrho_n n^2\gamma_n^2)
    \to 0.
$$
Thus, for large $n$, $X_n(\bc)$ and $Y_n$ are ``close'' to their respective expectations for all $\bc$.
\newline

\noindent
For the second step, recall that 
$$
    f(x,y)
    =\frac{x}{\sqrt{y(n^2-y)}}\frac{1}{\sqrt{\bar\Delta_n(1-\bar\Delta_n)}}.
$$
Then we want to show that $|f(x,y)-f(x^*,y^*)|$ is upper bounded by a small constant when $|x-x^*|<t_1$ and $|y-y^*|<t_2$. Consider the first-order Taylor expansion:
$$
    f(x,y)
    -f(x^*,y^*)
    =(x-x^*)\frac{\partial}{\partial x}f(x,y,n)\Big|_{x=x^*,y=y^*} + (y-y^*)\frac{\partial}{\partial y}f(x,y,n)\Big|_{x=x^*,y=y^*} + R_n,
$$
where
$$
    R_n \sim (x-x^*,y-y^*)
    \begin{pmatrix}
        \partial_x^2 f & \partial_{xy} f\\
        \partial_{xy}f & \partial_y^2 f
    \end{pmatrix}
    \begin{pmatrix}
        x-x^*\\y-y^*
    \end{pmatrix}.
$$
We have that $x^*= \mathcal O(\alpha_n \varrho_n n^2)$ and $y^*=\mathcal O(\varrho_n n^2)$. It's also easy to see that
$$
    \frac{\partial}{\partial x}f(x,y)\Big|_{x=x^*,y=y^*}
    = \frac{1}{\sqrt{y^*(n^2-y^*)}}\frac{1}{\sqrt{\bar\Delta_n(1-\bar\Delta_n)}}
    \sim \alpha_n^{-1/2}\varrho_n^{-1/2}n^{-2},
$$
and
$$
    \frac{\partial}{\partial y}f(x,y)\Big|_{x=x^*,y=y^*}
    =\frac{x^*(n^2-2y^*)}{2\{y^*(n^2-y^*)\}^{3/2}}\frac{1}{\sqrt{\bar\Delta_n(1-\bar\Delta_n)}}
    \sim \alpha_n^{1/2} \varrho_n^{-1/2} n^{-2}.
$$
From step 1, $|x-x^*|< t_1 = \alpha_n \varrho_n n^2\gamma_n$ and $|y-y^*| < t_2 =\varrho_n n^2\gamma_n$, so
$$
    f(x,y)
    -f(x^*,y^*)
    \sim \alpha_n\varrho_n n^2\gamma_n\frac1{\alpha_n^{1/2}\varrho_n^{1/2}n^2} + \varrho_n n^2\gamma_n \frac{\alpha_n^{1/2}}{\varrho_n^{1/2}n^2} + R_n
    \sim \alpha_n^{1/2} \varrho_n^{1/2} \gamma_n +R_n.
$$
Additionally, 
\begin{align*}
    R_n &\sim
    \frac{\{y^*(n^2-y^*)\}^{-5/2}} {\sqrt{\bar\Delta_n(1-\bar\Delta_n)}}
    (x-x^*,y-y^*)
    \begin{pmatrix}
        0 & -2(n^2-2y^*)\{y^*(n^2-y^*)\}\\
        -2(n^2-2y^*)\{y^*(n^2-y^*)\}& x^*(3n^4-8n^2y^*+8(y^*)^2)
    \end{pmatrix}
    \begin{pmatrix}
        x-x^*\\y-y^*
    \end{pmatrix}\\
    &\sim \alpha_n^{1/2} \varrho_n^{1/2}\gamma_n^2
\end{align*}
such that 
$$
    f(x,y) - f(x^*, y^*)
    \sim \alpha_n^{1/2} \varrho_n^{1/2}\gamma_n.
$$

We are now ready to show the desired result. Let
$$
    E= \left\{\bigcap_{\bc}\{|X(\bc)-\mu_x(\bc)|<t_1\}\right\}\bigcap \{|Y-\mu_y|<t_2\}.
$$
Then
\begin{multline*}
    \mathbb P(\max_{\bc}\{|T(A_n,\bc)-\rho(P_n,\bc)|\}>\varepsilon_n)
    =\mathbb P(\max_{\bc}\{|f(X,Y)-f(\mu_x,\mu_y)|>\varepsilon_n)\\
    \leq \mathbb P(\max_{\bc}\{|f(X,Y)-f(\mu_x,\mu_y)|>\varepsilon_n|E)\mathbb P(E) + \mathbb P(\max_{\bc}\{|f(X,Y)-f(\mu_x,\mu_y)|>\varepsilon_n|E^c)\mathbb P(E^C)\\
\end{multline*}
Now, by step 1, $\mathbb P(E^C) \lesssim \exp(-\varrho_n n^2\gamma_n^2) + \exp(n-\varrho_n\alpha_n n^2\gamma_n^2)$. Thus, 
\begin{multline*}
    \mathbb P(\max_{\bc}\{|T(A_n,\bc)-\rho(P_n,\bc)|\}>\varepsilon_n)
    \leq \mathbb P(\max_{\bc}\{|f(X,Y)-f(\mu_x,\mu_y)|>\varepsilon_n|E) + \mathbb P(E^C)\\
    \lesssim \exp(-\varrho_n n^2\gamma_n^2) + \exp(n-\varrho_n\alpha_n n^2\gamma_n^2)
   \to 0
\end{multline*}
by step 2 and A1. $\square$\newline

\subsection*{Lemma 2.} {\sc Lemma 2.} {\it Let $P_n$ be a two-block SBM where 
$P_n = \varrho_nP$ and $P_{ij}=\Omega_{c^*_i c^*_j}$ where}
$$
    \Omega=
    \begin{pmatrix}
        p_{11}&p_{12}\\
        p_{21}&p_{22}
    \end{pmatrix}
    \text{ and } p_{11} > p_{12} = p_{21} > p_{22}.
$$
{\it Then for any $\bc \neq \bc^*$ such that the size of the core is the same for $\bc$ and $ \bc^*$,
$
    \rho(P_n,\bc) < \rho(P_n,\bc^*),
$
and we have
$$
\rho(P_n,\bc^*) - \rho(P_n,\bc) = 
\xi_n(\bc) \sqrt{\frac{\varrho_n}{2\alpha_n}} \frac{p_{12}-p_{22}}{\sqrt{p_{22}}}  (1 + o(1))
.
$$
}
 {\it Proof.}
Consider
$$
    \rho(P_n,\bc)
    =\frac{\sum_{i<j} \varrho_n (P_{ij}-\bar p)(\Delta_{ij}-\bar\Delta)}{\sqrt{{n\choose 2} \varrho_n \bar p(1- \varrho_n\bar p){n\choose 2}\bar\Delta(1-\bar\Delta)}}.
$$
Now, let $\Delta(\bc^*)$ and $\Delta(\bc)$ denote the $\Delta$ matrices corresponding to the labels $\bc^*$ and $\bc$, respectively.
Since $\bc^*$ and $\bc$ have the same core size, 
we have
$$
1-\bar\Delta(\bc^*)
=
1-\bar\Delta(\bc)
=
\alpha_3 = 
\frac{(n-k)(n-k-1)}{n(n-1)}
= 1 + o(1)
$$
as $k=o(n)$ from A1, and
$$
\bar\Delta(\bc^*)
=
\bar\Delta(\bc)
=
1-\alpha_3
=
\frac{2k(n-k)-k(k-1)}{n(n-1)}
= 2 \alpha_n (1 + o(1)).
$$
Now, given $$
    \Omega=
    \begin{pmatrix}
        p_{11}&p_{12}\\
        p_{21}&p_{22}
    \end{pmatrix}
    \text{ and } p_{11} > p_{12} = p_{21} > p_{22},
$$ we have
$$
    \bar p 
    = \alpha^{(1)}_{n,k} p_{11} + \alpha^{(2)}_{n,k} p_{12} + \alpha^{(3)}_{n,k} p_{22},
$$
where
$$
\alpha^{(1)}_{n,k} = \frac{{k \choose 2}}{{n \choose 2}},
\alpha^{(2)}_{n,k} = \frac{k (n-k)}{{n \choose 2}},
\text{ and }
\alpha^{(3)}_{n,k} = \frac{{n-k \choose 2}}{{n \choose 2}}.
$$
For notational simplicity, we will write $\alpha^{(i)}_{n,k}$ as $\alpha_i$ for $i=1,2,3$.
Note that $\alpha_i > 0$ for $i=1,2,3$ and $ \sum_{i=1}^3 \alpha_i = 1$.
Also note that under A1, $\bar p 
    = p_{22}(1 + o(1))$.
Therefore, the denominator of $\rho(P_n,\bc)$ and $\rho(P_n,\bc^*)$ is given by
$$
{\sqrt{{n\choose 2} \varrho_n \bar p(1- \varrho_n\bar p){n\choose 2}\bar\Delta(1-\bar\Delta)}}
=
{n \choose 2} \sqrt{2 \varrho_n \alpha_n p_{22} (1-\varrho_np_{22})} (1 + o(1)).
$$
Now consider the numerators of $\rho(P_n,\bc)$ and $\rho(P_n,\bc^*)$ and note that
$\Delta_{ij}(\cdot)$ is the only term that changes between $\rho(P_n,\bc)$ and $\rho(P_n,\bc^*)$.
Therefore,
$$
    \rho(P_n,\bc^*) - \rho(P_n,\bc)
    =\frac{\sum_{i<j} \varrho_n (P_{ij}-\bar p)(\Delta_{ij}(\bc^*)-\Delta_{ij}(\bc))}{{n \choose 2} \sqrt{2 \varrho_n \alpha_n p_{22}(1-\varrho_np_{22})}}  (1 + o(1)).
$$
Note that
$$
    \Delta_{ij}(\bc^*)-\Delta_{ij}(\bc)
    =\begin{cases}
        +1&\{c_i=0\}\cap\{c_j=0\}\cap\{ \{c^*_i=1\} \cup \{c^*_j=1\} \}  \\
        -1&\{c^*_i=0\}\cap\{c^*_j=0\}\cap\{ \{c_i=1\} \cup \{c_j=1\} \}  \\
        0&\text{otherwise}
    \end{cases}.
$$
Also,
$$
    P_{ij}-\bar p
    =\begin{cases}
        \alpha_2 (p_{11}-p_{12}) + \alpha_3 (p_{11}-p_{22})  &c^*_i=c^*_j=1\\
         \alpha_3 (p_{12}-p_{22}) - \alpha_1 (p_{11}-p_{12}) &\{c^*_i=1, c^*_j=0\} \cup \{c^*_i=0, c^*_j=1\}\\
        -\alpha_1 (p_{11}-p_{22}) - \alpha_2 (p_{12}-p_{22})&c^*_i=c^*_j=0
    \end{cases}.
$$
Now consider the pairs $(i,j)$ such that $i<j$ and $(P_{ij}-\bar p)(\Delta_{ij}(\bc^*)-\Delta_{ij}(\bc)) \neq 0$.
This can only happen when the pair $(i,j)$ belongs to one of the following three sets:
\begin{itemize}
    \item $S_1=
\{i<j\}\cap\{c_i=0\}\cap\{c_j=0\}\cap\{ c^*_i=c^*_j=1 \}$
\item $S_2=
\{i<j\}\cap\{c_i=0\}\cap\{c_j=0\}\cap\{ \{c^*_i=1, c^*_j=0\} \cup \{c^*_i=0, c^*_j=1\} \}$
\item $S_3=
\{i<j\}\cap\{c^*_i=0\}\cap\{c^*_j=0\}\cap\{ \{c_i=1\} \cup \{c_j=1\} \}$
\end{itemize}
Therefore,
$$
    (P_{ij}-\bar p)(\Delta_{ij}(\bc^*)-\Delta_{ij}(\bc))
    =\begin{cases}
        \alpha_2 (p_{11}-p_{12}) + \alpha_3 (p_{11}-p_{22})&(i,j) \in S_1 \\
        \alpha_3 (p_{12}-p_{22}) - \alpha_1 (p_{11}-p_{12})&(i,j) \in S_2 \\
        \alpha_1 (p_{11}-p_{22}) + \alpha_2 (p_{12}-p_{22})&(i,j) \in S_3 \\
        0&\text{otherwise}
    \end{cases}.
$$
Since $\alpha_3 = O(1)$ but $\alpha_1 = O(\alpha_n^2) = o(1)$, we have $\alpha_3 (p_{12}-p_{22}) - \alpha_1 (p_{11}-p_{12}) > 0$, which means that $(P_{ij}-\bar p)(\Delta_{ij}(\bc^*)-\Delta_{ij}(\bc)) > 0$ for $(i,j) \in S_1 \cup S_2 \cup S_3$.
Therefore 
$
\sum_{i<j} (P_{ij}-\bar p)(\Delta_{ij}(\bc^*)-\Delta_{ij}(\bc)) > 0
$
for any $\bc \neq \bc^*$,
which proves that 
$
    \rho(P_n,\bc^*) > \rho(P_n,\bc)
$
for all $\bc \neq \bc^*$.

Next, to precisely characterize $\rho(P_n,\bc^*) - \rho(P_n,\bc)$,  we need to count the number of pairs $(i,j)$ in the sets $S_1$, $S_2$, and $S_3$.
From the definition of $\xi_n({\bc})$, there are $n\xi_n({\bc})$ disagreements between $\bc$ and $\bc^*$.
Let $m$ be the number of vertices such that $c^*_i=1$ and $c_i=0$.
Since $\bc$ and $\bc^*$ have the same core size, 
$$
m = \frac{n\xi_n({\bc})}{2}.
$$

Without loss of generality, suppose $c^*_i=1$ for $i=1, \ldots, k$ and
$c^*_i=0$ for $i=k+1, \ldots, n$.
Then there are $m$ indices, $1\le i_1 < \ldots < i_m \le k$, such that 
$c^*_{i}=1$ and $c_{i}=0$ for $i = i_1 , \ldots , i_m$.
Fix $i=i_1$ and consider the pairs of the form $(i_1,j)$ such that $(i_1,j) \in
S_1$.
This means $c_j=0$ and $c^*_j=1$, which happens for $m$ distinct values of $j$.
But we should only count the pairs where $i<j$, which means we should exclude $j=i_1$, giving us $(m-1)$ pairs of the form $(i_1,j)$ such that  $(i_1,j) \in
S_1$.
Next, fix $i=i_2$.
The same argument plays out, except now we should exclude $j=i_1$ and $j=i_2$ to ensure that $i<j$.
Therefore there are $(m-2)$ pairs of the form $(i_2,j)$ such that $(i_2,j) \in
S_1$.
Proceeding in this manner through $i_1 , \ldots , i_m$, the total number of $(i,j)$ pairs such that $(i,j) \in
S_1$
is given by
$$
(m-1) + (m-2) + \ldots + 0 = \frac{m(m-1)}{2}.
$$

Next, fix $i=i_1$ and consider the pairs of the form $(i_1,j)$ such that $(i_1,j) \in
S_2$.
Since $c^*_i=1$ for $i=1, \ldots, k$ and
$c^*_i=0$ for $i=k+1, \ldots, n$, we do not need to consider the set $\{c^*_i=0, c^*_j=1\}$ because $i$ is always greater than $j$ in this set.
This means that $c_j=0$ and $c^*_j=0$, the vertex $j$ is a periphery vertex that is not misclassified.
This happens for $(n-k-m)$ distinct values of $j$.
Proceeding in this manner through $i_1 , \ldots , i_m$, the total number of $(i,j)$ pairs such that $(i,j) \in
S_2$
is given by
$$
m (n-k -m).
$$

Finally, consider the pairs of the form $(i,j)$ such that $(i,j) \in
S_3$.
The positive and negative entries of $\Delta_{ij}(\bc^*)-\Delta_{ij}(\bc)$ must cancel out, which means that the size of $S_3$ is the same as the combined size of $S_1$ and $S_2$.
Therefore, the number of $(i,j)$ pairs such that $(i,j) \in
S_3$
is given by
$$
(n-k-1) + (n-k-2) + \ldots + (n-k-m) = m \left(n-k - \frac{m+1}{2} \right).
$$

$$
    (P_{ij}-\bar p)(\Delta_{ij}(\bc^*)-\Delta_{ij}(\bc))
    =\begin{cases}
        \alpha_2 (p_{11}-p_{12}) + \alpha_3 (p_{11}-p_{22})&(i,j) \in S_1 \\
        \alpha_3 (p_{12}-p_{22}) - \alpha_1 (p_{11}-p_{12})&(i,j) \in S_2 \\
        \alpha_1 (p_{11}-p_{22}) + \alpha_2 (p_{12}-p_{22})&(i,j) \in S_3 \\
        0&\text{otherwise}
    \end{cases}.
$$

Therefore, 
\begin{align*}
    \sum_{i<j} (P_{ij}-\bar p)(\Delta_{ij}(\bc^*)-\Delta_{ij}(\bc))
    &=
    \sum_{\{i,j\} \in S_1} (P_{ij}-\bar p)(\Delta_{ij}(\bc^*)-\Delta_{ij}(\bc))\\
     &+
    \sum_{\{i,j\} \in S_2} (P_{ij}-\bar p)(\Delta_{ij}(\bc^*)-\Delta_{ij}(\bc))\\
    &+
    \sum_{\{i,j\} \in S_3} (P_{ij}-\bar p)(\Delta_{ij}(\bc^*)-\Delta_{ij}(\bc))\\
    &=
    \frac{m(m-1)}{2} 
    \left[\alpha_2 (p_{11}-p_{12}) + \alpha_3 (p_{11}-p_{22})\right]\\
    &+
    m (n-k -m)\left[\alpha_3 (p_{12}-p_{22}) - \alpha_1 (p_{11}-p_{12})\right]\\
    &+
    m \left(n-k - \frac{m+1}{2} \right)\left[\alpha_1 (p_{11}-p_{22}) + \alpha_2 (p_{12}-p_{22})\right]\\
    &=
    \frac{n^2\xi_n({\bc})}{2}(p_{12}-p_{22}) (1+o(1))
\end{align*}
since 
$\alpha_1 = O(\alpha_n^2)$,
$\alpha_2 = O(\alpha_n)$,
and $m={n\xi_n({\bc})}/{2}$.
Putting the numerator and denominator together,
$$
    \rho(P_n,\bc^*) - \rho(P_n,\bc)
    =\frac{\varrho_n (p_{12}-p_{22})n^2\xi_n({\bc})}{2{n \choose 2} \sqrt{2 \varrho_n \alpha_n p_{22} (1-\varrho_np_{22})}}  (1 + o(1))
    =\frac{p_{12}-p_{22}}{\sqrt{p_{22}}} \sqrt{\frac{\varrho_n}{2\alpha_n(1-\varrho_np_{22})}} \xi_n({\bc})  (1 + o(1)).
$$
This completes the proof. $\square$\newline

\noindent
Comment: Lemma 2 is a deterministic result about the population parameter that does not involve any probabilistic statements. The first part of the lemma states that the population parameter is uniquely maximized at the true labels. The second part characterizes the gap between the true labels and any other set of labels (with the same core size) with respect to the corresponding values of the population parameter.
The proof is based on a counting argument that quantifies a precise connection between the error rate, $\xi_n(\bc)$, and the corresponding difference in the population parameter,  $\rho(P_n,\bc^*) - \rho(P_n,\bc)$.

\subsection*{Lemma 3.} {\sc Lemma 3.} {\it Let $P_n$ be a Chung-Lu model where 
$P_n = \varrho_nP$ and $P_{ij}=\theta_i \theta_j$, where $\theta_1, \ldots, \theta_n$ are the degree parameters.}
{\it Let $\theta_{(1)} \geq \theta_{(2)} \geq \ldots \geq \theta_{(k)} \geq \theta_{(k+1)} \geq \ldots \geq \theta_{(n-1)} \geq \theta_{(n)}$ be the ordered values of the degree parameters.
 Define the ``true'' core-periphery labels as $c^*_i = 1$ for the $k$ vertices with the largest $\theta_i$ values and $c^*_i =0$ for the remaining $(n-k)$ vertices. Then for any $\bc \neq \bc^*$ such that the size of the core is the same for $\bc$ and $ \bc^*$,
and under the assumption that $\theta_{(k)} \theta_{(n)} > \theta_{(k+1)}\theta_{(k+2)}$,
we have
$
    \rho(P_n,\bc) < \rho(P_n,\bc^*),
$
and 
$$
\rho(P_n,\bc^*) - \rho(P_n,\bc) \geq 
\xi_n(\bc) \sqrt{\frac{\varrho_n}{2\alpha_n}} \frac{\theta_{(k)} \theta_{(n)} - \theta_{(k+1)}\theta_{(k+2)}}{\bar{\theta}}  (1 + o(1))
.
$$
}
 {\it Proof.}
  We broadly follow the arguments from the proof of Lemma 2.
Consider
$$
    \rho(P_n,\bc)
    =\frac{\sum_{i<j} \varrho_n (P_{ij}-\bar p)(\Delta_{ij}-\bar\Delta)}{\sqrt{{n\choose 2} \varrho_n \bar p(1- \varrho_n\bar p){n\choose 2}\bar\Delta(1-\bar\Delta)}}.
$$
Now, let $\Delta(\bc^*)$ and $\Delta(\bc)$ denote the $\Delta$ matrices corresponding to the labels $\bc^*$ and $\bc$, respectively.
Since $\bc^*$ and $\bc$ have the same core size, 
we have
$$
1-\bar\Delta(\bc^*)
=
1-\bar\Delta(\bc)
=
\alpha_3 = 
\frac{(n-k)(n-k-1)}{n(n-1)}
= 1 + o(1)
$$
as $k=o(n)$ from A1, and
$$
\bar\Delta(\bc^*)
=
\bar\Delta(\bc)
=
1-\alpha_3
=
\frac{2nk-(k^2+k)}{n(n-1)}
= 2 \alpha_n (1 + o(1)).
$$
Next, we have
$$
    \bar p 
    =  \frac{1}{{n \choose 2}} \sum_{i <j} \theta_i \theta_j
    \sim \bar{\theta}^2.
$$
Therefore, the denominator of $\rho(P_n,\bc)$ and $\rho(P_n,\bc^*)$ is given by
$$
{\sqrt{{n\choose 2} \varrho_n \bar p(1- \varrho_n\bar p){n\choose 2}\bar\Delta(1-\bar\Delta)}}
=
{n \choose 2} \sqrt{2 \varrho_n \alpha_n \bar{\theta}^2 (1-\varrho_n\bar{\theta}^2)} (1 + o(1)).
$$
Now consider the numerators of $\rho(P_n,\bc)$ and $\rho(P_n,\bc^*)$ and note that
$\Delta_{ij}(\cdot)$ is the only term that changes between $\rho(P_n,\bc)$ and $\rho(P_n,\bc^*)$.
Therefore,
$$
    \rho(P_n,\bc^*) - \rho(P_n,\bc)
    =\frac{\sum_{i<j} \varrho_n (P_{ij}-\bar p)(\Delta_{ij}(\bc^*)-\Delta_{ij}(\bc))}
    {{n \choose 2} \sqrt{2 \varrho_n \alpha_n \bar{\theta}^2 (1-\varrho_n\bar{\theta}^2)}}  (1 + o(1)).
$$
Note that
$$
    \Delta_{ij}(\bc^*)-\Delta_{ij}(\bc)
    =\begin{cases}
        +1&\{c_i=0\}\cap\{c_j=0\}\cap\{ \{c^*_i=1\} \cup \{c^*_j=1\} \}  \\
        -1&\{c^*_i=0\}\cap\{c^*_j=0\}\cap\{ \{c_i=1\} \cup \{c_j=1\} \}  \\
        0&\text{otherwise}
    \end{cases}.
$$
Also,
$$
    P_{ij}-\bar p = \theta_i \theta_j - \bar{\theta}^2.
$$
Now consider the pairs $(i,j)$ such that $i<j$ and $(P_{ij}-\bar p)(\Delta_{ij}(\bc^*)-\Delta_{ij}(\bc)) \neq 0$.
This can only happen when the pair $(i,j)$ belongs to one of the following three sets:
\begin{itemize}
    \item $S_1=
\{i<j\}\cap\{c_i=0\}\cap\{c_j=0\}\cap\{ c^*_i=c^*_j=1 \}$
\item $S_2=
\{i<j\}\cap\{c_i=0\}\cap\{c_j=0\}\cap\{ \{c^*_i=1, c^*_j=0\} \cup \{c^*_i=0, c^*_j=1\} \}$
\item $S_3=
\{i<j\}\cap\{c^*_i=0\}\cap\{c^*_j=0\}\cap\{ \{c_i=1\} \cup \{c_j=1\} \}$
\end{itemize}
Next, to precisely characterize $\rho(P_n,\bc^*) - \rho(P_n,\bc)$,  we need to count the number of pairs $(i,j)$ in the sets $S_1$, $S_2$, and $S_3$.
From the definition of $\xi_n({\bc})$, there are $n\xi_n({\bc})$ diasgreements between $\bc$ and $\bc^*$.
Let $m$ be the number of vertices such that $c^*_i=1$ and $c_i=0$.
Since $\bc$ and $\bc^*$ have the same core size, 
$$
m = \frac{n\xi_n({\bc})}{2}.
$$

Without loss of generality, suppose $c^*_i=1$ for $i=1, \ldots, k$ and
$c^*_i=0$ for $i=k+1, \ldots, n$.
Then there are $m$ indices, $1\le i_1 < \ldots < i_m \le k$, such that 
$c^*_{i}=1$ and $c_{i}=0$ for $i = i_1 , \ldots , i_m$.
Fix $i=i_1$ and consider the pairs of the form $(i_1,j)$ such that $(i_1,j) \in
S_1$.
This means $c_j=0$ and $c^*_j=1$, which happens for $m$ distinct values of $j$.
But we should only count the pairs where $i<j$, which means we should exclude $j=i_1$, giving us $(m-1)$ pairs of the form $(i_1,j)$ such that  $(i_1,j) \in
S_1$.
Next, fix $i=i_2$.
The same argument plays out, except now we should exclude $j=i_1$ and $j=i_2$ to ensure that $i<j$.
Therefore there are $(m-2)$ pairs of the form $(i_2,j)$ such that $(i_2,j) \in
S_1$.
Proceeding in this manner through $i_1 , \ldots , i_m$, the total number of $(i,j)$ pairs such that $(i,j) \in
S_1$
is given by
$$
(m-1) + (m-2) + \ldots + 0 = \frac{m(m-1)}{2}.
$$

Next, fix $i=i_1$ and consider the pairs of the form $(i_1,j)$ such that $(i_1,j) \in
S_2$.
Since $c^*_i=1$ for $i=1, \ldots, k$ and
$c^*_i=0$ for $i=k+1, \ldots, n$, we do not need to consider the set $\{c^*_i=0, c^*_j=1\}$ because $i$ is always greater than $j$ in this set.
This means that $c_j=0$ and $c^*_j=0$, the vertex $j$ is a periphery vertex that is not misclassified.
This happens for $(n-k-m)$ distinct values of $j$.
Proceeding in this manner through $i_1 , \ldots , i_m$, the total number of $(i,j)$ pairs such that $(i,j) \in
S_2$
is given by
$$
m (n-k -m).
$$

Finally, consider the pairs of the form $(i,j)$ such that $(i,j) \in
S_3$.
The positive and negative entries of $\Delta_{ij}(\bc^*)-\Delta_{ij}(\bc)$ must cancel out, which means that the size of $S_3$ is the same as the combined size of $S_1$ and $S_2$.
Therefore, the number of $(i,j)$ pairs such that $(i,j) \in
S_3$
is given by
$$
(n-k-1) + (n-k-2) + \ldots + (n-k-m) = m \left(n-k - \frac{m+1}{2} \right).
$$

Note that $\Delta_{ij}(\bc^*)-\Delta_{ij}(\bc) = 1$ when $(i,j) \in S_1 \cup S_2$ and 
$\Delta_{ij}(\bc^*)-\Delta_{ij}(\bc) = -1$ when $(i,j) \in S_3$.
Also, the number of $(i,j)$ pairs in $S_1 \cup S_2$ is the same as the number of $(i,j)$ pairs in $S_3$.
Finally for any $(i,j) \in S_1 \cup S_2$ we have $P_{ij} \geq \theta_{(k)} \theta_{(n)}$ and 
for any $(i,j) \in S_3$ we have $P_{ij} \le \theta_{(k+1)}\theta_{(k+2)}$.
Therefore, 
\begin{align*}
    \sum_{i<j} (P_{ij}-\bar p)(\Delta_{ij}(\bc^*)-\Delta_{ij}(\bc))
    &=
    \sum_{\{i,j\} \in S_1 \cup S_2} (P_{ij}-\bar p)
    -
    \sum_{\{i,j\} \in S_3} (P_{ij}-\bar p)\\
    &=
    \sum_{\{i,j\} \in S_1 \cup S_2} P_{ij} - \sum_{\{i,j\} \in S_3} P_{ij}\\
    &\geq
    m (n-k -m)\left(\theta_{(k)} \theta_{(n)} - \theta_{(k+1)}\theta_{(k+2)}\right)\\
      &=
    \frac{n^2\xi_n({\bc})}{2}\left(\theta_{(k)} \theta_{(n)} - \theta_{(k+1)}\theta_{(k+2)}\right) (1+o(1)).
\end{align*}
Putting the numerator and denominator together,
\begin{align*}
    \rho(P_n,\bc^*) - \rho(P_n,\bc)
    &\geq\frac{\varrho_n 
    (\theta_{(k)} \theta_{(n)} - \theta_{(k+1)}\theta_{(k+2)})n^2\xi_n({\bc})}{2{n \choose 2} \sqrt{2 \varrho_n \alpha_n \bar{\theta}^2 (1-\varrho_n \bar{\theta}^2)}}  (1 + o(1))\\
    &=\frac{\theta_{(k)} \theta_{(n)} - \theta_{(k+1)}\theta_{(k+2)}}{\bar{\theta}} \sqrt{\frac{\varrho_n}{2\alpha_n(1-\varrho_n\bar{\theta}^2)}} \xi_n({\bc})  (1 + o(1)).
\end{align*}
This completes the proof. $\square$\newline

\noindent
Comment: The extra assumption $\theta_{(k)} \theta_{(n)} > \theta_{(k+1)}\theta_{(k+2)}$ ensures that the product of the smallest degree parameters in the core and periphery is larger than the product of the two largest degree parameters in the periphery. Moreover, notice that if $\theta_{(1)}=\cdots=\theta_{(k)}$ and $\theta_{(k+1)}=\cdots=\theta_{(n)}$, then the CL model becomes the SBM model and we recover the result in Lemma 1.

\subsection*{Lemma 4.} {\sc Lemma 4.} {\it Let $P_n$ be a CP-DCBM model where 
$P_n = \varrho_nP$ and $P_{ij}=\theta_i \Omega_{c^*_i c^*_j} \theta_j$, where $\theta_1, \ldots, \theta_n$ are the degree parameters, and }
$$
    \Omega=
    \begin{pmatrix}
        p_{11}&p_{12}\\
        p_{21}&p_{22}
    \end{pmatrix}
    \text{ and } p_{11} > p_{12} = p_{21} > p_{22}.
$$
{\it Let $\theta_c = \min\{\theta_i: c^*_i = 1 \}$ be the lowest degree parameter in the core set, and $\theta_{p, max} = \max\{\theta_i: c^*_i =0\}$ and $\theta_{p, min} = \min\{\theta_i: c^*_i =0\}$ be the highest and lowest degree parameters in the periphery set, respectively.
Then for any $\bc \neq \bc^*$ such that the size of the core is the same for $\bc$ and $ \bc^*$,
and under the assumption that 
$$
 \frac{p_{12}}{p_{22}} > \frac{\theta_{p, max}^2}{\theta_c \theta_{p, min}} ,
$$
we have
$
    \rho(P_n,\bc) < \rho(P_n,\bc^*),
$
and 
$$
\rho(P_n,\bc^*) - \rho(P_n,\bc) \geq 
\xi_n(\bc) \sqrt{\frac{\varrho_n}{2\alpha_n}} \frac{p_{12} \theta_c \theta_{p, min} - p_{22}\theta_{p, max}^2}{\sqrt{\bar{p}}}  (1 + o(1))
.
$$
}
 {\it Proof.}
  We broadly follow the arguments from the proofs of Lemma 2 and Lemma 3.
Consider
$$
    \rho(P_n,\bc)
    =\frac{\sum_{i<j} \varrho_n (P_{ij}-\bar p)(\Delta_{ij}-\bar\Delta)}{\sqrt{{n\choose 2} \varrho_n \bar p(1- \varrho_n\bar p){n\choose 2}\bar\Delta(1-\bar\Delta)}}.
$$
Now, let $\Delta(\bc^*)$ and $\Delta(\bc)$ denote the $\Delta$ matrices corresponding to the labels $\bc^*$ and $\bc$, respectively.
Since $\bc^*$ and $\bc$ have the same core size, 
we have
$$
1-\bar\Delta(\bc^*)
=
1-\bar\Delta(\bc)
=
\alpha_3 = 
\frac{(n-k)(n-k-1)}{n(n-1)}
= 1 + o(1)
$$
as $k=o(n)$ from A1, and
$$
\bar\Delta(\bc^*)
=
\bar\Delta(\bc)
=
1-\alpha_3
=
\frac{2nk-(k^2+k)}{n(n-1)}
= 2 \alpha_n (1 + o(1)).
$$
Next, we have
$$
    \bar p 
    =  \frac{1}{{n \choose 2}} \sum_{i <j} P_{ij}.
$$
Therefore, the denominator of $\rho(P_n,\bc)$ and $\rho(P_n,\bc^*)$ is given by
$$
{\sqrt{{n\choose 2} \varrho_n \bar p(1- \varrho_n\bar p){n\choose 2}\bar\Delta(1-\bar\Delta)}}
=
{n \choose 2} \sqrt{2 \varrho_n \alpha_n \bar p (1-\varrho_n\bar p)} (1 + o(1)).
$$
Now consider the numerators of $\rho(P_n,\bc)$ and $\rho(P_n,\bc^*)$ and note that
$\Delta_{ij}(\cdot)$ is the only term that changes between $\rho(P_n,\bc)$ and $\rho(P_n,\bc^*)$.
Therefore,
$$
    \rho(P_n,\bc^*) - \rho(P_n,\bc)
    =\frac{\sum_{i<j} \varrho_n (P_{ij}-\bar p)(\Delta_{ij}(\bc^*)-\Delta_{ij}(\bc))}
    {{n \choose 2} \sqrt{2 \varrho_n \alpha_n \bar p (1-\varrho_n\bar p)}}  (1 + o(1)).
$$
Note that
$$
    \Delta_{ij}(\bc^*)-\Delta_{ij}(\bc)
    =\begin{cases}
        +1&\{c_i=0\}\cap\{c_j=0\}\cap\{ \{c^*_i=1\} \cup \{c^*_j=1\} \}  \\
        -1&\{c^*_i=0\}\cap\{c^*_j=0\}\cap\{ \{c_i=1\} \cup \{c_j=1\} \}  \\
        0&\text{otherwise}
    \end{cases}.
$$
Now consider the pairs $(i,j)$ such that $i<j$ and $(P_{ij}-\bar p)(\Delta_{ij}(\bc^*)-\Delta_{ij}(\bc)) \neq 0$.
This can only happen when the pair $(i,j)$ belongs to one of the following three sets:
\begin{itemize}
    \item $S_1=
\{i<j\}\cap\{c_i=0\}\cap\{c_j=0\}\cap\{ c^*_i=c^*_j=1 \}$
\item $S_2=
\{i<j\}\cap\{c_i=0\}\cap\{c_j=0\}\cap\{ \{c^*_i=1, c^*_j=0\} \cup \{c^*_i=0, c^*_j=1\} \}$
\item $S_3=
\{i<j\}\cap\{c^*_i=0\}\cap\{c^*_j=0\}\cap\{ \{c_i=1\} \cup \{c_j=1\} \}$
\end{itemize}
Next, to precisely characterize $\rho(P_n,\bc^*) - \rho(P_n,\bc)$,  we need to count the number of pairs $(i,j)$ in the sets $S_1$, $S_2$, and $S_3$.
From the definition of $\xi_n({\bc})$, there are $n\xi_n({\bc})$ diasgreements between $\bc$ and $\bc^*$.
Let $m$ be the number of vertices such that $c^*_i=1$ and $c_i=0$.
Since $\bc$ and $\bc^*$ have the same core size, 
$$
m = \frac{n\xi_n({\bc})}{2}.
$$

Without loss of generality, suppose $c^*_i=1$ for $i=1, \ldots, k$ and
$c^*_i=0$ for $i=k+1, \ldots, n$.
Then there are $m$ indices, $1\le i_1 < \ldots < i_m \le k$, such that 
$c^*_{i}=1$ and $c_{i}=0$ for $i = i_1 , \ldots , i_m$.
Fix $i=i_1$ and consider the pairs of the form $(i_1,j)$ such that $(i_1,j) \in
S_1$.
This means $c_j=0$ and $c^*_j=1$, which happens for $m$ distinct values of $j$.
But we should only count the pairs where $i<j$, which means we should exclude $j=i_1$, giving us $(m-1)$ pairs of the form $(i_1,j)$ such that  $(i_1,j) \in
S_1$..
Next, fix $i=i_2$.
The same argument plays out, except now we should exclude $j=i_1$ and $j=i_2$ to ensure that $i<j$.
Therefore there are $(m-2)$ pairs of the form $(i_2,j)$ such that $(i_2,j) \in
S_1$.
Proceeding in this manner through $i_1 , \ldots , i_m$, the total number of $(i,j)$ pairs such that $(i,j) \in
S_1$
is given by
$$
(m-1) + (m-2) + \ldots + 0 = \frac{m(m-1)}{2}.
$$

Next, fix $i=i_1$ and consider the pairs of the form $(i_1,j)$ such that $(i_1,j) \in
S_2$.
Since $c^*_i=1$ for $i=1, \ldots, k$ and
$c^*_i=0$ for $i=k+1, \ldots, n$, we do not need to consider the set $\{c^*_i=0, c^*_j=1\}$ because $i$ is always greater than $j$ in this set.
This means that $c_j=0$ and $c^*_j=0$, the vertex $j$ is a periphery vertex that is not misclassified.
This happens for $(n-k-m)$ distinct values of $j$.
Proceeding in this manner through $i_1 , \ldots , i_m$, the total number of $(i,j)$ pairs such that $(i,j) \in
S_2$
is given by
$$
m (n-k -m).
$$

Finally, consider the pairs of the form $(i,j)$ such that $(i,j) \in
S_3$.
The positive and negative entries of $\Delta_{ij}(\bc^*)-\Delta_{ij}(\bc)$ must cancel out, which means that the size of $S_3$ is the same as the combined size of $S_1$ and $S_2$.
Therefore, the number of $(i,j)$ pairs such that $(i,j) \in
S_3$
is given by
$$
(n-k-1) + (n-k-2) + \ldots + (n-k-m) = m \left(n-k - \frac{m+1}{2} \right).
$$

Note that $\Delta_{ij}(\bc^*)-\Delta_{ij}(\bc) = 1$ when $(i,j) \in S_1 \cup S_2$ and 
$\Delta_{ij}(\bc^*)-\Delta_{ij}(\bc) = -1$ when $(i,j) \in S_3$.
Also, the number of $(i,j)$ pairs in $S_1 \cup S_2$ is the same as the number of $(i,j)$ pairs in $S_3$.
Finally for any $(i,j) \in S_1 \cup S_2$ we have
 $P_{ij} \geq p_{12} \theta_c \theta_{p, min}$ 
and 
for any $(i,j) \in S_3$ we have 
$P_{ij} \le p_{22}\theta_{p, max}^2$.
Therefore, 
\begin{align*}
    \sum_{i<j} (P_{ij}-\bar p)(\Delta_{ij}(\bc^*)-\Delta_{ij}(\bc))
    &=
    \sum_{\{i,j\} \in S_1 \cup S_2} (P_{ij}-\bar p)
    -
    \sum_{\{i,j\} \in S_3} (P_{ij}-\bar p)\\
    &=
    \sum_{\{i,j\} \in S_1 \cup S_2} P_{ij} - \sum_{\{i,j\} \in S_3} P_{ij}\\
    &\geq
    m (n-k -m)\left(p_{12} \theta_c \theta_{p, min} - p_{22}\theta_{p, max}^2\right)\\
      &=
    \frac{n^2\xi_n({\bc})}{2}\left(p_{12} \theta_c \theta_{p, min} - p_{22}\theta_{p, max}^2\right) (1+o(1)).
\end{align*}
Putting the numerator and denominator together,
\begin{align*}
    \rho(P_n,\bc^*) - \rho(P_n,\bc)
    &\geq
    \frac{\varrho_n 
    (p_{12} \theta_c \theta_{p, min} - p_{22}\theta_{p, max}^2)n^2\xi_n({\bc})}{2{n \choose 2} \sqrt{2 \varrho_n \alpha_n \bar p (1-\varrho_n \bar p)}}  (1 + o(1))\\
    &=
    \frac{p_{12} \theta_c \theta_{p, min} - p_{22}\theta_{p, max}^2}{\sqrt{\bar p}} \sqrt{\frac{\varrho_n}{2\alpha_n(1-\varrho_n\bar{ p)}}} \xi_n({\bc})  (1 + o(1)).
\end{align*}
This completes the proof. $\square$\newline

\noindent
Comment: Notice that $p_{12}\theta_c\theta_{p,min}$ is the minimum core-periphery edge probability, and $p_{22}\theta^2_{p,max}$ is the largest periphery-periphery edge probability. Thus, this extra assumption ensures that all core-periphery edge probabilities are larger than all periphery-periphery edge probabilities, as sensible requirement for a CP model. Also, this condition reduces to $p_{12}>p_{22}$ in the SBM case ($\theta_1=\cdots\theta_n$) and $\theta_c\theta_{p,min}>\theta_{p,max}^2$ in the CL case $(p_{12}=p_{22})$, recovering our previous results.

\subsection*{Proof of Theorem 2.1}
\subsubsection*{SBM}
\textit{{ Let $P_n$ be a two-block SBM where 
$P_n = \varrho_nP$ and $P_{ij}=\Omega_{c^*_i c^*_j}$ where}
$$
    \Omega=
    \begin{pmatrix}
        p_{11}&p_{12}\\
        p_{21}&p_{22}
    \end{pmatrix}
    \text{ and } p_{11} > p_{12} = p_{21} > p_{22}.
$$
{Under A1, 
 we have}
$$
\mathbb P\left(\xi_n({\hat{\bc}}) < \frac{\sqrt{p_{22}}\alpha_n }{p_{12}-p_{22}} \sqrt{\frac{(8+\eta) \log(n\varrho_n\alpha_n)}{n\varrho_n}}\right)
\rightarrow 1 
\text{ for any } \eta >0
\text{ as }  n\to\infty
.
$$
{
Furthermore, 
the optimal labels are weakly consistent under A1, i.e., for any $\delta > 0$},
\begin{equation}
    \mathbb P(\xi_n(\hat{\bc})>\delta)
    \to0 \text{ as }  n\to\infty,
\end{equation}
{ and under the additional assumption that $
{\alpha_n}/{\sqrt{\varrho_n}} = o \left({1}/{\sqrt{n \log n }} \right)
$, the labels are strongly consistent, i.e.,} 
\begin{equation}
    \mathbb P\left({\hat{\bc}} = \bc^* \right)
    \rightarrow 1 \text{ as }  n\to\infty.
\end{equation}}

\noindent
\textit{Proof:}
Recall that 
$$
\hat{\bc} = \arg \max_{\bc} T(A_n,\bc),
$$
which means that $T(A_n,\hat{\bc}) \geq T(A_n,{\bc^*})$.
Consider the event
$$
E = \{T(A_n,\hat{\bc}) \le \rho(P_n,\hat{\bc}) + \varepsilon_n \} \cap \{T(A_n,{\bc^*}) \geq \rho(P_n,{\bc^*})-\varepsilon_n \}
$$
where $\varepsilon_n=\alpha_n^{1/2}\varrho_n^{1/2}
\sqrt{\frac{\log(n\varrho_n\alpha_n)}{n\varrho_n\alpha_n}}$.
Then from Lemma 1, we have $\mathbb P(E) \rightarrow 1$ as  $n\to\infty$.
Note that $E$ implies that $\rho(P_n,\bc^*) - \rho(P_n,\hat{\bc}) < 2 \varepsilon_n$.
But, from Lemma 2, we know that 
$$
    \rho(P_n,\bc^*) - \rho(P_n,\hat{\bc})
        =\frac{p_{12}-p_{22}}{\sqrt{p_{22}}} \sqrt{\frac{\varrho_n}{2\alpha_n(1-\varrho_np_{22})}} \xi_n({\bc})  (1 + o(1)).
$$
This means that $E$ can happen only if 
$$
\frac{p_{12}-p_{22}}{\sqrt{p_{22}}} \sqrt{\frac{\varrho_n}{2\alpha_n(1-\varrho_np_{22})}} \xi_n({\bc})
< 2 \varepsilon_n  (1 + o(1))
\Rightarrow
\xi_n({\hat{\bc}}) < \frac{\sqrt{p_{22}}\alpha_n }{p_{12}-p_{22}} \sqrt{\frac{8(1-\varrho_nb) \log(n\varrho_n\alpha_n)}{n\varrho_n}}(1 + o(1)).
$$
Therefore, for any $\eta >0$
$$
\mathbb P\left(\xi_n({\hat{\bc}}) < 
\frac{\sqrt{p_{22}}\alpha_n }{p_{12}-p_{22}} \sqrt{\frac{(8+\eta)\log(n\varrho_n\alpha_n)}{n\varrho_n}}\right)
\rightarrow 1 \text{ as }  n\to\infty.
$$

\noindent \textbf{Weak consistency:}
Under A1, as $n\to\infty$, 
$$n\varrho_n\alpha_n\to\infty  
\Rightarrow
\frac{\log(n\varrho_n\alpha_n)}{n\varrho_n\alpha_n}\to 0
\Rightarrow
\frac{\sqrt{p_{22}}\alpha_n }{p_{12}-p_{22}} \sqrt{\frac{(8+\eta)\log(n\varrho_n\alpha_n)}{n\varrho_n}}\to 0.
$$
Now fix any $\delta >0$.
There is some $n_0$ such that
$$
\delta > \frac{\sqrt{p_{22}}\alpha_n }{p_{12}-p_{22}} \sqrt{\frac{(8+\eta)\log(n\varrho_n\alpha_n)}{n\varrho_n}}
$$
for all $n > n_0$.
Therefore,
$$
\lim_{n\to\infty} 
\mathbb P\left(\xi_n({\hat{\bc}}) < \delta\right)
\geq 
\lim_{n\to\infty} 
\mathbb P\left(\xi_n({\hat{\bc}}) < \frac{\sqrt{p_{22}}\alpha_n }{p_{12}-p_{22}} \sqrt{\frac{(8+\eta)\log(n\varrho_n\alpha_n)}{n\varrho_n}}\right)
= 1.
$$

\noindent \textbf{Strong consistency:}
Suppose that 
$$
\frac{\alpha_n}{\sqrt{\varrho_n}} = o \left(\frac{1}{\sqrt{n \log n }} \right).
$$
Then 
$$
\frac{\sqrt{p_{22}}\alpha_n }{p_{12}-p_{22}} \sqrt{\frac{(8+\eta)\log(n\varrho_n\alpha_n)}{n\varrho_n}}
=
o \left(\sqrt{\frac{\log(n\varrho_n\alpha_n)}{n^2 \log n }}  \right)
=
o\left(\frac{1}{n} \right).
$$
Therefore,
$$
\mathbb P\left({\hat{\bc}} = \bc^* \right)
=
\mathbb P\left(\xi_n({\hat{\bc}}) < \frac{1}{n} \right)
\rightarrow 1
\text{ as } n\to\infty.
$$
This completes the proof.
$\square$

\subsubsection*{Chung-Lu}
\textit{{\it Let $P_n$ be a Chung-Lu model where 
$P_n = \varrho_nP$ and $P_{ij}=\theta_i \theta_j$, where $\theta_1, \ldots, \theta_n$ are the degree parameters.
Recall the definition of $ \bc^*$ from Lemma 3.}
{Under A1 and the assumption in Lemma 3, 
 we have}
$$
\mathbb P\left(\xi_n({\hat{\bc}}) < \frac{\bar{\theta}\alpha_n }{\theta_{(k)} \theta_{(n)} - \theta_{(k+1)}\theta_{(k+2)}} \sqrt{\frac{(8+\eta) \log(n\varrho_n\alpha_n)}{n\varrho_n}}\right)
\rightarrow 1 
\text{ for any } \eta >0
\text{ as }  n\to\infty
.
$$
{
Furthermore, 
the optimal labels are weakly consistent under A1 and the assumption in Lemma 3, i.e., for any $\delta > 0$},
\begin{equation}
    \mathbb P(\xi_n(\hat{\bc})>\delta)
    \to0 \text{ as }  n\to\infty,
\end{equation}
{ and under the additional assumption that $
{\alpha_n}/{\sqrt{\varrho_n}} = o \left({1}/{\sqrt{n \log n }} \right)
$, the optimal labels are strongly consistent, i.e.,} 
\begin{equation}
    \mathbb P\left({\hat{\bc}} = \bc^* \right)
    \rightarrow 1 \text{ as }  n\to\infty.
\end{equation}}

\textit{Proof:}
 We broadly follow the arguments from the proofs of Theorem 2.1. Recall that 
$$
\hat{\bc} = \arg \max_{\bc} T(A_n,\bc),
$$
which means that $T(A_n,\hat{\bc}) \geq T(A_n,{\bc^*})$.
Consider the event
$$
E = \{T(A_n,\hat{\bc}) \le \rho(P_n,\hat{\bc}) + \varepsilon_n \} \cap \{T(A_n,{\bc^*}) \geq \rho(P_n,{\bc^*})-\varepsilon_n \}
$$
where $\varepsilon_n=\alpha_n\varrho_n^{1/2}
\sqrt{\frac{\log(n\varrho_n\alpha_n)}{n\varrho_n\alpha_n}}$.
Then from Lemma 1, we have $\mathbb P(E) \rightarrow 1$ as  $n\to\infty$.
Note that $E$ implies that $\rho(P_n,\bc^*) - \rho(P_n,\hat{\bc}) < 2 \varepsilon_n$.
But, from Lemma 3, we know that 
$$
\rho(P_n,\bc^*) - \rho(P_n,\bc) \geq 
\xi_n(\bc) \sqrt{\frac{\varrho_n}{2\alpha_n}} 
\frac{\theta_{(k)} \theta_{(n)} - \theta_{(k+1)}\theta_{(k+2)}}{\bar{\theta}}  
(1 + o(1))
.
$$
Let
$$
\beta = \frac{\theta_{(k)} \theta_{(n)} - \theta_{(k+1)}\theta_{(k+2)}}{\bar{\theta}}.
$$
Then $E$ can happen only if 
$$
\beta \sqrt{\frac{\varrho_n}{2\alpha_n}} \xi_n({\bc})
< 2 \varepsilon_n  (1 + o(1))
\Rightarrow
\xi_n({\hat{\bc}}) < 
\frac{\alpha_n}{\beta} 
\sqrt{\frac{8\log(n\varrho_n\alpha_n)}{n\varrho_n}}(1 + o(1)).
$$
Therefore, for any $\eta >0$
$$
\mathbb P\left(\xi_n({\hat{\bc}}) < 
\frac{\alpha_n}{\beta}  \sqrt{\frac{(8+\eta)\log(n\varrho_n\alpha_n)}{n\varrho_n}}\right)
\rightarrow 1 \text{ as }  n\to\infty.
$$

\noindent \textbf{Weak consistency:}
Under A1, as $n\to\infty$, 
$$n\varrho_n\alpha_n\to\infty  
\Rightarrow
\frac{\log(n\varrho_n\alpha_n)}{n\varrho_n\alpha_n}\to 0
\Rightarrow
\frac{\alpha_n}{\beta} \sqrt{\frac{(8+\eta)\log(n\varrho_n\alpha_n)}{n\varrho_n}}\to 0.
$$
Now fix any $\delta >0$.
There is some $n_0$ such that
$$
\delta > \frac{\alpha_n}{\beta} \sqrt{\frac{(8+\eta)\log(n\varrho_n\alpha_n)}{n\varrho_n}}
$$
for all $n > n_0$.
Therefore,
$$
\lim_{n\to\infty} 
\mathbb P\left(\xi_n({\hat{\bc}}) < \delta\right)
\geq 
\lim_{n\to\infty} 
\mathbb P\left(\xi_n({\hat{\bc}}) < \frac{\alpha_n}{\beta} \sqrt{\frac{(8+\eta)\log(n\varrho_n\alpha_n)}{n\varrho_n}}\right)
= 1.
$$

\noindent \textbf{Strong consistency:}
Suppose that 
$$
\frac{\alpha_n}{\sqrt{\varrho_n}} = o \left(\frac{1}{\sqrt{n \log n }} \right).
$$
Then 
$$
\frac{\alpha_n}{\beta} \sqrt{\frac{(8+\eta)\log(n\varrho_n\alpha_n)}{n\varrho_n}}
=
o \left(\sqrt{\frac{\log(n\varrho_n\alpha_n)}{n^2 \log n }}  \right)
=
o\left(\frac{1}{n} \right).
$$
Therefore,
$$
\mathbb P\left({\hat{\bc}} = \bc^* \right)
=
\mathbb P\left(\xi_n({\hat{\bc}}) < \frac{1}{n} \right)
\rightarrow 1
\text{ as } n\to\infty.
$$
This completes the proof.
$\square$

{\it Let $P_n$ be a CP-DCBM model where 
$P_n = \varrho_nP$ and $P_{ij}=\theta_i \Omega_{c^*_i c^*_j} \theta_j$, where $\theta_1, \ldots, \theta_n$ are the degree parameters, and }
$$
    \Omega=
    \begin{pmatrix}
        p_{11}&p_{12}\\
        p_{21}&p_{22}
    \end{pmatrix}
    \text{ and } p_{11} > p_{12} = p_{21} > p_{22}.
$$
{\it Let $\theta_c = \min\{\theta_i: c^*_i = 1 \}$ be the lowest degree parameter in the core set, and $\theta_{p, max} = \max\{\theta_i: c^*_i =0\}$ and $\theta_{p, min} = \min\{\theta_i: c^*_i =0\}$ be the highest and lowest degree parameters in the periphery set, respectively.
Then for any $\bc \neq \bc^*$ such that the size of the core is the same for $\bc$ and $ \bc^*$,
and under the assumption that 
$$
 \frac{p_{12}}{p_{22}} > \frac{\theta_{p, max}^2}{\theta_c \theta_{p, min}} ,
$$
we have
$
    \rho(P_n,\bc) < \rho(P_n,\bc^*),
$
and 
$$
\rho(P_n,\bc^*) - \rho(P_n,\bc) \geq 
\xi_n(\bc) \sqrt{\frac{\varrho_n}{2\alpha_n}} \frac{p_{12} \theta_c \theta_{p, min} - p_{22}\theta_{p, max}^2}{\sqrt{\bar{p}}}  (1 + o(1))
.
$$
}

\subsubsection*{DCBM}
\textit{{\it Let $P_n$ be a CP-DCBM model where 
$P_n = \varrho_nP$ and $P_{ij}=\theta_i \Omega_{c^*_i c^*_j} \theta_j$, where $\theta_1, \ldots, \theta_n$ are the degree parameters, and }
$$
    \Omega=
    \begin{pmatrix}
        p_{11}&p_{12}\\
        p_{21}&p_{22}
    \end{pmatrix}
    \text{ and } p_{11} > p_{12} = p_{21} > p_{22}.
$$
{Under A1 and the assumption in Lemma 4, 
 we have}
$$
\mathbb P\left(\xi_n({\hat{\bc}}) < \frac{\sqrt{\bar{p}}\alpha_n }
{p_{12} \theta_c \theta_{p, min} - p_{22}\theta_{p, max}^2} 
\sqrt{\frac{(8+\eta) \log(n\varrho_n\alpha_n)}{n\varrho_n}}\right)
\rightarrow 1 
\text{ for any } \eta >0
\text{ as }  n\to\infty
.
$$
{
Furthermore, 
the optimal labels are weakly consistent under A1 and the assumption in Lemma 3, i.e., for any $\delta > 0$},
\begin{equation}
    \mathbb P(\xi_n(\hat{\bc})>\delta)
    \to0 \text{ as }  n\to\infty,
\end{equation}
{ and under the additional assumption that $
{\alpha_n}/{\sqrt{\varrho_n}} = o \left({1}/{\sqrt{n \log n }} \right)
$, the optimal labels are strongly consistent, i.e.,} 
\begin{equation}
    \mathbb P\left({\hat{\bc}} = \bc^* \right)
    \rightarrow 1 \text{ as }  n\to\infty.
\end{equation}}

\textit{Proof:}
 We broadly follow the arguments from the proofs of Theorem 2.1. Recall that 
$$
\hat{\bc} = \arg \max_{\bc} T(A_n,\bc),
$$
which means that $T(A_n,\hat{\bc}) \geq T(A_n,{\bc^*})$.
Consider the event
$$
E = \{T(A_n,\hat{\bc}) \le \rho(P_n,\hat{\bc}) + \varepsilon_n \} \cap \{T(A_n,{\bc^*}) \geq \rho(P_n,{\bc^*})-\varepsilon_n \}
$$
where $\varepsilon_n=\alpha_n\varrho_n^{1/2}
\sqrt{\frac{\log(n\varrho_n\alpha_n)}{n\varrho_n\alpha_n}}$.
Then from Lemma 1, we have $\mathbb P(E) \rightarrow 1$ as  $n\to\infty$.
Note that $E$ implies that $\rho(P_n,\bc^*) - \rho(P_n,\hat{\bc}) < 2 \varepsilon_n$.
But, from Lemma 4, we know that 
$$
\rho(P_n,\bc^*) - \rho(P_n,\bc) \geq 
\xi_n(\bc) \sqrt{\frac{\varrho_n}{2\alpha_n}} 
\frac{p_{12} \theta_c \theta_{p, min} - p_{22}\theta_{p, max}^2}{\sqrt{\bar{p}}} 
(1 + o(1))
.
$$
Let
$$
\beta = \frac{p_{12} \theta_c \theta_{p, min} - p_{22}\theta_{p, max}^2}{\sqrt{\bar{p}}}.
$$
Then $E$ can happen only if 
$$
\beta \sqrt{\frac{\varrho_n}{2\alpha_n}} \xi_n({\bc})
< 2 \varepsilon_n  (1 + o(1))
\Rightarrow
\xi_n({\hat{\bc}}) < 
\frac{\alpha_n}{\beta} 
\sqrt{\frac{8\log(n\varrho_n\alpha_n)}{n\varrho_n}}(1 + o(1)).
$$
Therefore, for any $\eta >0$
$$
\mathbb P\left(\xi_n({\hat{\bc}}) < 
\frac{\alpha_n}{\beta}  \sqrt{\frac{(8+\eta)\log(n\varrho_n\alpha_n)}{n\varrho_n}}\right)
\rightarrow 1 \text{ as }  n\to\infty.
$$

\noindent \textbf{Weak consistency:}
Under A1, as $n\to\infty$, 
$$n\varrho_n\alpha_n\to\infty  
\Rightarrow
\frac{\log(n\varrho_n\alpha_n)}{n\varrho_n\alpha_n}\to 0
\Rightarrow
\frac{\alpha_n}{\beta} \sqrt{\frac{(8+\eta)\log(n\varrho_n\alpha_n)}{n\varrho_n}}\to 0.
$$
Now fix any $\delta >0$.
There is some $n_0$ such that
$$
\delta > \frac{\alpha_n}{\beta} \sqrt{\frac{(8+\eta)\log(n\varrho_n\alpha_n)}{n\varrho_n}}
$$
for all $n > n_0$.
Therefore,
$$
\lim_{n\to\infty} 
\mathbb P\left(\xi_n({\hat{\bc}}) < \delta\right)
\geq 
\lim_{n\to\infty} 
\mathbb P\left(\xi_n({\hat{\bc}}) < \frac{\alpha_n}{\beta} \sqrt{\frac{(8+\eta)\log(n\varrho_n\alpha_n)}{n\varrho_n}}\right)
= 1.
$$

\noindent \textbf{Strong consistency:}
Suppose that 
$$
\frac{\alpha_n}{\sqrt{\varrho_n}} = o \left(\frac{1}{\sqrt{n \log n }} \right).
$$
Then 
$$
\frac{\alpha_n}{\beta} \sqrt{\frac{(8+\eta)\log(n\varrho_n\alpha_n)}{n\varrho_n}}
=
o \left(\sqrt{\frac{\log(n\varrho_n\alpha_n)}{n^2 \log n }}  \right)
=
o\left(\frac{1}{n} \right).
$$
Therefore,
$$
\mathbb P\left({\hat{\bc}} = \bc^* \right)
=
\mathbb P\left(\xi_n({\hat{\bc}}) < \frac{1}{n} \right)
\rightarrow 1
\text{ as } n\to\infty.
$$
This completes the proof.
$\square$
\clearpage

\section{Technical proofs for hypothesis tests}
\subsection*{Proof of Theorem 3.1}

\noindent
{\sc Theorem 3.1.} \textit{Under $H_0$, i.e., when the network is generated from the ER model, the Type I error rate converges to 0, i.e., for any $\eta>0$,}
$
    \lim_{n\to\infty} \mathbb P\left[\{T_1(A) > C_1\}\cap \{T_2(A)>C_2\}\right] < \eta,
$
\textit{where the rejection thresholds are given by}
\begin{equation}
C_1 = \sqrt{\frac{\log (n \varrho_n \alpha_n)}{n}} 
\text{ and } C_2 = \frac{2\sqrt{2} \varrho_n \log(n)}{\sqrt{k}}.
\label{eq:inter_er}
\end{equation}
{\it Under the assumption that} 
$
\alpha_n \geq \frac{(\log (n \log n))^2}{n}
$,
\textit{ suppose that any of the following alternative hypotheses hold:}
\begin{itemize}
    \item $H_1:$ $P$ is a CP-SBM where 
$p_{11} > p_{12} = p_{21} > p_{22}$
\item $H_1:$ $P$ is a Chung-Lu model where 
$\theta_{(k)} \theta_{(n)} > \theta_{(k+1)}\theta_{(k+2)}$
and $\theta_{(k-1)}>\theta_{(k+1)}$.
\item $H_1:$ $P$ is a CP-DCBM where 
$
 \frac{p_{12}}{p_{22}} > \frac{\theta_{p, max}^2}{\theta_c \theta_{p, min}}
$
and $\frac{p_{11}}{p_{12}} > \frac{\theta_{p, max}}{\theta_c}$.

\end{itemize}
\textit{Under each of these alternatives, the power of the test goes to one, i.e., for any $\eta>0$,}
$$
    \lim_{n\to\infty}\mathbb P\left[\{T_1(A) > C_1\}\cap \{T_2(A)>C_2\}\right] > 1-\eta.
$$

\subsubsection*{Under $H_0$:}
Under the null, i.e., when the network is generated from the ER model, we want to prove that
$$
\mathbb P [T_1 > C_1] \rightarrow 0 \text{ and } \mathbb P [T_2 > C_2] \rightarrow 0.
$$
First consider $T_1(A) = T(A, \hat \bc)$.
Recall that from Lemma 1,
$$
\mathbb P [T(A, \hat \bc) - \rho(P_n, \hat \bc) > \varepsilon_n] \rightarrow 0
$$
Note that $\rho(P_n,\bc) = 0$ for all $\bc$, which means $\rho(P_n, \hat \bc) = 0$.
Also note that $C_1 = \varepsilon_n$,
which means 
$$
\mathbb P [T_1 > C_1] 
=
\mathbb P [T_1 > \varepsilon_n]
\rightarrow 0.
$$
Note that we do not need to analyze the second part of the intersection test, since the first part is sufficient to prove that type-1 error converges to zero.

\subsubsection*{Under $H_1$:}
Under the alternative, we want to prove that
$$
\mathbb P [T_1 > C_1] \rightarrow 1 \text{ and } \mathbb P [T_2 > C_2] \rightarrow 1.
$$
First consider $T_1$ and note that, from the definition of $\hat \bc$,
$$
T_1(A) = T(A, \hat \bc) \geq T(A, \bc^*),
$$
and that from Lemma 1,
$$
\mathbb P [T(A, \bc^*) > \rho(P_n, \bc^*) - \varepsilon_n] \rightarrow 1.
$$
We claim that under each of the alternatives listed in the theorem statement,
\begin{equation}
    \rho(P_n,\bc^*) = \mathcal{O} \left(\sqrt{\alpha_n \varrho_n} \right).
    \label{eq-ernullH1}
\end{equation}
whereas, from assumption A1,
$$
\varepsilon_n
=
\sqrt{\alpha_n \varrho_n}
\sqrt{\frac{\log(n\varrho_n\alpha_n)}{n\alpha_n\varrho_n}}
=
o(\sqrt{\alpha_n \varrho_n}),
$$
which implies that
$$
\rho(P_n,\bc^*) > 2 \varepsilon_n.
$$
Therefore,
$$
\mathbb P [T_1 > C_1] 
\geq 
\mathbb P [T(A, \bc^*) >\varepsilon_n] 
\geq
\mathbb P [T(A, \bc^*) > \rho(P_n, \bc^*) - \varepsilon_n]
\rightarrow 1.
$$
It remains to prove that the claim in \eqref{eq-ernullH1} is true.
Recall that
$$
\rho(P_n,\bc^*)
= \frac{\sum_{i<j} \varrho_n (P_{ij}-\bar p)(\Delta_{ij}(\bc^*)-\bar\Delta(\bc^*))}{\sqrt{{n\choose 2} \varrho_n \bar p(1- \varrho_n\bar p){n\choose 2}\bar\Delta(\bc^*) (1-\bar\Delta(\bc^*))}}
    =\frac{\sum_{i<j} \varrho_n (P_{ij}-\bar p)\Delta_{ij}(\bc^*)}{\sqrt{{n\choose 2} \varrho_n \bar p(1- \varrho_n\bar p){n\choose 2}\bar\Delta(\bc^*) (1-\bar\Delta(\bc^*))}}.
$$
Now consider the three alternatives listed in the theorem statement:
\begin{itemize}
    \item $H_1:$ $P_n$ is a two-block CP-SBM where 
$p_{11} > p_{12} = p_{21} > p_{22}$.
Then,
  $$
  {\sum_{i<j} \varrho_n (P_{ij}-\bar p)\Delta_{ij}(\bc^*)}
    \geq
    {n^2 \alpha_n \varrho_n (p_{12}-p_{22})}
     = 
    \mathcal{O} \left(n^2 {\alpha_n \varrho_n} \right).$$

\item $H_1:$ $P_n$ is a a Chung-Lu model where 
$\theta_{(k)} \theta_{(n)} > \theta_{(k+1)}\theta_{(k+2)}$
and $\theta_{(k-1)}>\theta_{(k+1)}$. Then,
  $$
  {\sum_{i<j} \varrho_n (P_{ij}-\bar p)\Delta_{ij}(\bc^*)}
    \geq
    {n^2 \alpha_n \varrho_n (\theta_{(k)} \theta_{(n)} -\theta_{(k+1)}\theta_{(k+2)})}
     = 
    \mathcal{O} \left(n^2 {\alpha_n \varrho_n} \right).$$
    
\item $H_1:$ $P_n$ is a a CP-DCBM where 
$
 \frac{p_{12}}{p_{22}} > \frac{\theta_{p, max}^2}{\theta_c \theta_{p, min}}
$
and $\frac{p_{11}}{p_{12}} > \frac{\theta_{p, max}}{\theta_c}$.
Then,
  $$
  {\sum_{i<j} \varrho_n (P_{ij}-\bar p)\Delta_{ij}(\bc^*)}
    \geq
    {n^2 \alpha_n \varrho_n (p_{12}\theta_c \theta_{p, min} -  p_{22}\theta_{p, max}^2)}
     = 
    \mathcal{O} \left(n^2 {\alpha_n \varrho_n} \right).$$
\end{itemize}
Therefore, in each case, we have
$$
\rho(P_n,\bc^*)
= \frac{\sum_{i<j} \varrho_n (P_{ij}-\bar p)\Delta_{ij}(\bc^*)}{\sqrt{{n\choose 2} \varrho_n \bar p(1- \varrho_n\bar p){n\choose 2}\bar\Delta(\bc^*) (1-\bar\Delta(\bc^*))}}
    =\frac{\mathcal{O} \left(n^2 {\alpha_n \varrho_n} \right)}{\mathcal{O} \left(n^2 \sqrt{\alpha_n \varrho_n} \right)}
    =
    \mathcal{O} \left(\sqrt{\alpha_n \varrho_n} \right),
        $$
which proves the claim made in \eqref{eq-ernullH1}.

Next, consider $T_2 = \hat{p}_{11} - \hat{p}_{12}$.
Define
$$
\bar{p}_{11} = \frac{1}{{k \choose 2}} \sum_{i,j:  c^*_i =  c^*_j =1} P_{ij}, \; \;
\bar{p}_{12} = \frac{1}{{k \choose 2}} \sum_{i,j:  c^*_i =1,  c^*_j =0} P_{ij},
$$
and note that
\begin{equation}
    {k \choose 2} \hat{p}_{11}
= 
\left[
\sum_{i,j: \hat c_i = \hat c_j =1} A_{ij} -
\varrho_n \sum_{i,j: \hat c_i = \hat c_j =1} P_{ij}
\right ]+
\varrho_n\left[\sum_{i,j: \hat c_i = \hat c_j =1} P_{ij} -
\sum_{i,j:  c^*_i =  c^*_j =1} P_{ij}
\right ]+
\varrho_n{k \choose 2} \bar{p}_{11}.
\label{eq:ERnullT2}
\end{equation}
Note that the third term in the above expression is $\mathcal{O} \left(n^2 \alpha^2_n \varrho_n\right)$. We will show that with probability going to one, the third term on the right hand side of \eqref{eq:ERnullT2} dominates the first and second terms.

Consider the first term on the right hand side of \eqref{eq:ERnullT2}.
Similar to the analysis of $\sum_{i,j: \hat c_i = \hat c_j =1} A_{ij}$ under $H_0$,  using Bernstein's inequality and union bound, we obtain
$$
\mathbb P \left[ \max_{\bc} \sum_{i,j: \hat c_i = \hat c_j =1} A_{ij} -
\sum_{i,j: \hat c_i = \hat c_j =1} P_{ij} > - \frac{k^{1.5}}{\sqrt{2}} \varrho_n \log n \right]
\rightarrow 1.
$$
Since $
\alpha_n \geq \frac{(\log (n \log n))^2}{n}
$,
we have ${k^{1.5}} \varrho_n \log n = o(n^2 \alpha^2_n \varrho_n)$, which means that the third term dominates the first term with probability going to one.

\begin{figure}
    \centering
    \includegraphics[width=0.75\linewidth]{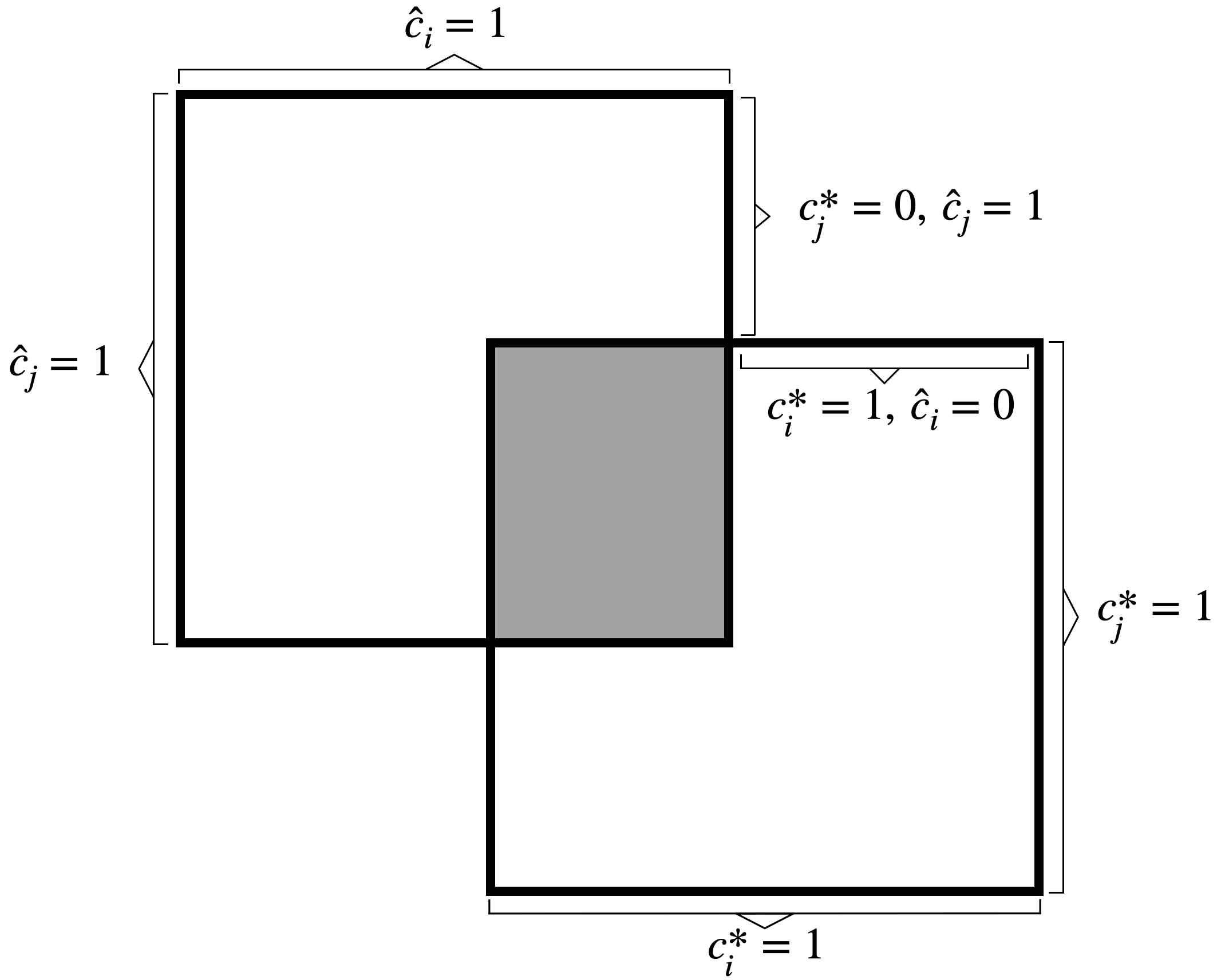}
    \caption{Schematic diagram for the second term}
    \label{fig:term2}
\end{figure}

Next consider the second term on the right hand side of \eqref{eq:ERnullT2}, and refer to the schematic diagram in Figure \ref{fig:term2}.
The shaded part represents the $(i,j)$ pairs where $\hat \bc$ and $\bc^*$ agree.
The summation in the second term is over the $(i,j)$ pairs where $\hat \bc$ and $\bc^*$ disagree, i.e., the unshaded parts of Figure \ref{fig:term2}.
From the definition of $\xi_n({\hat \bc})$, there are $n\xi_n({\hat \bc})$ disagreements between $\hat \bc$ and $\bc^*$.
Let $m$ be the number of vertices such that $c^*_i \neq \hat c_i$.
Since $\hat \bc$ and $\bc^*$ have the same core size, 
$$
m = \frac{n\xi_n({\hat \bc})}{2}.
$$
Therefore, the second term is given by
$$
\varrho_n\left[\sum_{i,j: \hat c_i = \hat c_j =1} P_{ij} -
\sum_{i,j:  c^*_i =  c^*_j =1} P_{ij}
\right ]
=
\mathcal{O} \left(\varrho_n mk \right)
=
\mathcal{O} \left(\varrho_n k n\xi_n({\hat \bc}) \right)
=
\mathcal{O} \left(n^2 \alpha_n \varrho_n \xi_n({\hat \bc}) \right)
=
o_p \left(n^2 \alpha^2_n \varrho_n \right),
$$
where the final step follows from  Theorem 2.1, where we proved that
$$
\xi_n({\hat \bc}) = o_p(\alpha_n)
$$
under each of the three alternatives. Therefore,
the second term is $o_p(n^2 \alpha^2_n \varrho_n)$ under each of the three alternatives.
Since the third term is $\mathcal{O} \left( n^2 \alpha^2_n \varrho_n \right)$, the third term dominates the second term with probability going to one.

Summarizing the above, we conclude that with probability going to one,
$\varrho_n{k \choose 2} \bar{p}_{11}$
dominates the expression for ${k \choose 2}\hat{p}_{11}$.
Therefore, with probability going to one,
$$
\hat{p}_{11} - \varrho_n \bar{p}_{11}
=
{ o(\varrho_n\bar{p}_{11})}
= 
o(\varrho_n)
.
$$
Using the same approach we can show that with probability going to one,
$$
\varrho_n\bar{p}_{12} - \hat{p}_{12} 
=
o(\varrho_n)
.
$$
We skip details given that the proof is very similar.
Next, we show that $\varrho_n (\bar{p}_{11} - \bar{p}_{12})
\geq
\mathcal{O}(\varrho_n)
$ under each of the three alternatives listed in the theorem statement:
\begin{itemize}
    \item $H_1:$ $P_n$ is a two-block CP-SBM where 
$p_{11} > p_{12} = p_{21} > p_{22}$.
Then,
  $$
  \varrho_n (\bar{p}_{11} - \bar{p}_{12})
    =
    \varrho_n ({p}_{11} - {p}_{12})
     = 
   \mathcal{O}(\varrho_n).$$

\item $H_1:$ $P_n$ is a a Chung-Lu model where 
$\theta_{(k)} \theta_{(n)} > \theta_{(k+1)}\theta_{(k+2)}$
and $\theta_{(k-1)}>\theta_{(k+1)}$. Then,
   $$
  \varrho_n (\bar{p}_{11} - \bar{p}_{12})
    \geq
    \varrho_n (\theta_{(k)} \theta_{(k-1)} - \theta_{(k)}\theta_{(k+1)})
     = 
     \varrho_n \theta_{(k)} (\theta_{(k-1)} - \theta_{(k+1)})
     = 
   \mathcal{O}(\varrho_n).$$
    
\item $H_1:$ $P_n$ is a a CP-DCBM where 
$
 \frac{p_{12}}{p_{22}} > \frac{\theta_{p, max}^2}{\theta_c \theta_{p, min}}
$
and $\frac{p_{11}}{p_{12}} > \frac{\theta_{p, max}}{\theta_c}$.
Then,
 $$
  \varrho_n (\bar{p}_{11} - \bar{p}_{12})
    \geq
    \varrho_n (p_{11} \theta^2_c- {p}_{12}\theta_c\theta_{p, max})
     =
     \varrho_n \theta_c(p_{11} \theta_c- {p}_{12}\theta_{p, max})
     =
   \mathcal{O}(\varrho_n).$$
\end{itemize}
Now, under the assumption that 
$
\alpha_n \geq \frac{(\log (n \log n))^2}{n}
$, we have
$
C_2 = o(\varrho_n),
$
whereas
$\varrho_n (\bar{p}_{11} - \bar{p}_{12})
=
\mathcal{O}(\varrho_n)
$
in each case.
This means 
$$
C_2 - \varrho_n (\bar{p}_{11} - \bar{p}_{12})
\le
- \frac{\varrho_n (\bar{p}_{11} - \bar{p}_{12})}{2},
$$
and
$$
\mathbb P \left[ 
\left\{
\left(\hat{p}_{11} - \varrho_n \bar{p}_{11} \right) 
> - \frac{\varrho_n (\bar{p}_{11} - \bar{p}_{12})}{4} \right\}\right]
\rightarrow 1,
\mathbb P \left[ 
\left\{
\left(\varrho_n \bar{p}_{12} - \hat{p}_{12} \right) 
> - \frac{\varrho_n (\bar{p}_{11} - \bar{p}_{12})}{4} \right\}
\right]
\rightarrow 1.
$$
Therefore, 
 \begin{align*}
\mathbb P \left[ T_2(A) > C_2 \right]
&=
\mathbb P \left[ \hat{p}_{11} - \hat{p}_{12} > C_2 \right]
\\
&=
\mathbb P \left[ 
\left(\hat{p}_{11} - \varrho_n\bar{p}_{11} \right) 
+ \left(\varrho_n\bar{p}_{12} - \hat{p}_{12} \right)
+ \varrho_n\left(\bar{p}_{11} - \bar{p}_{12} \right)
> C_2 
\right]
\\
&=
\mathbb P \left[ 
\left(\hat{p}_{11} - \varrho_n\bar{p}_{11} \right) 
+ \left(\varrho_n\bar{p}_{12} - \hat{p}_{12} \right)
> C_2 - \varrho_n\left(\bar{p}_{11} - \bar{p}_{12} \right)
\right]
\\
&\geq
\mathbb P \left[ 
\left(\hat{p}_{11} - \varrho_n\bar{p}_{11} \right) 
+ \left(\varrho_n\bar{p}_{12} - \hat{p}_{12} \right)
> - \frac{\varrho_n (\bar{p}_{11} - \bar{p}_{12})}{2} 
\right]
\\
&\geq
\mathbb P \left[ 
\left\{
\left(\hat{p}_{11} - \varrho_n\bar{p}_{11} \right) 
> - \frac{\varrho_n (\bar{p}_{11} - \bar{p}_{12})}{4} \right\}
\cap
\left\{
\left(\varrho_n\bar{p}_{12} - \hat{p}_{12} \right) 
> - \frac{\varrho_n (\bar{p}_{11} - \bar{p}_{12})}{4} \right\}
\right]
\\
&\rightarrow
1.
\end{align*}
This completes the proof.


\subsection*{Proof of Theorem 3.2}

\noindent
{\sc Theorem 3.2.} \textit{Under $H_0$, i.e., when the network is generated from the CL model, the Type I error rate converges to 0, i.e., for any $\eta>0$,}
$
    \lim_{n\to\infty} \mathbb P\left[\{T_1(A) > C_1\}\cap \{T_2(A)>C_2\}\right] < \eta,
$
\textit{where the rejection thresholds are given by}
\begin{equation}
    C_1 = \rho(\hat{P}, \hat c) + \tilde \epsilon_n + \varepsilon'_n,
    \text{ where }
\tilde{\epsilon}_n = 
\frac{\sqrt{\alpha_n} \log (n \alpha_n)}{n^{1.5}\sqrt{\varrho_n}}
, \text{ and }
\varepsilon'_n =\frac{\sqrt{\log (n \alpha_n \varrho_n)}}{n}.    
    \label{eq:inter_cl}
\end{equation}
{\it Under $H_0$, i.e., when the network is generated from the CP-DCBM, 
consider the three following assumptions.
\begin{enumerate}
    \item $\varepsilon'_n + \tilde \epsilon_n = o ({\alpha_n^{1.5} \varrho_n^{0.5}})$,
where 
$
\tilde{\epsilon}_n = 
\frac{\sqrt{\alpha_n} \log (n \alpha_n)}{n^{1.5}\sqrt{\varrho_n}},
\varepsilon'_n =\frac{\sqrt{\log (n \alpha_n \varrho_n)}}{n}
$.
\item There exists $\epsilon >0$ such that 
$
\frac{p^2_{12}-p_{11}p_{22}}{n^4 \alpha^2_n \varrho_n}
\left(
S_{10}^2 + S_{01}^2 - S_{11}S_{00} 
\right)
> \epsilon
$,
where
$
S_{kl} = \sum_{{(i,j):c^*_i = k,c^*_j = l}} \theta_i \theta_j,
$
for $k, l = 0,1$.
\item Similar to Theorem 2.1, let $\theta_c = \min\{\theta_i: c^*_i = 1 \}$ be the lowest degree parameter in the core set, and $\theta_{p, max} = \max\{\theta_i: c^*_i =0\}$ and $\theta_{p, min} = \min\{\theta_i: c^*_i =0\}$ be the highest and lowest degree parameters in the periphery set, respectively.
Then $
 \frac{p_{12}}{p_{22}} > \frac{\theta_{p, max}^2}{\theta_c \theta_{p, min}}
$.
\end{enumerate}
Under these three assumptions, the power of the test goes to one, i.e., for any $\eta>0$,}
$$
    \lim_{n\to\infty}\mathbb P\left[\{T_1(A) > C_1\}\cap \{T_2(A)>C_2\}\right] > 1-\eta.
$$

To prove this theorem, we first prove two results, \eqref{eq:thm3.2} and \eqref{eq:thm3.2_2}, that apply to both the null and the alternative.
Recall that we estimate the degree parameters as
$$
\hat {\theta}_i = \frac{d_i}{\sqrt{2m}},
$$
where $2m = \sum_i \sum_j A_{ij}$ is the total degree of the network. 

To make the derivations convenient to follow, we will use three ``short cuts.''
First, note that $2m = \sum_i \sum_j A_{ij}$ is very tightly concentrated around its mean, $\sum_i \sum_j P_{n,ij}$, and the error bound for $2m$ (or its scaled versions, such as $\hat p$) is much smaller than error bounds for the other (appropriately scaled) random variables.
Therefore, in what follows, we will ignore the randomness of $2m$, i.e., consider it to be approximately equal to $\sum_i \sum_j P_{n,ij}$.
Second, for any $i \neq j$ we will consider $d_i$ and $d_j$ to be independent for pedagogical convenience, although they share a common $A_{ij}$.
This is justified because the correlation between $d_i$ and $d_j$ goes to zero at the rate of $\mathcal{O}(1/n)$.
Whenever we use the first or second short cut we will flag it with the ``$\approx$'' sign.
Third, we will drop the subscript $n$ from expressions such as $P_{n,ij}$ for notational convenience.

We first derive the expectation of $\hat{P}_{ij} = \hat {\theta}_i \hat {\theta}_j$ under $H_0$ and $H_1$.
Under $H_0$, we have
$$
\mathbb{E}[\hat {\theta}_i \hat {\theta}_j] 
=
\mathbb{E} \left[\frac{d_i d_j}{2m}\right]
\approx
\frac{\mathbb{E}[d_i] \mathbb{E}[d_j]}{\sum_i \sum_j P_{ij}} 
=
\frac{(\theta_i  \sum_{k=1}^n \theta_k)  (\theta_j  \sum_{k=1}^n \theta_k)}{(\sum_{k=1}^n \theta_k)^2} 
= \theta_i \theta_j,
$$
and under $H_1$, we have
$$
\mathbb{E}[\hat {\theta}_i \hat {\theta}_j] 
=
\mathbb{E} \left[\frac{d_i d_j}{2m}\right]
\approx
\frac{\mathbb{E}[d_i] \mathbb{E}[d_j]}{\sum_i \sum_j P_{ij}} 
=
\frac{(\theta_i  \sum_{k=1}^n p_{c^*_i c^*_k} \theta_k)  (\theta_j  \sum_{l=1}^n p_{c^*_j c^*_l} \theta_l)}{\sum_{k=1}^n\sum_{l=1}^n \theta_k \theta_l p_{c^*_k c^*_l}} 
.
$$
Now, define 
$$
\tilde{P}_{ij}
=
\begin{cases}
			\theta_i \theta_j & \text{if $H_0$ is true}\\
          \theta_i \theta_j 
          \frac{\sum_{k=1}^n\sum_{l=1}^n \theta_k \theta_l p_{c^*_i c^*_k}p_{c^*_j c^*_l}}
          {\sum_{k=1}^n\sum_{l=1}^n \theta_k \theta_l p_{c^*_k c^*_l}}  
          & \text{if $H_1$ is true}
		 \end{cases}
$$
for $i,j = 1, \ldots, n$.
Then we have $
\mathbb{E}[\hat{P}_{ij}] = \tilde{P}_{ij}
$
under both $H_0$ and $H_1$.
Note that $\tilde{P}_{ij}={P}_{ij}$ for each $i,j$ under $H_0$ but not under $H_1$. 
However, $\sum_{ij}\tilde{P}_{ij}=\sum_{ij}{P}_{ij}$ under both $H_0$ and $H_1$.
This is a key observation that we will leverage in what follows.

Next, we state and prove the following result: under both $H_0$ and $H_1$,
\begin{equation}
    \mathbb P \left[|\rho(\hat{P}, \bc^*) - \rho(\tilde{P}, \bc^*)| \le \tilde{\epsilon}_n \right] \rightarrow 0
    \label{eq:thm3.2}
\end{equation}
as $n \rightarrow \infty$, where 
$$
\tilde{\epsilon}_n = \mathcal{O}\left(
\frac{\sqrt{\alpha_n} \log (n \alpha_n)}{n^{1.5}\sqrt{\varrho_n}}
\right).
$$
To prove \eqref{eq:thm3.2},
write
$$
\rho(\hat{P}, \bc^*) = \frac{X_1 - X_2}{X_3},
$$
where
$$
X_1 = \sum_{i<j} \hat{P}_{ij} \Delta_{ij}
=
\sum_{i<j} \frac{d_i d_j}{2m} \Delta_{ij}
\approx
 \frac{\sum_{i<j} d_i d_j \Delta_{ij}}{\sum_i \sum_j P_{ij}},
$$
$$
X_2 = {n \choose 2} \bar\Delta \hat {p}
\approx
\sum_{i<j} {n \choose 2} \bar\Delta \bar {p},
\text{ and }
$$
$$
X_3 = {n \choose 2} \sqrt{\bar\Delta (1-\bar\Delta) \hat {p} (1-\hat {p})}
\approx
{n \choose 2} \sqrt{\bar\Delta (1-\bar\Delta) \bar {p} (1-\bar {p})}.
$$
Similarly, write
$$
\rho(\tilde{P}, \bc^*) = \frac{\mu_1 - \mu_2}{\mu_3},
$$
where
$$
\mu_1 = \sum_{i<j} \tilde{P}_{ij} \Delta_{ij},
\mu_2 = 
\sum_{i<j} {n \choose 2} \bar\Delta \bar {p},
\text{ and }
\mu_3 = 
{n \choose 2} \sqrt{\bar\Delta (1-\bar\Delta) \bar {p} (1-\bar {p})}.
$$
Note that $\mathbb{E}[X_1] \approx \mu_1$, $X_2 \approx \mu_2$, and $X_3 \approx \mu_3$. 
Therefore,
\begin{equation}
    |\rho(\hat{P}, \bc^*) - \rho(\tilde{P}, \bc^*)|
=
\left| \frac{X_1 - X_2}{X_3} - \frac{\mu_1 - \mu_2}{\mu_3} \right|
\approx
\left| \frac{X_1 - \mathbb{E}[X_1]}{\mu_3} \right|.
\label{eq:thm3.2_1}
\end{equation}
Define
$Y = X_1 \sum_i \sum_j P_{ij} 
= 
\sum_{i<j} d_i d_j \Delta_{ij}
$.
Now, without loss of generality, suppose that $c^*_i = 1$ for $i=1, \ldots, k$ and
$c^*_i =0$ for $i=k+1, \ldots, n$.
Then we can write
\begin{align*}
    Y
&=
\sum_{i<j} d_i d_j \Delta_{ij}\\
&=
d_1(d_2 + \ldots + d_n) +
d_2(d_3 + \ldots + d_n) + \ldots +
d_k(d_{k+1} + \ldots + d_n)\\
&=
U_1 + \ldots + U_k,
\end{align*}
where $U_i = d_i (d_{i+1} + \ldots + d_n)$ for $i=1, \ldots, k$.
Note that while the $U_i$'s are dependent across $i=1, \ldots, k$, each $U_i$ is a sum of independent Bernoulli random variables, since we are working under the approximation that $d_i$ and $d_j$ are independent for $i\neq j$.
Therefore for each $i=1, \ldots, k$, using Bernstein's inequality,
$$
\mathbb P \left[ |U_i - \mathbb E(U_i)| > t \right]
\le
2 \exp\left(- \frac{t^2/2}{\mathbb E(U_i) + t/3} \right).
$$
Note that $\mathbb E(U_i) = \mathcal{O}(n^3 \varrho_n^2)$.
Therefore, by setting $t =  n^{1.5} \varrho_n \log (n \alpha_n)$ and using union bound over $i=1, \ldots, k$, we obtain that
\begin{align*}
    \cup_{i=1}^k \mathbb P \left[ |U_i - \mathbb E(U_i)| > t \right]
&\le
2 \sum_{i=1}^k \exp\left(- \frac{t^2/2}{\mathbb E(U_i) + t/3} \right)\\
&=
2 \exp\left(- \frac{\mathcal{O}(n^3 \varrho_n^2 \log^2 (n \alpha_n))}
{\mathcal{O}(n^3 \varrho_n^2) + \mathcal{O}(n^{1.5} \varrho_n \log (n \alpha_n))} + \log k\right)\\
&=
2 \exp\left(- \frac{\mathcal{O}(n^3 \varrho_n^2 \log^2 (n \alpha_n))}
{\mathcal{O}(n^3 \varrho_n^2) + \mathcal{O}(n^{1.5} \varrho_n \log (n \alpha_n))} + \log (n\alpha_n) \right)\\
&\rightarrow 0,
\end{align*}
which implies that
$$
\mathbb P \left[ |Y - \mathbb E(Y)| > kt \right] \rightarrow 0
\Rightarrow
\mathbb P \left[ |X_1 - \mathbb E(X_1)| > \frac{kt}{\mathcal{O}(n^2 \varrho_n)} \right] \rightarrow 0.
$$
Recall that from \eqref{eq:thm3.2_1},
$$
|\rho(\hat{P}, \bc^*) - \rho(\tilde{P}, \bc^*)|
\approx
\left| \frac{X_1 - \mathbb{E}[X_1]}{\mu_3} \right|
=
\left| \frac{X_1 - \mathbb{E}[X_1]}{\mathcal{O}(n^2 \sqrt{\alpha_n \varrho_n})} \right|.
$$
Therefore,
with probability going to 1,
$$
|\rho(\hat{P}, \bc^*) - \rho(\tilde{P}, \bc^*)|
\le
\frac{kt}{\mathcal{O}(n^2 \varrho_n)\mathcal{O}(n^2 \sqrt{\alpha_n \varrho_n})}
=
\frac{n^{2.5} \alpha_n \varrho_n \log (n \alpha_n)}
{\mathcal{O}(n^4 \varrho_n\sqrt{\alpha_n \varrho_n})}
= 
\mathcal{O}\left(
\frac{\sqrt{\alpha_n} \log (n \alpha_n)}{n^{1.5}\sqrt{\varrho_n}}
\right).
$$
This completes the proof of \eqref{eq:thm3.2}.

Next, we state and prove the following result: under both $H_0$ and $H_1$,
\begin{equation}
    \mathbb P \left[|T(A, \bc^*) - \rho({P}, \bc^*)| \le \varepsilon'_n \right] \rightarrow 0
    \label{eq:thm3.2_2}
\end{equation}
as $n \rightarrow \infty$, where 
$$
\varepsilon'_n =\frac{\sqrt{\log (n \alpha_n \varrho_n)}}{n}.
$$
To prove \eqref{eq:thm3.2_2},
we will essentially mimic the proof of Lemma 1.
Let $\bc = \bc^*$ and $X_n(\bc)=\sum_{i<j}A_{ij}(\Delta_{ij}-\bar\Delta_n)$ and $Y_n=\sum_{i<j}A_{ij}$. Then
$$
    T(A_n,\bc) = \frac{X_n}{\sqrt{Y_n({n\choose 2}-Y_n)}}\frac{1}{\sqrt{\bar\Delta_n(1-\bar\Delta_n)}}
    := f(X_n, Y_n). 
$$
We will prove the desired result using two pieces. First, we show that $X_n(\bc)$ and $Y_n$ are ``close'' to their respective expectations, $\mu_x(\bc)$ and $\mu_y$, for all $\bc$. Then we will show that $f(x,y)$ is ``close'' to $f(x^*,y^*)$ when $x\approx x^*$ and $y\approx y^*$.
\newline

\noindent
First, Bernstein's inequality states that if $\mathsf{E}(X_i)=0$ and $|X_i|\leq M$ for $i=1, \ldots, n$, then
$$
    \mathbb P\left(\sum_{i=1}^n X_i >t\right)
    \leq \exp\left(-\frac{\frac12t^2}{\sum_{i=1}^n \mathsf{E}(X_i^2)+\frac13 Mt}\right).
$$
Thus, 
$$
    \mathbb P\left(|X_n(\bc)-\mu_x(\bc)| >t_1\right)
    \leq 2\exp\left(-\frac{\frac12t_1^2}{\sum_{i<j} P_{ij}(1-P_{ij})(\Delta_{ij}-\bar\Delta_n)^2+\frac13t_1}\right)
    \sim \exp\left(-\frac{t_1^2}{\varrho_n\alpha_n n^2 + t_1}\right)
$$
since $P_{ij}(1-P_{ij})=\mathcal O(\varrho_n)$ and $\sum_{i<j}(\Delta_{ij}-\bar\Delta_n)^2=\mathcal O(\alpha_nn^2)$. 
Let 
$t_1=n \sqrt{\varrho_n\alpha_n\log(n\varrho_n\alpha_n)}$. 
Then 
$$
    \mathbb P(|X_n-\mu_x(\bc)| > t_1)
    \leq 2\exp\left(-\frac{n^2 {\varrho_n\alpha_n}\log(n\varrho_n\alpha_n)}{\varrho_n\alpha_nn^2}\right) 
    \leq 2\exp( - \log(n\varrho_n\alpha_n))
    \to 0
$$
since $n\varrho_n\alpha_n\to\infty$ by A1.
Additionally, 
$$
    \mathbb P\left(|Y_n-\mu_y| >t_2\right)
    \leq 2\exp\left(-\frac{\frac12t_2^2}{\sum_{i<j} P_{ij}(1-P_{ij})+\frac13t_2}\right)
    \sim 2\exp\left(-\frac{t_2^2}{\varrho_n n^2 + t_2}\right).
$$
Additionally, letting $t_2=n\sqrt{\varrho_n \log(n\varrho_n\alpha_n)}$,
$$
    \mathbb P(|Y_n-\mu_y| > n\sqrt{\varrho_n \log(n\varrho_n\alpha_n)})
    \leq 2\exp\left(-\frac{n^2{\varrho_n \log(n\varrho_n\alpha_n)}}{\varrho_nn^2 +  n\sqrt{\varrho_n \log(n\varrho_n\alpha_n)}}\right) 
    \leq 2\exp(- \log(n\varrho_n\alpha_n))
    \to 0.
$$
Thus, for large $n$, $X_n(\bc)$ and $Y_n$ are ``close'' to their respective expectations for all $\bc$.
\newline

\noindent
For the second step, recall that 
$$
    f(x,y)
    =\frac{x}{\sqrt{y(n^2-y)}}\frac{1}{\sqrt{\bar\Delta_n(1-\bar\Delta_n)}}.
$$
Then we want to show that $|f(x,y)-f(x^*,y^*)|$ is upper bounded by a small constant when $|x-x^*|<t_1$ and $|y-y^*|<t_2$. Consider the first-order Taylor expansion:
$$
    f(x,y)
    -f(x^*,y^*)
    =(x-x^*)\frac{\partial}{\partial x}f(x,y,n)\Big|_{x=x^*,y=y^*} + (y-y^*)\frac{\partial}{\partial y}f(x,y,n)\Big|_{x=x^*,y=y^*} + R_n,
$$
where
$$
    R_n \sim (x-x^*,y-y^*)
    \begin{pmatrix}
        \partial_x^2 f & \partial_{xy} f\\
        \partial_{xy}f & \partial_y^2 f
    \end{pmatrix}
    \begin{pmatrix}
        x-x^*\\y-y^*
    \end{pmatrix}.
$$
We have that $x^*= \mathcal O(\alpha_n \varrho_n n^2)$ and $y^*=\mathcal O(\varrho_n n^2)$. It's also easy to see that
$$
    \frac{\partial}{\partial x}f(x,y)\Big|_{x=x^*,y=y^*}
    = \frac{1}{\sqrt{y^*(n^2-y^*)}}\frac{1}{\sqrt{\bar\Delta_n(1-\bar\Delta_n)}}
    \sim \alpha_n^{-1/2}\varrho_n^{-1/2}n^{-2},
$$
and
$$
    \frac{\partial}{\partial y}f(x,y)\Big|_{x=x^*,y=y^*}
    =\frac{x^*(n^2-2y^*)}{2\{y^*(n^2-y^*)\}^{3/2}}\frac{1}{\sqrt{\bar\Delta_n(1-\bar\Delta_n)}}
    \sim \alpha_n^{1/2} \varrho_n^{-1/2} n^{-2}.
$$
From step 1, $|x-x^*|< t_1 = n \sqrt{\varrho_n\alpha_n\log(n\varrho_n\alpha_n)}$ and $|y-y^*| < t_2 =n\sqrt{\varrho_n \log(n\varrho_n\alpha_n)}$ with high probability, whuich means that with probability going to one,
$$
    |f(x,y)
    -f(x^*,y^*)|
    \le \frac{n \sqrt{\varrho_n \alpha_n \log(n\varrho_n\alpha_n)}}{n^2 \sqrt{\alpha_n\varrho_n}} +
    \frac{n\sqrt{\alpha_n \varrho_n \log(n\varrho_n\alpha_n)}}{\varrho_n^{1/2}n^2}
    \sim \frac{\sqrt{\log(n \alpha_n \varrho_n)}}{n}
    = \varepsilon'_n.
$$
This completes the proof of \eqref{eq:thm3.2_2}.


Next, we will use  \eqref{eq:thm3.2} and \eqref{eq:thm3.2_2} to prove the desired convergence of type-1 error and power under $H_0$ and $H_1$, respectively.

\subsubsection*{Under $H_0$:}
Note that since $
{\alpha_n}/{\sqrt{\varrho_n}} = o \left({1}/{\sqrt{n \log n }} \right),
$ strong consistency is achieved as per Theorem 2.1.
Recall that
$C_1 = \rho(\hat{P}, \bc^*) + \varepsilon'_n + \tilde \epsilon_n$.
Under $H_0$, we have
\begin{align*}
 T_1 
&=
T(A, \hat \bc)\\
&=
T(A, \bc^*) \text{ with probability going to one, since strong consistency holds}\\
&\le
\rho(P,\bc^*) + \varepsilon'_n \text{ with probability going to one, by \eqref{eq:thm3.2_2}}\\
&=
\rho(\tilde P,\bc^*) + \varepsilon'_n \text{ since $\tilde P = P$ under $H_0$}\\
&\le
\rho(\hat{P}, \bc^*) + \varepsilon'_n + \tilde \epsilon_n \text{ with probability going to one, by \eqref{eq:thm3.2}}\\
&=\rho(\hat{P}, \hat \bc) + \varepsilon'_n + \tilde \epsilon_n \text{ with probability going to one, since strong consistency holds}\\
&=
C_1,
\end{align*}
which completes the proof of convergence of type-1 error.
Note that we do not need to analyze the second part of the intersection test, since the first part is sufficient to prove that type-1 error converges to zero.

\subsubsection*{Under $H_1$:}
Note that since $
{\alpha_n}/{\sqrt{\varrho_n}} = o \left({1}/{\sqrt{n \log n }} \right),
$ strong consistency is achieved as per Theorem 2.3.
Here we will need to show that both $\mathbb{P} [T_1 > C_1]$ and $\mathbb{P} [T_2 > C_2]$ go to one, which will prove that the power of the intersection test goes to one.

First, consider $T_1$ and
recall that
$C_1 = \rho(\hat{P}, \bc^*) + \varepsilon'_n + \tilde \epsilon_n$.
Also recall that under $H_1$, 
$
\mathbb{E}[\hat{P}_{ij}] = \tilde{P}_{ij}
\neq {P}_{ij}$.
The proof hinges on this key observation, and we start by carefully quantifying the difference between ${P}_{ij}$ and $\tilde{P}_{ij}$.
To this end, note that $\sum_{ij}\tilde{P}_{ij}=\sum_{ij}{P}_{ij}$, which means
$$
\rho(P,\bc^*) - \rho(\tilde{P},\bc^*)
=
\frac{\sum_{i<j} \left(P_{ij} - \tilde{P}_{ij}\right)\Delta_{ij}}{\frac12n(n-1)\{\bar P(1-\bar P)\bar\Delta(1-\bar\Delta)\}^{1/2}}.
$$
Recall that 
$
S_{kl} = \sum_{{(i,j):c^*_i = k,c^*_j = l}} \theta_i \theta_j,
$
for $k, l = 1,2$, and consider the numerator of the above expression.
\begin{align*}
    \sum_{i<j} \left(P_{ij} - \tilde{P}_{ij}\right)\Delta_{ij}
    &=
    \sum_{i<j}
      \left(       \theta_i \theta_j p_{c^*_i c^*_j}
    -
    \theta_i \theta_j 
          \frac{\sum_{k=1}^n\sum_{l=1}^n \theta_k \theta_l p_{c^*_i c^*_k}p_{c^*_j c^*_l}}
          {\sum_{k=1}^n\sum_{l=1}^n \theta_k \theta_l p_{c^*_k c^*_l}}
    \right)\Delta_{ij}\\
    &=
    \frac{1}{2}\sum_{i,j}
    \theta_i \theta_j p_{c^*_i c^*_j}\Delta_{ij}
     \left(1-\frac{\sum_{k=1}^n\sum_{l=1}^n \theta_k \theta_l p_{c^*_i c^*_k}p_{c^*_j c^*_l}}
          {\sum_{k=1}^n\sum_{l=1}^n \theta_k \theta_l p_{c^*_k c^*_l}p_{c^*_i c^*_j}}
    \right)\\
 &=
    \frac{1}{2}\sum_{i,j}
    \theta_i \theta_j p_{c^*_i c^*_j}\Delta_{ij}
     \left(\frac{\sum_{k,l=1}^n \theta_k \theta_l 
     \left(p_{c^*_k c^*_l}p_{c^*_i c^*_j} - p_{c^*_i c^*_k}p_{c^*_j c^*_l} 
     \right)}
          {\sum_{k,l=1}^n \theta_k \theta_l p_{c^*_k c^*_l}p_{c^*_i c^*_j}}
    \right)\\
    &=
    \frac{1}{2\sum_{k,l=1}^n \theta_k \theta_l p_{c^*_k c^*_l}}\sum_{i,j}
    \theta_i \theta_j \Delta_{ij}
     \left({\sum_{k,l=1}^n \theta_k \theta_l 
     \left(p_{c^*_k c^*_l}p_{c^*_i c^*_j} - p_{c^*_i c^*_k}p_{c^*_j c^*_l} 
     \right)}
          {}
    \right)\\
    &=
    \frac{1}{2n^2 \bar p}
    \sum_{i,j}
    \theta_i \theta_j \Delta_{ij}
     \left({\sum_{k,l=1}^n \theta_k \theta_l 
     \left(p_{c^*_k c^*_l}p_{c^*_i c^*_j} - p_{c^*_i c^*_k}p_{c^*_j c^*_l} 
     \right)}
          {}
    \right)\\
&=
 \frac{p^2_{12}-p_{11}p_{22}}{2n^2 \bar p}
\left(
S_{12}^2 + S_{21}^2 - S_{11}S_{22} 
\right) \\
&>
\epsilon n^2 \alpha^2_n \varrho_n
\end{align*}
from the condition in the theorem statement.
This implies that
\begin{equation}
\rho(P,\bc^*) - \rho(\tilde{P},\bc^*)
=
\frac{\sum_{i<j} \left(P_{ij} - \tilde{P}_{ij}\right)\Delta_{ij}}{{n \choose 2}\{\bar P(1-\bar P)\bar\Delta(1-\bar\Delta)\}^{1/2}}
>
\frac{\epsilon n^2 \alpha^2_n \varrho_n}{\mathcal{O}(n^2 \sqrt{\alpha_n \varrho_n})}
>
\mathcal{O}({\alpha_n^{1.5} \varrho_n^{0.5}})
>> \varepsilon'_n + \tilde \epsilon_n.
\label{eq:thm3.2_3}    
\end{equation}
where the last two steps follow from the assumptions in the theorem.
Therefore, under $H_1$,
\begin{align*}
T_1 
&=
T(A, \hat \bc)\\
&\geq
T(A, \bc^*) \text{ by definition of $\hat \bc$}\\
&\geq
\rho(P,\bc^*) - \varepsilon'_n \text{ with probability going to one, by \eqref{eq:thm3.2_2}}\\
&=
\left(\rho(\hat{P}, \bc^*) + \tilde \epsilon_n + \varepsilon'_n\right)  
+
\left(\rho(P,\bc^*) - \rho(\tilde P,\bc^*) - (2\tilde \epsilon_n + \varepsilon'_n)\right) 
+ 
\left(\rho(\tilde P,\bc^*) - \rho(\hat{P}, \bc^*) \right) 
 \\
&\geq
\left(\rho(\hat{P}, \bc^*) + \tilde \epsilon_n + \varepsilon'_n \right)  
 +
\left(\rho(P,\bc^*) - \rho(\tilde P,\bc^*) - (2\tilde \epsilon_n + \varepsilon'_n)\right)
- \tilde \epsilon_n
\text{ with prob $\rightarrow 1$, by \eqref{eq:thm3.2}}\\
&=
\left(\rho(\hat{P}, \bc^*) + \tilde \epsilon_n + \varepsilon'_n \right) 
 +
\left(\rho(P,\bc^*) - \rho(\tilde P,\bc^*) - 2(\tilde \epsilon_n + \varepsilon'_n)\right)
\\
&>
\rho(\hat{P}, \bc^*) + \tilde \epsilon_n + \varepsilon'_n \text{ by \eqref{eq:thm3.2_3}}\\
&=
\rho(\hat{P}, \hat \bc) + \tilde \epsilon_n + \varepsilon'_n \text{ with probability going to one, since strong consistency holds} \\
&=
C_1,
\end{align*}
which completes the proof of convergence of power for the first part of the intersection test.


For the second part of the intersection test,
we will essentially mimic the second part of the proof of Theorem 3.1.
Consider $T_2 = \hat{p}_{11} - \hat{p}_{12}$, and note that when $\hat \bc = \bc^*$,
$$
{k \choose 2} \hat{p}_{11}
= 
\sum_{i,j: c^*_i = c^*_j =1} (A_{ij} -
P_{ij})
+
\sum_{i,j: c^*_i = c^*_j =1} P_{ij}
.
$$
For the first term, using Bernstein's inequality we obtain
$$
\mathbb P \left[ |\sum_{i,j: c^*_i = c^*_j =1} (A_{ij} -
 P_{ij}| > n \alpha_n \sqrt{\varrho_n \log (n \alpha_n \varrho_n)} \right]
\rightarrow 1.
$$
Note that $\sum_{i,j: c^*_i = c^*_j =1} P_{ij} = \mathcal{O}(n^2 \alpha_n^2 \varrho_n) >> n \alpha_n \sqrt{\varrho_n \log (n \alpha_n \varrho_n)}$, which means that the second term dominates the first term with probability going to one.
Therefore, with probability going to one,
$$
\hat{p}_{11} - \bar {p}_{11} 
=
o(\bar {p}_{11})
=
o(\varrho_n),
$$
where
$$
\bar {p}_{11} = 
\frac{1}{{k \choose 2}} \sum_{i,j:  c^*_i =  c^*_j =1} P_{ij}.
$$
Using the same approach we can show that with probability going to one,
$$
\bar{p}_{12} - \hat{p}_{12} 
=
o(\bar{p}_{12})
=
o(\varrho_n),
$$
where
$$
\bar {p}_{11} = 
\frac{1}{{k (n-k)}} \sum_{i,j:  c^*_i = 1,  c^*_j =0} P_{ij}.
$$
We skip details given that the proof is very similar.
Finally, note that under $H_1$ and the assumption that 
$
\alpha_n \geq \frac{(\log (n \log n))^2}{n}
$, we have
$
C_2 = o(\varrho_n),
$
whereas
$(\bar{p}_{11} - \bar{p}_{12})
=
\mathcal{O}(\varrho_n)
$
from the third assumption in the theorem.
This means 
$$
C_2 - (\bar{p}_{11} - \bar{p}_{12})
\le
- \frac{(\bar{p}_{11} - \bar{p}_{12})}{2},
$$
and
$$
\mathbb P \left[ 
\left\{
\left(\hat{p}_{11} - \bar{p}_{11} \right) 
> - \frac{(\bar{p}_{11} - \bar{p}_{12})}{4} \right\}\right]
\rightarrow 1,
\mathbb P \left[ 
\left\{
\left(\bar{p}_{12} - \hat{p}_{12} \right) 
> - \frac{(\bar{p}_{11} - \bar{p}_{12})}{4} \right\}
\right]
\rightarrow 1.
$$
Therefore, 
 \begin{align*}
\mathbb P \left[ T_2(A) > C_2 \right]
&=
\mathbb P \left[ \hat{p}_{11} - \hat{p}_{12} > C_2 \right]
\\
&=
\mathbb P \left[ 
\left(\hat{p}_{11} - \bar{p}_{11} \right) 
+ \left(\bar{p}_{12} - \hat{p}_{12} \right)
+ \left(\bar{p}_{11} - \bar{p}_{12} \right)
> C_2 
\right]
\\
&=
\mathbb P \left[ 
\left(\hat{p}_{11} - \bar{p}_{11} \right) 
+ \left(\bar{p}_{12} - \hat{p}_{12} \right)
> C_2 - \left(\bar{p}_{11} - \bar{p}_{12} \right)
\right]
\\
&\geq
\mathbb P \left[ 
\left(\hat{p}_{11} - \bar{p}_{11} \right) 
+ \left(\bar{p}_{12} - \hat{p}_{12} \right)
> - \frac{(\bar{p}_{11} - \bar{p}_{12})}{2} 
\right]
\\
&\geq
\mathbb P \left[ 
\left\{
\left(\hat{p}_{11} - \bar{p}_{11} \right) 
> - \frac{ (\bar{p}_{11} - \bar{p}_{12})}{4} \right\}
\cap
\left\{
\left(\bar{p}_{12} - \hat{p}_{12} \right) 
> - \frac{ (\bar{p}_{11} - \bar{p}_{12})}{4} \right\}
\right]
\\
&\rightarrow
1.
\end{align*}
This completes the proof.

\end{document}